\documentclass[aps,twocolumn,nofootinbib,preprintnumbers,superscriptaddress]{revtex4}

\usepackage{color}
\usepackage{bbm}
\usepackage{tikz}
\usepackage{braket}
\usepackage{comment}

\usepackage[
pagebackref=false,
colorlinks=true,
linkcolor=blue,
urlcolor=blue,
filecolor=black,
citecolor=red,
pdfstartview=FitV,
pdftitle={},
pdfauthor={},
pdfsubject={},
pdfkeywords={},
pdfpagemode=None,
bookmarksopen=true
]{hyperref}

\usepackage[normalem]{ulem}
\usepackage{amsmath, amssymb, bm}
\usepackage{enumerate}
\usepackage{amsfonts}
\usepackage{epsfig}
\usepackage{mathbbol}
\usepackage{mathrsfs}
\definecolor{darkgreen}{rgb}{0.0, 0.5, 0.0}
\newcommand{\dg}{\dagger}
\newcommand{\ra}{\rightarrow}
\newcommand{\tb}{\quad}

\begin{document}

\title{\textbf{Multipartitioning topological phases by vertex states and 
quantum entanglement 
}}

\author{Yuhan Liu}
\affiliation{Kadanoff Center for Theoretical Physics, University of Chicago, Chicago, IL~60637, USA}
\affiliation{James Franck Institute, University of Chicago, Chicago, Illinois 60637, USA}

\author{Ramanjit Sohal}
\affiliation{ Department of Physics, Princeton University, Princeton, New Jersey, 08540, USA}

\author{Jonah Kudler-Flam}
\affiliation{Kadanoff Center for Theoretical Physics, University of Chicago, Chicago, IL~60637, USA}
\affiliation{James Franck Institute, University of Chicago, Chicago, Illinois 60637, USA}

\author{Shinsei Ryu}
\affiliation{ Department of Physics, Princeton University, Princeton, New Jersey, 08540, USA}

\begin{abstract}
We discuss multipartitions of the gapped ground states of (2+1)-dimensional topological
liquids into three (or more)
spatial regions that are adjacent to each other and meet at points.
By considering the reduced density matrix 
obtained by tracing over a subset of the regions, 
we compute various correlation measures, such as
entanglement negativity, reflected entropy, and associated spectra.
We utilize the bulk-boundary
correspondence to show that such multipartitions can be achieved by 
using what we call \textit{vertex states}
in (1+1)-dimensional conformal field theory -- these
are a type of state used to define an interaction vertex in string field theory
and can be thought of as a proper generalization of conformal boundary states.
This approach allows an explicit construction
of the reduced density matrix near the
entangling boundaries.
We find the fingerprints of topological liquid in these quantities, such as
(universal pieces in) the scaling of the entanglement negativity, and a
non-trivial distribution of the spectrum of the partially transposed density
matrix.
For reflected entropy, we test the recent
claim that 
states the difference between reflected entropy and mutual information is given, 
once short-range correlations are properly removed,
by
$(c/3)\ln 2$ where $c$ is 
the central charge of the topological liquid
that measures ungappable
edge degrees of freedom.
As specific examples, we consider topological chiral $p$-wave superconductors and Chern insulators. We also study a specific lattice fermion model realizing Chern insulator phases and calculate the 
correlation measures numerically, both in its gapped phases and at critical points separating them.
\end{abstract}

%
\maketitle
\tableofcontents

%
%
%

\section{Introduction}
\label{intro}

``Quantum entanglement is not \textit{one} but \textit{the} characteristic trait of quantum mechanics, the one that enforces its entire departure from classical lines of thought'' \cite{schrodinger_1935}.
Entanglement also plays a central role in
understanding various phenomena and phases in many-body quantum physics. For example, the scaling of the entanglement entropy defined for a given subregion is a useful probe to understand different phases of matter and renormalization group flows connecting them \cite{2003PhRvL..90v7902V,2009JPhA...42X4005C,levin2006detecting,kitaev2006topological,2007JSMTE..08...24H,2007JPhA...40.7031C, 2018RvMP...90c5007N}.   
Modern approaches to many-body quantum problems, such as the density matrix renormalization group and tensor networks, are based on the concept of quantum entanglement 
\cite{Fradkin:2013sab, 2013NJPh...15b5002G, LAFLORENCIE20161, zeng2018quantum, Verstraete_2008}.

Quantum entanglement is particularly useful
for characterizing 
topological phases of matter, 
which lack conventional order parameters.
One of the simplest settings to consider is
a bipartition of the ground state of a topological liquid
into two spatial subregions, $A$ and its complement $\bar{A}$, say. 
We can then study the scaling of the entanglement entropy
as a function of the size of the subregion $A$,
which allows us to extract
the topological entanglement entropy
of the topologically ordered ground state
\cite{levin2006detecting,kitaev2006topological}.
One can also study 
the entanglement spectrum, which also serves as a
probe of different topological orders and symmetry-protected
topological phases
\cite{Ryu_2006, PhysRevLett.101.010504, Pollmann_2010}.

\begin{figure}
\centering
\includegraphics[width=0.8\columnwidth]{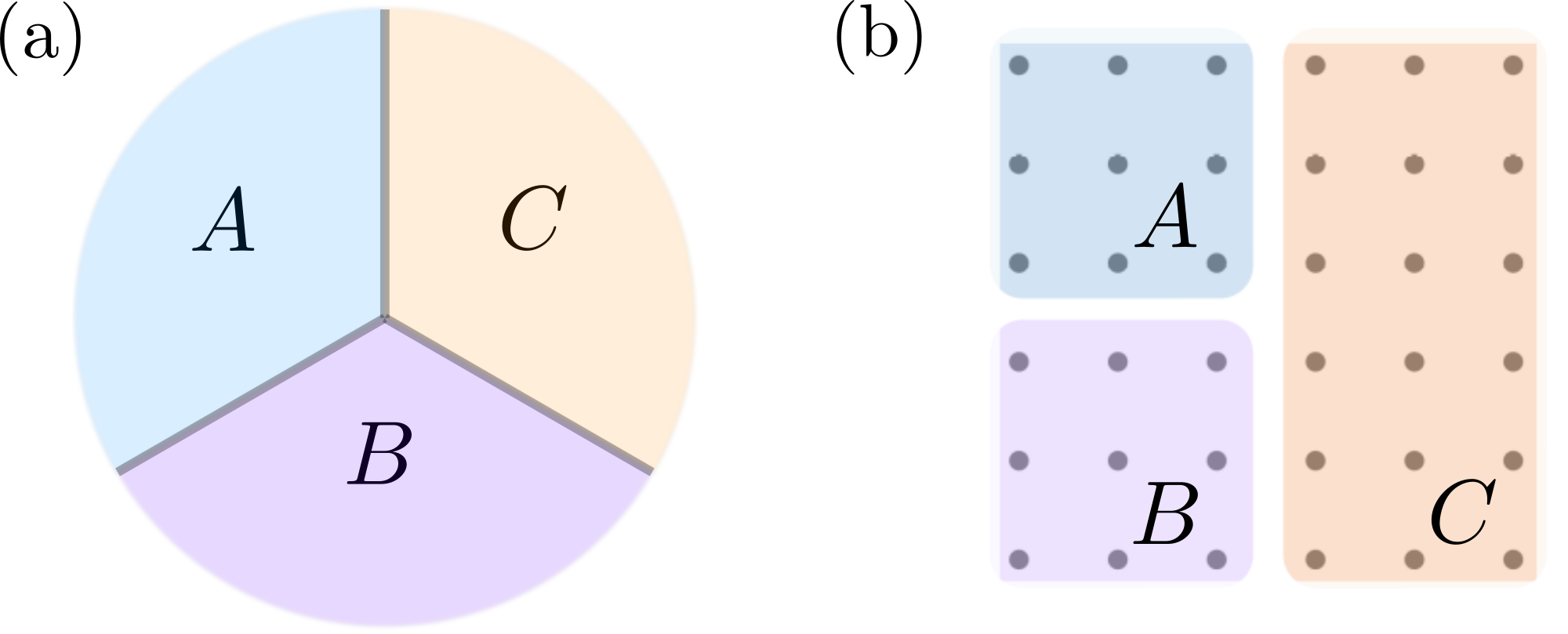}
\caption{
\label{partition}
  Tripartition of topological liquid on a two-dimensional plane (a)
  and two-dimensional square lattice (b).
}
\end{figure}

In this paper, we move beyond bipartitions and consider
multipartitions of the ground states of (2+1)-dimensional topological
liquids. 
Specifically, 
we consider a tripartition (multipartition)
in which the boundaries between the three subregions $A$, $B$, and $C$ meet at
a junction, as shown in Fig.\ \ref{partition}.
We note that this partitioning is analogous to
the one first considered in Ref.\ \cite{kitaev2006topological}.
A similar setup was also used 
recently
in \cite{2021arXiv211006932K,kim2021modular}
to 
derive a formula for the chiral central 
charge in terms of the modular commutator.
%

This multipartition setting allows us to define
and compute various  correlation measures.
For example, 
when one of the three subregions, say $C$,
is traced out, we are left with the reduced density matrix for
$A\cup B$, which is now mixed.
We can then discuss mixed state 
correlation measures,
such as the entanglement negativity
\cite{Zyczkowski:1998yd, 
Vidal:2002zz, 
Peres:1996dw, 1999JMOp...46..145E, 2005PhRvL..95i0503P, 2000PhRvL..84.2726S, 1996PhLA..223....1H}
and reflected entropy
\cite{dutta2021canonical}. 
We can also study the associated spectra, such as the
spectrum of the partially transposed density matrix. 
These entanglement measures may capture universal data
related to multipartite entanglement
of topologically-ordered ground states, which are not accessible in bipartition
settings.
(For previous studies  on multipartite correlations
in topological liquid, see, for example, \cite{2016PhRvA..93b2317K}.)

The entanglement negativity and reflected
entropy have been previously studied 
in the context of topologically-ordered phases in setups different from ours 
\cite{2013PhRvA..88d2318L, castelnovo2013negativity,
wen2016edge,
wen2016surgery,
lim2021disentangling,  
berthiere2021}. 
We give a brief overview of the previous results
in Sec.\ \ref{sec:measure}.
As for the reflected entropy,
for the tripartition setup above, 
it was recently 
claimed
\cite{BerkeleyPaper}
that
the difference between 
the reflected entropy
and mutual information
is
given by
$(c/3)\ln 2 +{\cal O}(e^{-\ell/\xi})$
where $c$ is 
the central charge of the topological liquid,
$\xi$ is the correlation length and $\ell$ is
the length scale for the three regions.
(To obtain the above universal value 
non-universal short-range correlations
must be removed by 
 a proper local unitary -- see 
 Sec.\ \ref{sec:measure}.)
As this multiparty entanglement
quantity
may capture the central charge,
the vanishing of this quantity may
be a prerequisite of having a  
PEPS (projected entangled pair state)
representation of the topological liquid
with finite bond dimension.
(Or non-vanishing of this quantity may be an obstruction
to having a PEPS representation with finite bond dimension.)
We will review this claim in Sec.\ \ref{sec:measure}.
These observations suggest
that there is much yet to be understood regarding topological phases from the lens of entanglement.


We study the tripartition of topological phases using two different approaches. First, we employ the edge theory or ``cut-and-glue" approach for computing the entanglement of topological phases \cite{qi2012general,lundgren2013cutandglue,cano2015interfaces,wen2016edge,sohal2020nonabelian},
in which one approximates the entanglement between the bulk regions as arising purely from entanglement of the gapped chiral edge modes along the entanglement cuts between the bulk regions.
This approach is not limited to non-interacting phases
(e.g. integer quantum Hall or Chern insulator phases)
but rather is also applicable to generic topologically-ordered phases.
We recall that for the case of bipartitioning a topological liquid, 
the entanglement entropy (and other related quantities)
can be obtained from conformal boundary states (Ishibashi states)
\cite{qi2012general,wen2016edge,wong2018note,fliss2017interface}.
(See Sec.\ \ref{Review of bipartition solution}.)
In this work, we will extend this approach to the case of a
multipartition (tripartition)
by considering what we call ``vertex states,''
which will be introduced in
Sec.\ \ref{subsec:outline}.
What the vertex states do for the case of tripartitioning
is quite analogous to
what Ishibashi states do for the case of bipartitioning.
We emphasize that the construction of these vertex states is a nontrivial extension of the corresponding computation for a bipartition, even for the case of free fermions. Indeed, 
with some minor differences,
states similar to vertex states have been considered
in the context of string field theory
\cite{gross1987field,gross1987operator,gross1987operator2,leclair1989string}.
They also resemble 
open boundary states or rectangular states
in conformal field theory
\cite{imamura2006boundary,imamura2008boundary,
bondesan2012conformal,bondesan2013rectangular}. 
We will construct these vertex
states using two methods: the Neumann coefficient method,
which makes use of conformal mappings to fix the form of the vertex state, and a direct
calculation method, in which we directly diagonalize the boundary conditions
defining the vertex state.
We check their equivalence numerically.

In the second approach, we consider the tripartite entanglement of
a specific non-interacting lattice fermion model
that realizes a Chern insulator phase.
The many-body ground state is given by
a Gaussian state (namely, a Slater determinant state), which allows us to make use of the ``correlator method'' developed in Refs.\
\cite{peschel2003calculation,chung2001density}
to compute 
various correlation measures.
In contrast to the edge theory calculation, which is only applicable for a
system deep in the topological phase, here we can study how the correlation 
measures of interest change as we tune across the phase transition between the
topological and trivial phases.

This paper is organized as follows.
In Sec.\ \ref{sec:measure},
we introduce the correlation measures of interest
and the correlator method.
In Subsec.\ \ref{subsec:outline},
after reviewing the edge theory approach to computing entanglement in bipartition settings,
  we introduce vertex states for multipartition.
We demonstrate how
to obtain the vertex state using the Neumann coefficient method for
both a Chern insulator and a chiral superconductor. As a warm-up, in Sec.\ \ref{subsec:bipartition} we compute the entanglement entropy for a bipartition and obtain a new topological contribution in the sector with nontrivial topological flux piercing the entanglement cut.
In Sec.\ \ref{Tripartition solution}, we present the tripartite vertex state solutions in different sectors, 
namely in the presence of
nontrivial topological fluxes,
and extract new fingerprints 
of the underlying topological state in entanglement. 
In particular,
we discuss the scaling of the entanglement negativity,
the spectra of the entanglement negativity and partially transposed
density matrix.
We also test the conjecture on the reflected entropy in 
Ref.\ \cite{BerkeleyPaper}.
In Sec.\ \ref{lattice},
we study the entanglement measures numerically 
in the lattice Chern insulator model.
By comparing the results between
vertex state and Chern insulator ground state,
we demonstrate the bulk-boundary correspondence for tripartitioned topological states.
We also gain access to the spatial structure of entanglement by calculating negativity contour. 

We collect the technical details
in Appendices.
In Appendix \ref{subsec:direct},
we give the detailed derivation of the vertex states
by the direct calculation method,
which is complementary to the Neumann coefficient method. 
In Appendix \ref{app:Neumann},
we provide the technical details of the Neumann coefficient method.
Finally, in Appendix \ref{app:correlation},
we show how to apply the correlator method to vertex states
to compute various entanglement measures.

\section{Correlation measures of interest}
\label{sec:measure}

In this section, we introduce the correlation
measures that will be discussed in this paper.
Some of the correlation measures, 
the entanglement entropy for 
the case of pure states,
and the entanglement negativity 
for generic mixed states,
are also entanglement measures,
while others such as mutual information and reflected entropy 
are not.
Here, entanglement measures are those quantity that 
capture quantum correlations and 
monotonically 
decrease under
local operations and classical communications (LOCCs).

\paragraph{Entanglement entropy}
When bipartitioning the total system
into two subregions $A$ and $\bar{A}$,
after tracing out subregion $\bar{A}$,
the reduced density matrix on $A$ is $\rho_A:=\mathrm{Tr}_{\bar{A}}\, \rho$.
The (von Neumann) entanglement entropy is defined as
\begin{align}
  S(\rho_A)
  :=
S_A:=-\mathrm{Tr}\left(\rho_A\ln{\rho_A}\right).
\end{align}
The entanglement entropy is also given by the $n \to 1$ limit of the R\'enyi entropies, defined as
$S_A^{(n)} :=
\ln \mathrm{Tr}\, \left(\rho_A^n\right)/(1-n)$.
We recall that for gapped ground states of two-dimensional Hamiltonians, 
$\rho = |{\it GS} \rangle\langle {\it GS}|$,
the entanglement entropy satisfies an area law, $S_A = \alpha L - \gamma$, where $\alpha$ is a nonuniversal constant, $L$ the length of the entanglement cut, and $\gamma$ the topological entanglement entropy. Since the topological phases we consider 
(chiral $p$-wave superconductor and Chern insulator)
are not topologically ordered (i.e.\ do not support anyon excitations), we will have $\gamma = 0$ in the absence of non-trivial fluxes. We can also combine entanglement entropy 
in different regions to form other correlation measures including
{the mutual information, $I_{A:B}=S_A+S_B-S_{A\cup B}$, and the}
{tripartite mutual information, $I_3=S_A+S_B+S_C-S_{AB}-S_{BC}-S_{AC}+S_{ABC}$. Note that tripartite mutual information is directly related to topological entanglement entropy. }

\paragraph{Entanglement negativity}

Let us now consider 
sub Hilbert spaces $A$ and $B$,
and the density matrix $\rho_{A\cup B}$ supported on
$A \cup B$.
For mixed states,
the entanglement entropy 
is not a proper entanglement measure
in that it does not decrease monotonically 
under LOCCs.
Instead, 
one can consider the entanglement negativity,
\begin{align}
  \mathcal{E}_{A:B}
  =\ln \mathrm{Tr}\, ||\rho_{A\cup B}^{T_A}||_1
  =\ln \mathrm{Tr}\, \big(
  \sqrt{\rho_{A\cup B}^{T_A}(\rho_{A\cup B}^{T_A})^\dagger}
  \big)
  \label{eqn:negativity}
\end{align}
with $T_A$ being the partial transpose on subregion $A$. 
When $\rho_{A\cup B}$ is pure, $\mathcal{E}_{A:B}=S_A^{(1/2)}$.
For bosonic systems, 
the partial transpose is defined as
\begin{align}
\langle e_i^{A} e_j^{B}|\rho^{T_A}_{A\cup B}|e_k^{A}e_l^{B}\rangle
=\langle e_k^{A} e_j^{B}|\rho^{\ }_{A\cup B}|e_i^{A}e_l^{B}\rangle ,
\end{align}
where
$\{|e^{A/B}_i\rangle\}$
are complete bases of states for subregions $A/B$, respectively.
We note that 
by introducing 
the normalized composite density operator as
$\rho_\times=\rho^{T_A}_{A\cup B}\big(\rho^{T_A}_{A\cup B}\big)^\dag/Z_\times$,
we can express the negativity as
\begin{align}
  \mathcal{E}_{A:B}
  &=\ln{\left[Z_\times^{1/2}
    \mathrm{Tr}\, \big(\rho_\times^{1/2}\big) \right]}
    \nonumber \\
  &=\ln{\mathrm{Tr}\, \big(\rho_\times^{1/2}\big)}
    +\frac{1}{2}\ln{\mathrm{Tr}\, \big(\rho^2_{A\cup B}\big)},
    \label{eqn:negativity-rewrite}
\end{align}
where $Z_\times:=\mathrm{Tr}\,
\big[\rho^{T_A}_{A\cup B}\big(\rho^{T_A}_{A\cup B}\big)^\dag\big]
=\mathrm{Tr}\, \big(\rho^2_{A\cup B}\big)$.

On the other hand, for fermionic systems,
the definition of the partial transpose has to take Fermi statistics 
into account properly
\cite{shapourian2019twisted}. 
If we use the Majorana basis and expand
a density matrix $\rho_{A\cup B}$
in terms of Majorana fermion operators $a$ and $b$
defined on $\mathcal{H}_A$ and $\mathcal{H}_B$, respectively,
\begin{align}
  \rho_{A\cup B}
  =\sum_{k_1,k_2}^{k_1+k_2=\text{even}}
  &\rho_{p_1,\cdots,p_{k_1},q_1,\cdots, q_{k_2}}
  \nonumber \\
  &\quad
    \times
  a_{p_1}\cdots a_{p_{k_1}}b_{q_1}\cdots b_{q_{k_2}},
\end{align}
then 
the partial transpose of $\rho_{A\cup B}$
with respect to subregion $A$ is defined as
\begin{align}
  \rho^{T_A}_{A\cup B}
  =\sum_{k_1,k_2}^{k_1+k_2=\text{even}}
    &\rho_{p_1,\cdots,p_{k_1},q_1,\cdots, q_{k_2}} i^{k_1}
    \nonumber \\
  &\quad
    \times
    a_{p_1}\cdots a_{p_{k_1}}b_{q_1}\cdots b_{q_{k_2}}.
\end{align}
Entanglement negativity 
in fermionic systems,
when formulated by using 
the fermionic partial transpose
above, 
is monotone under LOCC
preserving the local fermion-number parity
\cite{2019PhRvA..99b2310S, 2020arXiv201202222S}.

The entanglement negativity 
has been previously studied 
in the context of topologically-ordered phases in setups different from ours 
\cite{2013PhRvA..88d2318L, castelnovo2013negativity,
wen2016edge,
wen2016surgery,
lim2021disentangling,  
berthiere2021}. 
The entanglement negativity 
for topologically-ordered ground states
has been shown to obey an area law with subleading, universal corrections that are non-zero for topologically-ordered
ground states,
much like the entanglement entropy. 
However, unlike the entanglement entropy, the entanglement negativity appears to exhibit distinct behavior between Abelian and non-Abelian topological phases when computed in superpositions of topologically degenerate states on manifolds with non-zero genus for certain tripartitions \cite{wen2016edge,lim2021disentangling}. 
The entanglement negativity 
was also studied 
for topological phases of 
matter at finite temperatures,
and shown to detect
finite temperature transitions
\cite{2020PhRvL.125k6801L,
2018PhRvB..97n4410H}.

In the same way that the entanglement spectrum provides
more information than the entanglement entropy,
also of interest to us is the spectral decomposition 
of the entanglement negativity.  
Specifically, 
we will study two types of spectra, 
one associated with  
$\rho_{\times} = \rho^{T_A}_{A\cup B} 
(\rho^{T_A }_{A \cup B})^{\dag}/Z_{\times}
$
and the other with
$
\rho_{A \cup B}^{T_A}
$.
We note that 
for fermionic systems, 
$\rho^{T_A}_{A\cup B}$
may not be Hermitian. 
For conformal field theories and non-trivial SPT phases in (1+1) dimensions, the spectrum of 
$
\rho_{A \cup B}^{T_A}
$
shows an interesting pattern and is sensitive to
the spin structure
\cite{shapourian2019twisted, inamura2020non}. 

\paragraph{Reflected entropy}

Finally,
the reflected entropy $R_{A:B}$ also provides a correlation measure for tripartite Hilbert spaces. Given a reduced density  matrix $\rho_{A\cup B}$ supported on $A\cup B$, we can obtain its
canonical purification 
$|\sqrt{\rho}\rangle\!\rangle$ in the doubled Hilbert space $(A\cup B)\cup(\tilde{A}\cup\tilde{B})$,
where
$\tilde{A}$ and $\tilde{B}$
are identical copies of $A$ and $B$, respectively (with complex conjugation).
The reflected entropy $R_{A:B}$ is defined as the entanglement entropy of the purified state $|\sqrt{\rho}\rangle\!\rangle$ 
when tracing out the degrees of freedom in $B,\tilde{B}$: 
\begin{align}
R_{A:B} = S(\rho_{A\cup \tilde{A}}),
\quad
\rho_{A\cup \tilde{A}}= 
\mathrm{Tr}_{B\cup \tilde{B}}\,
\big(
| \sqrt{\rho}\rangle\!\rangle
\langle\!\langle \sqrt{\rho}|
\big).
\end{align}
The reflected entropy has been studied 
in various many-body quantum systems.
For example, in (1+1)d CFT, the reflected entropy
has been studied for the ground state
\cite{dutta2021canonical},
and for time-dependent states after 
quantum quench
\cite{2020JHEP...02..017K,2021PhLB..81436105K,2020JHEP...04..074K}.
The reflected entropy
was also computed
for multi-sided thermofield double states
in (non-chiral) (1+1)d CFT
(which has some similarly to vertex states
that we will introduce later)
\cite{2021arXiv210809366Z}.
The reflected 
entropy is
a more sensitive probe of multipartite entanglement than the von Neumann entropy
\cite{2020JHEP...04..208A, zou2021universal}.
The difference between the reflected entropy and mutual information
\begin{equation}
    h_{A:B} = R_{A:B} - I_{A:B} ,
\end{equation}
is bounded from below,
$h_{A:B}\ge 0$
\cite{dutta2021canonical},
and called the Markov gap 
in Ref.\ \cite{2021arXiv210700009H} as it is related to the
fidelity of a particular Markov recovery process on the canonical purification.
The difference $h_{A:B}$ is
proposed as a non-negative universal tripartite entanglement invariant 
\cite{zou2021universal}.
It was also shown that for the ground states of 
1d lattice quantum systems at conformal critical points
when the subregion $A$ and $B$ are adjacent to each other,
$h_{A:B}$ takes a universal value, 
$h_{A:B}=(c/3)\ln 2$, where $c$ is the (non-chiral) central charge \cite{zou2021universal}.

For the ground states of (2+1)d topological liquids,
it was recently conjectured in Ref.\ \cite{BerkeleyPaper}
that $h_{A:B}$, 
when computed for the tripartite 
setting in Fig.\ \ref{partition},
captures the chiral 
central charge of the topological liquid.
Specifically, 
from the topological ground state 
$|\Psi\rangle$,
we consider a state 
$U|\Psi\rangle$
where a local unitary $U$ acts 
near the junction. 
This unitary $U$ can be optimized
such that it removes non-universal,
short-range correlation near the
junction.
Then, the claim in 
\cite{BerkeleyPaper} is that 
the optimized version of 
$h_{A:B}$,
\begin{align}
h^{{\it IR}}_{A:B}=
\mathrm{min}_{U}
h_{A:B}( U|\Psi\rangle ), 
\end{align}
takes the universal value,
\begin{align}
\label{eq: conj for h}
h^{{\it IR}}_{A:B}=
\frac{c}{3}\ln 2 + 
\mathcal{O}(e^{ -\ell/\xi}),
\end{align}
where 
$\xi$ is the correlation length, 
$\ell$ is the length scale for the three regions,
and 
$c$ is the central charge of the topological liquid
that measures ungappable edge 
degrees of freedom,
i.e., $c_L+c_R$ where $c_{L/R}$ is the left/right
central charge.
This conjecture was tested in
Ref.\ \cite{BerkeleyPaper}
for sting-net models, for which $c=0$,
and for a non-interacting 
Chern insulator model 
with proper optimization over $U$.

\subsection{Fermionic Gaussian states}
\label{sec:fermonic-Gaussian}

When the (reduced) density matrix of interest is Gaussian, 
the above correlation measures can be efficiently computed 
by using 
the correlator (or covariance matrix) method 
\cite{peschel2003calculation, Peschel_2009,shapourian2019twisted,KudlerFlam2020contour}.
A Gaussian state $\rho_{A\cup B}$ is fully characterized
by the correlation matrices $C,F$, or,
equivalently, by the covariance matrix $\Gamma$,
\begin{align}
  C_{IJ}&:=\mathrm{Tr}\, \big(\rho_{A\cup B}\,  f^\dagger_I f^{\ }_J\big),
          \nonumber \\
  F_{IJ}&:=\mathrm{Tr}\,\big(\rho_{A\cup B}\,  f^\dagger_I f^\dagger_J\big),
          \nonumber \\
  \Gamma_{JK}&:=\frac{1}{2}\mathrm{Tr}\,\big(\rho_{A\cup B} \left[c_J, c_K\right]\big).
\end{align}
Here, $\lbrace f^\dag_I, f^{\ }_I\rbrace$ is a set of fermion
creation/annihilation operators where the indices $I,J$ run over all relevant
degrees of freedom, site, spin, orbital, etc. $c_I$ is the Majorana operator and
we adopt the convention
$c_{2J-1}=( f_J+ f_J^\dag)$, $c_{2J}=i( f_J- f_J^\dag)$. $\Gamma$ can be
expressed in terms of $C,F$ as
\begin{align}
  \Gamma&=(C-C^T)\otimes \mathbbm{1}+(\mathbbm{1}-C-C^T)\otimes \sigma_y
          \nonumber \\
        &
          \quad
          +(F+F^\dagger)\otimes \sigma_z-i(F-F^\dagger)\otimes\sigma_x,
          \label{eqn:GammaCF}
\end{align}
where the Pauli matrices act on the space of odd and even indices of the Majorana fermions. 

\paragraph{Entanglement entropy and negativity}
The von Neumann entropy for the density matrix $\rho_{A\cup B}$
is obtained from the eigenvalues $\gamma_k$ of the covariance matrix $\Gamma$:
\begin{align}
  S_{AB}
    =-\sum'_{k}
    &
    \Bigg[
    \Big(\frac{1}{2}+\frac{\gamma_k}{2}\Big)
    \ln \Big(\frac{1}{2}+\frac{\gamma_k}{2}\Big)
    \nonumber \\
  &\qquad
    +\Big(\frac{1}{2}-\frac{\gamma_k}{2}\Big)
    \ln
    \Big(\frac{1}{2}-\frac{\gamma_k}{2}\Big)
    \Bigg].
\end{align}
Here, the prime on $\sum$ means we only sum over one of the eigenvalues in the
$\pm\gamma_k$ pairs. In particle number conserving systems, the eigenvalues
$\gamma_k$ are related to the eigenvalues $\epsilon_k$ of the quadratic
entanglement Hamiltonian $H_E$,
defined as $\rho_{A\cup B}\propto 
\exp{(- \sum_{I,J} f_I^\dagger(H_E)_{IJ} f_J)}$), by $\epsilon_k=\ln{[(1-\gamma_k)/(1+\gamma_k)]}$. For $\eta_k$ being eigenvalues of $C$, $\epsilon_k$ can be expressed equivalently as $\epsilon_k=\ln{[(1-\eta_k)/\eta_k)]}$.
We call the set of eigenvalues $\lbrace\epsilon_k\rbrace$ the (single-particle)
entanglement spectrum (ES) of $\rho_{A\cup B}$.

Similar to the entanglement entropy,
the entanglement negativity for a fermionic Gaussian
state can also be computed from the covariance matrix.
In particular,
the covariance matrix associated to $\rho_{\times}$
can be constructed as follows.
Upon bipartitioning the Hilbert space,
$\mathcal{H}_{A\cup B}=\mathcal{H}_A\otimes\mathcal{H}_B$,
we can write the covariance matrix in a block matrix form,
\begin{equation}
\Gamma=\left(
\begin{array}{cc}
\Gamma_{AA} & \Gamma_{AB}\\
\Gamma_{BA} &\Gamma_{BB}
\end{array}
\right).
\end{equation}
Here, $\Gamma_{AA}$ and $\Gamma_{BB}$ denote the reduced covariance matrices of subsystems $\mathcal{H}_A$ and $\mathcal{H}_B$, respectively, whereas $\Gamma_{AB}$ and $\Gamma_{BA}$ contain the expectation values of mixed quadratic terms. 
The covariance matrix for the partially transposed density matrix
$\rho^{T_A}_{A\cup B}$ and its conjugate,
$(\rho^{T_A}_{A\cup B})^\dag$, can be constructed as
\begin{equation}
\Gamma_\pm=\left(
\begin{array}{cc}
-\Gamma_{AA} & \pm i\Gamma_{AB}\\
\pm i\Gamma_{BA} &\Gamma_{BB}
\end{array}
\right),
\end{equation}
respectively. Using the algebra of the product of Gaussian operators
\cite{fagotti2010entanglement}, the covariance matrix $\Gamma_\times$ associated
with the normalized composite density operator
$\rho_\times$ is given by 
\begin{equation}
\Gamma_{\times}=\mathbbm{1}-(\mathbbm{1}-\Gamma_-)(\mathbbm{1}+\Gamma_+\Gamma_-)^{-1}(\mathbbm{1}-\Gamma_+).
\end{equation}
In terms of the eigenvalues $\lbrace\gamma_k\rbrace$ and
$\lbrace\gamma_{\times k}\rbrace$
of the covariance matrices $\Gamma$ and $\Gamma_\times$,
 using Eq.\ \eqref{eqn:negativity-rewrite}, we can write
\begin{equation}
 \begin{aligned}
 &\mathcal{E}_{A:B}=\sum_{k}^{\prime}\left[h(\gamma_{\times k};1/2)+\frac{1}{2}h(\gamma_k;2)\right]\\
 &\mathrm{where}
 \quad
 h(\lambda;q)=\ln{\left[\left(\frac{1-\lambda}{2}\right)^q+\left(\frac{1+\lambda}{2}\right)^q\right]}.
 \end{aligned}
 \end{equation}
Again, only one eigenvalue in each of the $\pm\gamma_k$ and $\pm\gamma_{\times k}$ pairs
needs to be summed over. Analogous to the entanglement sectrum, the negativity
spectrum (NS) is defined as $\ln{[(1-\gamma_{\times k})/(1+\gamma_{\times k})]}$.

\paragraph{Spectrum of $\Gamma_+$}
The spectrum of $\rho^{T_A}_{A\cup B}$ can be constructed from the eigenvalues
of $\Gamma_+$, which appear in pairs $\lbrace \pm \zeta_k\rbrace$. We will also study the distribution of the eigenvectors associated with the eigenvalues $\zeta_k$. 

\paragraph{Negativity contour}

The negativity contour is a spatial decomposition of the negativity. While the negativity associates a number to two extended spatial regions, the contour, $e_{A:B}(\bm{r})$, is a function of the spatial coordinates of the regions which can be interpreted as the contribution of each degree of freedom to the negativity. 
The contour is constructed such that 
when summed over all positions 
it reproduces $\mathcal{E}_{A:B}$, 
$\sum_{\bm{r}} e_{A:B}(\bm{r})=\mathcal{E}_{A:B}$. 
This elucidates where the entanglement is coming from. For example, in ground states of gapped Hamiltonians, the contour is concentrated at the entangling surface, decaying exponentially in space, representing the area law. In, critical systems, the contour instead decays away from the entangling surface as a power law. For highly excited (thermal) states, the contour is finite and approximately constant, representing the thermal entropy.

For Gaussian states,
the negativity contour is defined 
using the eigenvectors of the covariance matrices
\begin{equation}
\label{neg cont}
\begin{aligned}
e_{A:B}(\bm{r})&=v_1(\bm{r})+v_2(\bm{r}),
 \\
v_1(\bm{r})&=\frac{1}{2}\sum_k |U_k(\bm{r})|^2  h(\gamma_{k,\times};1/2),
 \\
v_2(\bm{r})&=\frac{1}{4}\sum_k |V_k(\bm{r})|^2  h(\gamma_k;2),
\end{aligned}
\end{equation}
where $U_k(\bm{r})$ and $V_k(\bm{r})$ are eigen states of $\Gamma_\times$ and $\Gamma$ with eigenvalues $\gamma_{k,\times}$ and $\gamma_k$, respectively. 
(For the particle number conserving case, 
Eq.\ \eqref{neg cont}
reduces to Eq.\ (A61) of \cite{KudlerFlam2020contour}.) 


\paragraph{Reflected entropy}

Finally, the reflected entropy can also be computed conveniently using the covariance matrix method
\cite{2020JHEP...05..103B}.
Using the orthogonal transformation $O$ to bring $\rho$ and $\Gamma$ to canonical forms: 
\begin{equation}
    \begin{aligned}
        \rho &= \prod_k \frac{1}{2}(1+\gamma_k c'_{2k-1} c'_{2k})\quad \mathrm{where} \quad c'=Oc,
        \\
    \Gamma & = O^T \left[\oplus_k \left(
\begin{array}{cc}
0 & i  \gamma_k\\
-i \gamma_k & 0
\end{array}
\right)\right]O.
   \end{aligned}
\end{equation}
The purified state is given by
\begin{equation}
    \begin{aligned}
|\sqrt{\rho}\rangle\!\rangle
     = 
     \prod_k
     \left[
     \sqrt{\frac{1+\gamma_k}{2}}
     |0\rangle_k|\tilde{0}\rangle_k +  \sqrt{\frac{1-\gamma_k}{2}}
     |1\rangle_k|\tilde{1}\rangle_k
     \right],
\end{aligned} 
\end{equation}
where $|\tilde{0}\rangle_k,|\tilde{1}\rangle_k$ 
are states in the second copy of the Hilbert space for the $k$-th mode. The associated covariance matrix for $|\sqrt{\rho}\rangle\!\rangle$ is
\begin{equation}
    \Gamma_{\sqrt{\rho}}= O
\left[\oplus_k\left(
\begin{array}{cc}
 \gamma_k \sigma^y & -i\sqrt{1-\gamma_k^2} \mathbbm{1}\\
i\sqrt{1-\gamma_k^2} \mathbbm{1} & -\gamma_k \sigma^y
\end{array}
\right)\right] O^T.
\end{equation}
The reflected entropy $R_{A:B}$ is then computed as the von Neumann entanglement entropy using the $A,\tilde{A}$ blocks in $\Gamma_{\sqrt{\rho}}$.


\section{Edge theory approach}
\label{sec:CFT-approach}

We now proceed to compute the correlation measures introduced in the preceding section from the perspective of the boundary edge theories. 
We perform these computations for a chiral superconductor and Chern insulator
(or integer quantum Hall state), the edge theories of which consist of single
chiral Majorana and Dirac fermions, respectively.
As we will review in more detail below, in the edge theory or ``cut-and-glue'' approach \cite{qi2012general,lundgren2013cutandglue,cano2015interfaces,wen2016edge,sohal2020nonabelian,lim2021disentangling}, we compute the entanglement between subregions of a topological phase by first physically cutting the system along the entanglement cut, which gives rise to the aforementioned chiral edge states. We then ``glue" the system back together by introducing a tunneling interaction to gap out the edge states. Since the correlation length vanishes in the bulk, we can approximate the entanglement between the bulk subregions as arising solely from entanglement between the gapped edge modes. The first step in this computation is then to determine the ground state of this gapped interface along the entanglement cut.

For the case of a simple bipartition, this ground state is known to take the
form of a conformal boundary state, or more precisely,
an Ishibashi state \cite{qi2012general,wen2016edge}.
For the tripartitions of interest to us, in which the entanglement cut involves
a trijunction, a generic form for the ground state of the interface is not known
and is difficult to compute, even in the present case of free fermions.
Fortunately, similar interface configurations
have appeared in the string field theory 
literature, in which the conformal boundary states for such trijunctions are known as \emph{vertex states}. In the following, 
we will use the Neumann coefficient method from string field theory
\cite{gross1987field,gross1987operator2,gross1987operator,leclair1989string,witten1986non}
to compute the appropriate boundary or vertex states.
We introduce boundary and vertex states 
and outline the essential steps of 
the Neumann coefficient method in Sec.\ \ref{subsec:outline}.
With the vertex state in hand, we can then proceed to compute all desired entanglement measures.

\begin{figure*}
\centering
\includegraphics[width=0.7\textwidth]{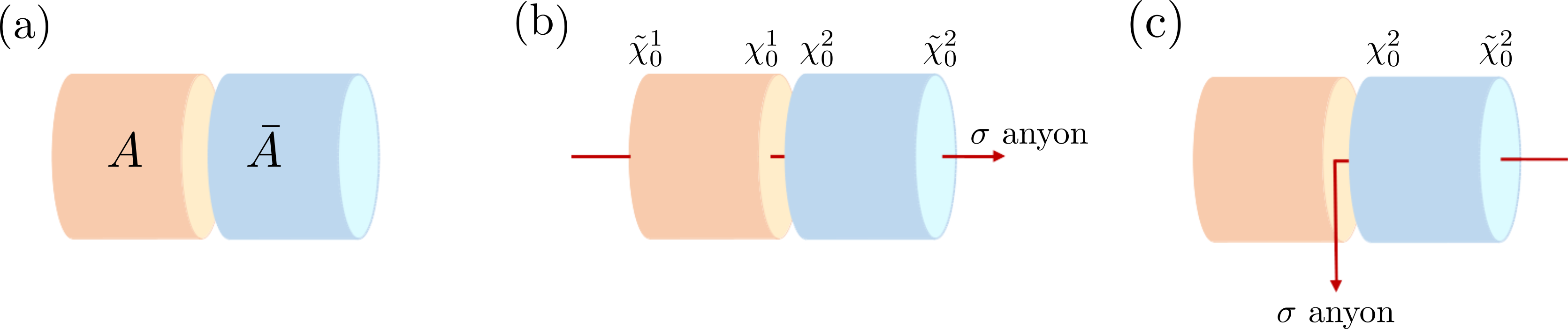}
\caption{Flux insertion configurations considered in computation of the bipartite entanglement on the cylinder geometry. (a) No fluxes are inserted. All edge fermions obey NS boundary conditions. (b) A single $\pi$-flux, corresponding to the insertion of a $\sigma$ anyon flux, through the cylinder. All edge fermions obey R boundary conditions. (c) A single $\pi$-flux is inserted through the right half of the cylinder, but exits through the entanglement cut. The edge fermions on the left (right) cylinder obey NS (R) boundary conditions. In (b) and (c), the zero modes on the inner (outer) edge are $\chi_0^1,\chi_0^2$ ($\bar{\chi}_0^1,\bar{\chi}_0^2$).
} \label{fig:bipartition-fluxes}
\end{figure*}

As a warm up, in Sec.\ \ref{subsec:bipartition}
we will first compute the entanglement for a topological phase 
on a cylinder and a bipartition cutting the cylinder in two, as shown in 
Fig.\ \ref{fig:bipartition-fluxes}(a). 
We use the Neumann function method to compute the boundary state, as an introduction to the technique. In particular, we compute the entanglement when we introduce a $\pi$ flux either passing through the cycle of the cylinder, or entering the cylinder through one end and exiting through the entanglement cut. For the chiral superconductor, these configurations are topologically equivalent, respectively, to computing the bipartite entanglement on a sphere, with a single Ising anyon ($\sigma$ anyon)
in each subregion and an Ising anyon in one subregion and the other on the entanglement cut, as depicted in Fig. \ref{fig:bipartition-fluxes}. 
At the level of the edge theory, this amounts to computing the 
boundary state $|B\rangle$ with three different choices of
boundary conditions for the chiral and anti-chiral fermions: NS-NS, R-R, NS-R
\cite{jevicki1988supersymmetry}. 
Here, NS (Neveu-Schwarz) and R (Ramond) denote anti-periodic and periodic boundary conditions, respectively. We note that the entanglement in the NS-R case -- in which an anyon lies \emph{on} the entanglement cut -- has not been considered before. 
Remarkably, we find a new quantized contribution to the entanglement in this configuration. 
With this framework in hand, we will move on to the focus of this work, the tripartitioning of a topological liquid, in the following section.




\subsection{Cut-and-glue approach and vertex states}
\label{subsec:outline}

We begin with a more detailed exposition of the cut-and-glue approach and explain the role of conformal boundary and vertex states, as well as how to construct them. For concreteness, we focus first the case of a chiral $p$-wave superconductor and then outline the simple extension of these methods to the case of a Chern insulator.

\subsubsection{Bipartition and Ishibashi boundary states}
\label{Review of bipartition solution}

The case of a bipartition was first considered in Ref.\ \cite{qi2012general}, which we review here. Let us consider a chiral superconductor on an infinite spatial cylinder with an entanglement cut, partitioning 
the total system
into two regions $A$ and $\bar{A}$ [Fig.\ \ref{fig:bipartition-fluxes}(a)]. As described above, we physically cut the system along the entanglement cut, resulting in gapless edge modes 
on the boundaries of regions $A$ and $\bar{A}$,
respectively.
For the case of the chiral $p$-wave superconductor,
they are described by chiral real (Majorana) fermion theories
with opposite chiralities, denoted by $\gamma_L$ and $\gamma_R$.
Their dynamics at low energies can be described by
\begin{align}
  H_0
  &=\int_0^{2\pi} d\sigma
    \big[\gamma_L i\partial_\sigma\gamma_L
    +\gamma_R(-i\partial_\sigma)\gamma^{\ }_R
    \big].
\end{align}
Here, we take the circumference of the cylinder to be
$L=2\pi$ for simplicity.
The Majorana fermion fields obey either
anti-periodic (Neveu-Schwarz, NS) or periodic (Ramond, R) boundary conditions. 
For later purposes,
it is convenient to introduce
\begin{align}
  \psi^{1}(\sigma)\equiv\gamma^{\ }_L(\sigma),
  \quad
  \psi^{2}(\sigma)\equiv \gamma^{\ }_R(2\pi-\sigma).
\end{align}
The edge state Hamiltonian is then written as 
\begin{align}
  H_0&=\int_0^{2\pi}d\sigma\,
       \sum_{I=1,2}
       \psi^I i\partial_\sigma \psi^I.
\end{align}
The chiral Majorana fermion field 
$\psi(\sigma)$ can be Fourier expanded as
\begin{align}
  &
    \psi(\sigma)=\sum_{s\in\mathbb{Z}+1/2}e^{-i\sigma s}\psi_s
    \nonumber \\
  &
    \mbox{where}
    \quad
    \psi_{-s}=\psi_s^\dag,
    \quad
    \lbrace\psi_s,\psi_{s'}\rbrace=\delta_{s,-s'}
\end{align}
in the NS sector.
The vacuum of the NS sector is defined by
\begin{equation}
  \psi_s|0\rangle=0
  \quad
  \mathrm{for}
  \quad  s>0
\end{equation}
We have a similar expansion for the R-sector with integer moding.

In order to ``glue" the system back together, we introduce a tunneling term which gaps out the chiral edge degrees of freedom. Explicitly, we describe the gapped edge with the Hamiltonian $H_0 + H_{int}$, where
\begin{align}
  H_{{\it int}}
  &=im \int_0^{2\pi}d\sigma\,
    \gamma_L \gamma^{\ }_R
   =im \int_0^{2\pi}d\sigma\,
        \psi^1(\sigma)\psi^{2}(2\pi-\sigma).
\end{align}
As described above, we identify the entanglement between $A$ and $\overline{A}$ as arising purely from the entanglement between the chiral and anti-chiral Majorana fermions in this gapped state (i.e. the ``left-right" entanglement \cite{das2015leftright}).

The gapped ground state is in fact related to a conformal boundary state, or more precisely, an Ishibashi state, $\ket{B}$, of the gapless theory described by $H_0$. For a general CFT, $\ket{B}$ is defined by the relation
\begin{align}
  \left[
  L_n - \bar{L}_{-n}
  \right]|B\rangle=0
  \quad
  (\forall n \in \mathbb{Z})
\end{align}
where $L_n$ ($\bar{L}_n$) is the Fourier component of the
energy-momentum tensor $T(\sigma)$ ($\bar{T}(\bar{\sigma})$)
of the edge theory.
For the case of the free fermion theory,
the Ishibashi state is defined by
\begin{align}
  \label{free fermion Ishibashi}
  &
    \quad
    \left[\gamma^{\ }_L(\sigma)
    \mp i\gamma^{\ }_R(\sigma)\right]|B\rangle=0.
\end{align}
Or in terms of $\psi^I$,
\begin{align}
  \label{overlap condition}
  [\psi^1(\sigma)\mp i\psi^2(2\pi-\sigma)]|B\rangle=0,
\end{align}
which is valid for the whole region $0\leq \sigma\leq 2\pi$
(this leads to $[\psi^2(\sigma)\pm i\psi^1(2\pi-\sigma)]|B\rangle=0$).
Indeed we see that $\ket{B}$ is the ground state of $H$ in the limit $|m| \to \infty$.
From the Ishibashi boundary state, 
we can approximate 
the ground state of the (2+1)d topological phase
near the entanglement boundary for large but finite $m$ with the regularized state, 
\begin{align}
  \label{GS edge}
|G\rangle=\mathcal{N}e^{-\epsilon H_{0}}|B\rangle.
\end{align}
Here, the regulator $\epsilon$ is inversely proportional to the bulk
energy gap.
The reduced density matrix can then be constructed
from $|G\rangle$
by tracing over $\bar{A}$,
$
\rho_A=\mathrm{Tr}_{\bar{A}}\, |G\rangle\langle G|.
$
We emphasize that, while we took the non-interacting fermion theory as an example, essentially the same construction of the reduced density matrix using the Ishibashi boundary state can be done for a much broader class of theories.

The condition \eqref{overlap condition},
$(\psi_r^1\mp i\psi_{-r}^2)|B\rangle=0$, 
for the free fermion boundary state can explicitly be solved.
For example, for the NS sector (the NS boundary condition), 
it is given in the form of a fermionic coherent state as:
\begin{align}
  \label{fermion ishibashi}
  |B\rangle = \exp\Big(
  i\sum_{r\geq 1/2}\psi_{-r}^1\psi_{-r}^2\Big)|0\rangle,
\end{align}
which has the form of Ishibashi state, as expected. 
Here $|0\rangle$ is the Fock vacuum defined by
$\psi^I_r|0\rangle=0$ for $r>0$.

\subsubsection{Multipartition and vertex states}

\begin{figure*}[!htb]
\centering
\includegraphics[width=0.75\textwidth]{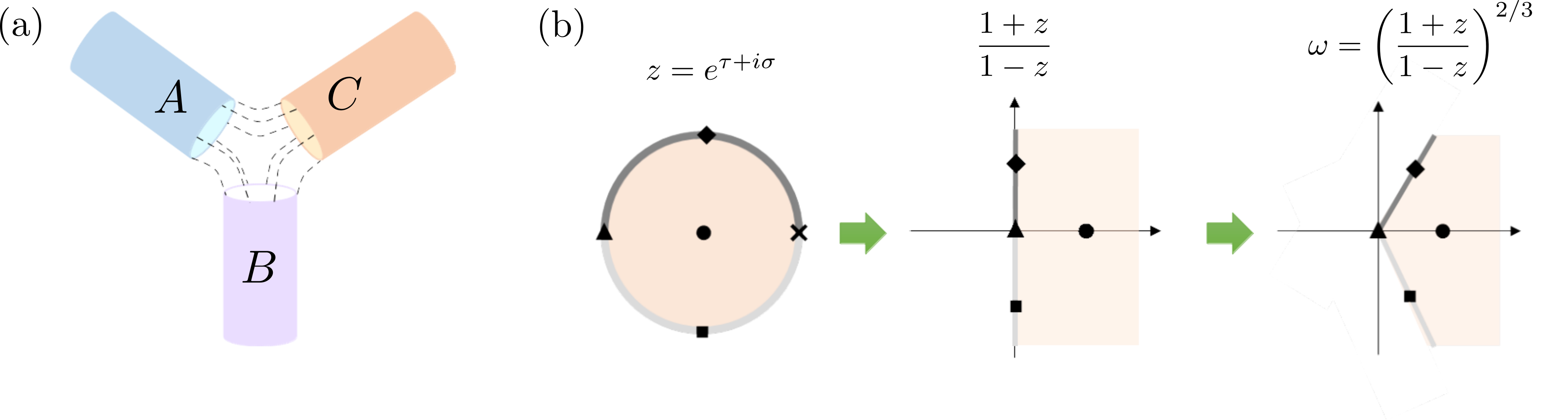}
\caption{
(a) Gluing three cylinders from edge theory point of view.  (b) The conformal map 
used to define vertex states
for tripartition.
One disk ($0\leq \sigma \leq 2\pi,\tau<0$) is mapped to the one-third of the whole plane. 
The past infinity point $\tau=-\infty$,
denoted by the black filled circle, is mapped to $\omega_{1,0}=e^{i\pi/3},\omega_{2,0}=e^{-i\pi/3},\omega_{3,0}=e^{-i\pi}$ for $I=1,2,3$ by Eq.~\eqref{eqn:conformal_map}. 
}
\label{fig:ct_close}
\end{figure*}

The bipartite setup
and the cut-and-glue method of
the reduced density matrix
presented above
can be extended to a
multipartition. 
In this section, 
we focus on a tripartition,
but the following discussion can readily be
extended to an $N$-partition ($N>3$).
We first note that the configuration 
in Fig.\ \ref{partition}(a)
is topologically equivalent to the one obtained by first
considering three cylinders, corresponding to the regions $A,B,C$, and then
gluing these cylinders together
[Fig.\ \ref{fig:ct_close}(a)]. 
As in the case of a bipartition, 
we cut open the system along the cut, resulting in an edge theory
comprising three
free Majorana fermions, as described by the Hamiltonian,
\begin{align}
  H_0
  &=\int_0^{2\pi}d\sigma
    \sum_{I=1}^3 \psi^I i\partial_{\sigma} \psi^I.
\end{align}
We again heal the cut by introducing tunneling terms of the form,
\begin{align}
  H_{{\it int} }
  &=im \int_0^{\pi} d\sigma
      \sum_I \psi^{I+1}(\sigma) \psi^{I}(2\pi-\sigma),
\end{align}
such that the total Hamiltonian is $H_0 + H_{{\it int}}$.
(Here and henceforth, 
we use the convention $\psi^{4}\equiv \psi^1$).
Analogously to the Ishibashi boundary state satisfying
the condition \eqref{overlap condition}, the ground state in the limit $|m| \to \infty$ is given by a conformal boundary state, $\ket{V}$, which satisfies
\begin{align}
  \big[\psi^{I+1}(\sigma)-i\psi^{I}(2\pi-\sigma)
  \big] |V\rangle=0,\tb 0\leq \sigma\leq \pi.
\label{eqn:bdy}
\end{align}
Solving the constraint, the state $|V\rangle$ is given in the form of a
fermionic coherent state. These types of states, which we will refer to as vertex states, have been considered in the
context of string field theory
\cite{gross1987field,gross1987operator,gross1987operator2,leclair1989string}
where they describe the interaction among strings. As before, we regularize this state and consider
$
|G \rangle=\mathcal{N} e^{-\epsilon H_0}|V\rangle
$,
which provides an approximation to the ground state of $H$ for large but finite $|m|$.
Once $|G\rangle$ is obtained, we can compute the reduced density matrices
$\rho_{A\cup B},\rho_{B\cup C}$, and $\rho_{C\cup A}$ as well as the entanglement measures.

Although Eq.~\eqref{eqn:bdy} uniquely defines the Majorana fermion vertex state, an equivalent and more general definition of vertex states, 
which also motivates the so-called Neumann coefficient approach to constructing them, proceeds as follows.
In the interest of generality, we consider the most general case of an $N$-junction, such that $N$ edge theories meet at a single point. Hence, we start with $N$ copies of chiral CFTs (edge theories)
defined on a spatial circle parameterized by
$0 \le \sigma \le 2\pi$.
Their Hilbert spaces are denoted by
$\mathcal{H}_{1,2,\ldots,N}$, respectively.
Together with the (imaginary) time direction $\tau$, we have a cylindrical spacetime. 
As usual, we can map each theory to the conformal plane through the coordinate transformation $z= e^{\tau +i \sigma}$, such that
the half of the cylinder $-\infty \le \tau \le 0$
is mapped to the unit disk, $|z|\le 1$.
We next consider conformal maps
$\omega_I$ from
the $I$-th unit disk to the complex plane $\mathbb{C}$ 
that are analytic inside the unit disk.
In particular, they 
map each disk to a separate wedge of the complex plane $\mathbb{C}$, 
with the requirement that the edges of each wedge are flush with one another 
so that
the desired boundary conditions are implemented.
This sequence of maps for one disk is 
illustrated for the case $N=3$ in Fig. \ref{fig:ct_close}(b).
We will elaborate more on this after
we present the explicit form of the conformal maps
momentarily.
Then, we define a vertex state
$|V\rangle \in
\mathcal{H}_1\otimes \mathcal{H}_{2}\otimes \cdots \otimes
\mathcal{H}_N$
by requiring it reproduce
correlation functions on the complex plane
as follows \cite{leclair1989string}:
\begin{gather}
  \langle  V|
    \,
  \big(
  O_\alpha|0\rangle_1
  \otimes
  O_\beta|0\rangle_{2}
    \otimes 
    \cdots
    \otimes 
  O_\gamma|0\rangle_{N}
  \big)
    \nonumber \\
  =
  \big\langle
    \,
  \omega_1[O_{\alpha}]
    \,
  \omega_2[O_{\beta}]
    \,
    \cdots
    \,
  \omega_N[O_{\gamma}]\, 
  \big\rangle_{\mathbb{C}}
  \label{def vertex 2}
\end{gather}
where $|0\rangle_{I}$ is the vacuum in $\mathcal{H}_{I}$,
$O_{\alpha, \beta, \cdots, \gamma}$
represents an arbitrary (primary)
operator acting on $\mathcal{H}_{1,2,\cdots, N}$,
$\omega_I[O]$ represents the transformation of
a primary operator $O$ by $\omega_I$,
$\omega_I[O(z)] = [\omega_I'(z)]^h O(\omega_I(z))$,
where $h$ is the conformal dimension of $O$. 
In order to fix the form of the conformal transformations $\omega_I$ which define the vertex state, we must impose additional constraints on $\ket{V}$. First, it is clear that, since the $N$ Hilbert space copies are equivalent, the vertex states must invariant under their cyclic permutation. That is to say, focusing on $N=3$, 
\begin{equation}
\langle V_{123}|=\langle V_{231}|=\langle V_{312}|, \label{eqn:cyclic-constraint}
\end{equation}
where the subscripts label the Hilbert space indices. Physically, this is just the statement that the trijunction is invariant under $120^{\circ}$ rotations. A second, less obvious requirement is given by, again focusing on $N=3$,
\begin{align}
    \langle V_{125}| \langle V_{5^\dg 34}|=\langle V_{235}|\langle V_{5^\dg 41}|,
\end{align}
The two sides of this expression correspond to gluing together two $N=3$ vertex states to obtain $N=4$ vertex states. This constraint expresses the fact that this $N=4$ vertex state must also be invariant under cyclic permutations of the Hilbert spaces (i.e. under $90^{\circ}$ rotations of the ``tetrajunction").\footnote{In the original string field theory context in which these vertex states first appeared, these cyclicity constraints follow from demanding gauge invariance of the string interaction vertex.} 
Enforcing these constraints restricts the choice of conformal transformations $\omega_I$, which in turn define the vertex state $\ket{V}$. We next describe choices of the $\omega_I$ satisfying these constraints, which then lead to vertex states satisfying Eq.~\eqref{eqn:bdy}.



For $N=2$, we can choose the following conformal maps~\cite{jevicki1988supersymmetry}:
\begin{gather}
    \omega_{I}(z)=\omega_{I,0} \frac{1+z}{1-z}; \quad \omega_{I,0} = -i e^{i\pi I}, \, I=1,2.
    \label{eqn:cfmap2}
\end{gather}
In this way, the first disk is mapped
to the upper half plane and the second to the lower half plane.
Note also that
the infinite past $\tau=-\infty$ is mapped to
$\pm i$, respectively. 
Here, we note that a quantum state at $\tau=0$ or $|z|=1$
can be obtained by a path integral from $\tau=-\infty$ or $|z|=0$
with possibly an insertion of an operator.
By the conformal maps $\omega_{1,2}$,
the $\tau=0$ slices of the disks are both mapped
to the real axis.
Hence, the field configurations for $\psi^1$ and $\psi^2$
are subject to the constraint in 
\eqref{overlap condition}; we will show this more explicitly in the following subsection.

Likewise, for $N=3$, we can choose $\omega_{1,2,3}$ as
\begin{gather}
  \label{eqn:conformal_map}
  \omega_I(z) =
    \omega_{I,0}
    \left(\frac{1+z}{1-z}\right)^{\frac{2}{3}};
    \,
    \omega_{I,0} = e^{\frac{4\pi i I}{3} - i\pi}, \, I=1,2,3.
\end{gather}
Note that $\omega_I(2\pi-\sigma)=\omega_{I+1}(\sigma)$ for $0\leq
\sigma\leq\pi$.
These conformal maps bring three disks
($0\leq \sigma \leq 2\pi,\tau<0$)
to the whole plane, 
such that each unit disk is mapped to a separate $120^{\circ}$ wedge of the conformal plane, as shown in Fig.~\ \ref{fig:ct_close}(b) and Fig.~\ref{fig:tripartition-map-RRRfluxes}(a). Here, the points at infinity are identified. 
We note that this construction is similar to, but slightly different from,
the conformal maps used in open string field theory by Witten \cite{witten1986non}; the CFTs we consider obey (potentially twisted) periodic boundary conditions. 
Though this alternative definition of the vertex states seems obtuse at first glance, we will see in the following that it provides an elegant way of deriving the explicit form of said states. 

\subsubsection{The Neumann coefficient method}
\label{The Neumann coefficient method}

Let us now move on to the methods of constructing vertex states.
On the one hand, 
the overlap condition \eqref{eqn:bdy}
can be solved directly, and the vertex state can be constructed
as a coherent state. 
We will discuss the direct construction in Appendix \ref{subsec:direct} 
and show the two methods give consistent results numerically.

On the other hand, 
the definition of vertex states \eqref{def vertex 2}
suggests the following strategy to construct vertex states,
which we call the the Neumann coefficient method.
For now, we focus on the NS sector for simplicity.  
We postulate the following Gaussian ansatz
for $|V\rangle$:
\begin{align}
  \label{gauss ansatz}
  |V\rangle=
  \exp
  \Big(
  \sum_{r,s\geq 1/2}\frac{1}{2}\psi_{-r}^{I}K_{rs}^{IJ} \psi_{-s}^J
  \Big)
  |0\rangle. 
\end{align}
(Here and henceforth, we adopt the convention in which repeated flavor indices $I,J,\cdots$ are summed over implicitly, unless otherwise stated.)
The coefficients $K_{rs}^{IJ}$ are chosen to reproduce the correlation function
on the right-hand side of \eqref{def vertex 2}.
Since $|V\rangle$ is Gaussian, it is sufficient to
consider the two point functions of the fermion fields.
We then consider, at $\tau=0$, the Neumann function
\begin{align}
  \label{eqn:neumann}
  K^{IJ}(\sigma,\sigma')
  &
    \equiv
    \langle \,
    \omega_I[\psi^I(\sigma)] \,
    \omega_J [\psi^J(\sigma')]\,
    \rangle_{\mathbb{C}}
    \nonumber \\
  &=
    \left(\frac{d\omega_I}{id\sigma}\right)^{1/2}
    \left(\frac{d\omega_J}{id\sigma'}\right)^{1/2}\frac{1}{\omega_I-\omega_J}.
\end{align}
(Here, $I,J$ are not summed on the right hand side.)
The Neumann coefficients $K_{rs}^{IJ}$ are related to 
the mode expansion of $K^{IJ}(\sigma,\sigma')$ as
\begin{equation}
K^{IJ}(\sigma,\sigma')=\sum_{r,s\geq 1/2} e^{ir\sigma}e^{is\sigma'}K^{IJ}_{rs}+\delta_{IJ}\sum_{r\geq 1/2}e^{-ir(\sigma-\sigma')}.
\end{equation}
Note that there are two contributions to $K^{IJ}$: the regular piece that contains
$K^{IJ}_{rs}$ and the singular piece
$\delta_{IJ}\sum_{r\geq 1/2}e^{-ir(\sigma-\sigma')}$.
The presence of the singular piece is
non-trivial, and needs to be verified case by case.

We now show the ansatz solution indeed satisfies
the boundary condition \eqref{eqn:bdy}.
We first note that,
with a proper choice of a branch in 
the conformal factor $(d\omega_I/i d\sigma)^{1/2}$, 
the Neumann function satisfies
\begin{align}
  \label{prop KIJ}
iK^{IJ}(2\pi-\sigma,\sigma')=K^{I+1,J}(\sigma,\sigma'),
\tb 0\leq\sigma\leq \pi,
\end{align}
which reflects the cyclic constraint of Eq.~\eqref{eqn:cyclic-constraint}. Using the mode expansion $\psi^I(\sigma)=\sum_r \psi_r^I e^{ir\sigma}$,
$\psi^I(\sigma)|V\rangle$ can be expressed as
\begin{align}
  &\psi^I(\sigma)|V\rangle
  \\
  &=
  \sum_{r\geq 1/2}\psi^I_{-r}e^{-ir\sigma}
  |V\rangle
  +
  \sum_{r,s\geq 1/2}
  e^{ir\sigma}K_{rs}^{IJ}\psi_{-s}^J
  |V\rangle
    \nonumber \\
  &= \int \frac{d\sigma'}{2\pi}
    K^{IJ}(\sigma,\sigma')\psi_{\mathrm{cr.}}^J(\sigma')|V\rangle,
\end{align}
where $\psi_{\mathrm{cr.}}^I(\sigma)=\sum_{r\geq 1/2}\psi^I_{-r}e^{-ir\sigma}$. 
Using the cyclic property of the Neumann function given in Eq.~\eqref{prop KIJ}, we find,
\begin{align}
  \psi^I(2\pi-\sigma)|V\rangle
  &=\int \frac{d\sigma'}{2\pi}K^{IJ}(2\pi-\sigma,\sigma')\psi_{\mathrm{cr.}}^J(\sigma')|V\rangle
    \nonumber \\
  &=(-i)\int \frac{d\sigma'}{2\pi}K^{I+1,J}(\sigma,\sigma')\psi_{\mathrm{cr.}}^J(\sigma')|V\rangle
    \nonumber \\
  &=-i \psi^{I+1}(\sigma)|V\rangle.
\end{align}
This completes the proof.
Note that it was crucial to carefully take into account the singular part of the Neumann function. 
The proof presented here applies for the NS sector,
and we leave the more complicated case of the R sector (Sec.\ \ref{Subsec:R-R-R})
to Appendix \ref{app:another-twist},\ \ref{app:boundary-NS-R}. 

The direct and Neumann coefficient methods complement one other.
When both methods can be applied, 
they give rise to the same (consistent) vertex states.
We demonstrate the equivalence of these 
methods in the NS-NS-NS sector in Appendix
\ref{subsec:direct}.
In other sectors, 
because of the presence of
zero modes, and 
because of the branch cuts,
sometimes 
one method has an advantage
over the other method.
In general, 
vertex states obtained from
these two methods 
are consistent, 
but may differ by 
an extra operator insertion 
at the junction
\cite{imamura2006boundary,imamura2008boundary}.

\subsubsection{Complex fermion}

We close this subsection by commenting on the
case of complex fermions, which parallels 
the treatment for real fermions.
Indeed, the desired vertex state is obtained by combining two copies of real fermions.
We consider complex fermion fields $f^I(\sigma), f^{I, \dag}(\sigma)$.
In the NS sector, they can be expanded as
\begin{align}
  &
     f(\sigma)=\sum_{s\in\mathbb{Z}+1/2}e^{-i\sigma s} f_s,
    \quad
     f^\dag(\sigma)=\sum_{s\in\mathbb{Z}+1/2}e^{i\sigma s} f_s^\dag,
    \nonumber \\
  &\qquad
    \mbox{with}
    \quad
    \lbrace f^{\ }_s, f_{s'}^\dag\rbrace=\delta_{s,s'}.
\end{align}
We have a similar mode expansion in the R-sector.
We consider a vertex state obeying the overlap condition,
\begin{align}
  \label{bdry complex}
  &
  \big[
  f^{I+1}(\sigma)-if^I(2\pi-\sigma)\big] |V\rangle=0,
  \nonumber \\
  &
    \big[
f^{I+1,\dg}(\sigma)-if^{I,\dg}(2\pi-\sigma)\big] |V\rangle=0.
\end{align}
The complex fermion field $f,f^\dag$
can be decomposed into two real fermion fields,
$\psi$ and $\varphi$ as 
$f=(\psi-i\varphi)/\sqrt{2}$,
$f^\dg=(\psi+i\varphi)/\sqrt{2}$.
Correspondingly, the Fourier modes of $f^{\dag}$ and $f$,
$f(\sigma)=\sum_r e^{ir \sigma} f_r$
($\lbrace f_r^\dg, f_s\rbrace=\delta_{r,s}$),
are
related to the Fourier modes of $\psi$ and $\varphi$ as
$
f_r = (\psi_r-i\varphi_r)/\sqrt{2},
$,
$
f_r^\dg = (\psi_{-r}+i\varphi_{-r})/\sqrt{2}.
$
The ansatz solution is then
\begin{equation}
\begin{aligned}
  |V\rangle&=\exp{\Big(
      \frac{1}{2}
      \sum_{r,s\geq 1/2}
    \psi_{-r}^I K_{rs}^{IJ}\psi_{-s}^J
    + \varphi_{-r}^I K_{rs}^{IJ}\varphi_{-s}^J\Big)}
  |0\rangle\\
  &=\exp{\Big(
      \sum_{r,s\geq 1/2}
      f_r^{I\dg} K_{rs}^{IJ} f_{-s}^J \Big)}|0\rangle.
\end{aligned}
\end{equation}
The treatment of the R sector follows similarly, although we need to
take into account the presence of zero modes properly, as we shall see in the following subsections.

\subsection{Bipartition}
\label{subsec:bipartition}

In this subsection,
we consider the bipartitions of
a chiral $p$-wave superconductor and a Chern insulator,
using the Neumann coefficient method described above.
As mentioned at the beginning of this section, we investigate the effect of inserting non-trivial $\pi$-fluxes through the 
cylinder on the entanglement. As shown in Fig.\ \ref{fig:bipartition-fluxes}, we consider the insertion of (a) no flux (b) $\pi$-flux through the cylinder, and (c) a $\pi$-flux through one end of the cylinder, which exits through the entanglement cut. 
For the chiral superconductor, a $\pi$-flux is an extrinsic defect which traps a Majorana zero-mode, forming an Ising anyon. 
Thus, (b) can be viewed as creating a pair of Ising anyons in the bulk and dragging them to opposite ends of the cylinder, while (c) results from dragging only one Ising anyon to an edge and leaving the other in the bulk. In the bulk language, the creation and manipulation of the Ising anyons leaves behind a Wilson line on the cylinder or, equivalently, an anyon flux through the cylinder. At the level of the edge theories, the braiding of the Majorana fermions around the Ising anyon flux results in a phase of $-1$. Hence, the three configurations in Fig.\ \ref{fig:bipartition-fluxes} are described by the boundary condition sectors of the edge theories: (a) NS-NS, in which all fermions obey anti-periodic boundary conditions (b) R-R, in which all fermions obey periodic boundary conditions, and (c) NS-R, in which the fermions on the left (right) cylinder obey anti-periodic (periodic) boundary conditions. We compute the entanglement in each sector in turn. As is well-established, we obtain an area law for case (a) and an area law term plus a subleading $\ln \sqrt{2}$ correction from the Ising anyons for case (b), which requires a careful treatment of the zero modes \cite{yao2010entanglement,sohal2020nonabelian,lim2021disentangling}. The case (c) has not been considered before and we find a novel subleading correction to the entanglement. 

\subsubsection{The NS-NS sector}

The setup of the calculation for the NS-NS sector is already outlined above; all that remains is to explicitly evaluate the Neumann functions.
Noting $\frac{d\omega_I}{id\sigma}=\frac{2z\omega_{I,0}}{(1-z)^2}$ and choosing
the branch cuts carefully
($\sqrt{\omega_{1,0}}=\sqrt{i},\sqrt{\omega_{2,0}}=i\sqrt{i}$, which leads to
$\sqrt{\omega_{1,0}\omega_{2,0}}=-1$),
we obtain:
\begin{equation}
\begin{aligned}
  K^{11}&=K^{22}=\frac{\sqrt{zz'}}{z-z'}=\sum_{r\geq 1/2}e^{-ir(\sigma-\sigma')},
  \\
K^{12}&=-K^{21}=\frac{i\sqrt{zz'}}{1-zz'}=i\sum_{r\geq 1/2}e^{ir(\sigma+\sigma')}.
\end{aligned}
\end{equation}
Note that $K^{11} = K^{22}$ yields the expected singularity. 
We also note that under
$\sigma\ra 2\pi-\sigma$
($z\ra 1/z$, $\sqrt{z}\ra -1/\sqrt{z}$),
the Neumann function satisfies 
$K^{1J}(2\pi-\sigma,\sigma')+iK^{2J}(\sigma,\sigma')=0$ and
$K^{2J}(2\pi-\sigma,\sigma')-iK^{1J}(\sigma,\sigma')=0$ for $0\leq \sigma\leq
2\pi$.
From the expansion of $K^{12}$, we conclude $K^{12}_{rs} = -K^{21}_{rs} = \delta_{rs}$. Plugging this into Eq.\ \eqref{gauss ansatz}, 
we obtain the Ishibashi state \eqref{fermion ishibashi}
as expected. 

\subsubsection{The R-R sector}
\label{subsubsec:RR}

Let us now consider the vertex state in the R-R sector.
We denote the fermion fields with the R boundary condition
as $\chi^I(\sigma)$. As before, the vertex state satisfies
\begin{align}
  \label{bdy 2 RR}
  &
  [\chi^1(\sigma)+i\chi^2(2\pi-\sigma)]|V\rangle
  \nonumber \\
  &
  =[\chi^2(\sigma)-i\chi^1(2\pi-\sigma)]|V\rangle=0
\end{align}
for  $0\leq\sigma\leq 2\pi$. 
In the bulk, 
this situation corresponds to
a flux or, Ising anyon Wilson line,
threading the hole of the cylinder
[Fig.\ \ref{fig:bipartition-fluxes} (b)].
From the edge theory point of view,
we need to include suitable twist operators  
to introduce branch cuts, which enforce periodic boundary conditions for the fermions. This will modify the Neumann function, which we now
denote as $R^{IJ}$. It is related to the Neumann function
in the NS sector via:
\begin{equation}
R^{IJ}(\sigma,\sigma')=K^{IJ}(\sigma,\sigma')g^{IJ}(\sigma,\sigma'), \label{eqn:branch-cut-neumann-fn}
\end{equation}
where $g^{IJ}$ is the new factor arising from the branch cuts. 
(Here, the summation convention
does not apply in the right hand side.)
We work with 
the following choice of 
the branch cuts, 
\begin{equation}
  g^{IJ}=\frac{1}{2}
  \left[\sqrt{\frac{(\omega_I-\omega_{1,0})(\omega'_J-\omega_{2,0})}{(\omega'_J-\omega_{1,0})(\omega_I-\omega_{2,0})}}+(\omega_I\leftrightarrow\omega'_J)
  \right],
\end{equation}
where we recall
$\omega_{1,0}=i,\omega_{2,0}=-i$.
Other choices are also possible
and give an identical vertex state,
as we demonstrate in Appendix 
\ref{app:another-twist}. 
%
Using the conformal map in Eq.\ \eqref{eqn:cfmap2}, the explicit form of $g^{IJ}$ is 
\begin{equation}
\begin{aligned}
  g^{11}&=g^{22}=\frac{1}{2}\left(\sqrt{\frac{z}{z'}}+\sqrt{\frac{z'}{z}}\right),
  \\
g^{12}&=g^{21}=\frac{1}{2}\left(\sqrt{zz'}+\frac{1}{\sqrt{zz'}}\right).
\end{aligned}
\end{equation}
These functions satisfy $g^{1J}(\sigma,\sigma')=-g^{2J}(2\pi-\sigma,\sigma')$ (using $z\ra 1/z, \sqrt{z}\ra -1/\sqrt{z}$). The Neumann function is:
\begin{equation}
\begin{aligned}
  R^{11}&=R^{22}=\frac{1}{2}\frac{z+z'}{z-z'}=\frac{1}{2}+\sum_{n\geq 1}e^{-in(\sigma-\sigma')},
  \\
R^{12}&=-R^{21}=\frac{i}{2}\frac{1+zz'}{1-zz'}=\frac{i}{2}+i\sum_{n\geq 1}e^{in(\sigma+\sigma')}.
\end{aligned}
\end{equation}
They satisfy $R^{1J}(2\pi-\sigma,\sigma')-iR^{2J}(\sigma,\sigma')= R^{2J}(2\pi-\sigma,\sigma')+iR^{1J}(\sigma,\sigma')=0$. Again, the correct singular terms show up 
in $R^{11}$ and $R^{22}$.  
The solution for $|V\rangle$ for real fermions in
the R-R sector is then
\begin{align}
  |V\rangle=
  \exp\Big(
  -i\sum_{n\geq 1}\chi_{-n}^1\chi_{-n}^2\Big)
  |\Omega\rangle,
\label{eqn:solutionRR}
\end{align}
with an additional requirement $[\chi_0^1+i\chi_0^2]|\Omega\rangle=0$.
One can verify that they satisfy the boundary condition \eqref{bdy 2 RR}.
The requirement that $|V\rangle$ has definite parity for the zero mode can also be understood from the $i/2$ term in $R^{12}$. 

The zero modes $\chi_0^1,\chi_0^2$ of the real fermion need to be handled with
extra care. $\chi_0^1,\chi_0^2$ live on the inner edges of the cylinders. To
have a well-defined Hilbert space, we also need to include the zero modes on the
outer edges of the cylinders, which we denote as
$\bar{\chi}_0^1,\bar{\chi}_0^2$, as shown in Fig.\ \ref{fig:bipartition-fluxes}(b).
Indeed, we recall that before making a physical cut along the entanglement cut, the cylinder with an Ising anyon flux passing through it is topologically equivalent to a sphere with a pair of Ising anyon defects. The anyons yield a double degeneracy, as each has quantum dimension $\sqrt{2}$. This corresponds to choosing whether the complex fermion formed from the corresponding zero modes, $\bar{\chi}_0^1+i\bar{\chi}_0^2$, is occupied or unoccupied. We must make a choice of which state in this degenerate subspace  we wish to compute the entanglement for.
For concreteness, we choose the state in which this fermion is unoccupied, which amounts to imposing
the boundary condition
$[\bar{\chi}_0^1+i\bar{\chi}_0^2]|\Omega\rangle=0$ for the outer edge zero modes. If we define the complex fermion 
\begin{equation}
g_i=\frac{1}{\sqrt{2}}(\chi_0^1+i\chi_0^2),\tb
g_o = \frac{1}{\sqrt{2}}(\bar{\chi}_0^1+i\bar{\chi}_0^2),
\end{equation}
the zero-mode vacuum state is $|\Omega\rangle=|0_i,0_o\rangle$. 
This completes the construction of the boundary state. 

Note that $g_i$ and $g_o$ mix the Hilbert spaces of the left and right cylinders. When we compute the entanglement we must trace out one of these cylinders, and so it is necessary to perform a change of basis to complex fermion modes localized on either the left or right cylinder:
$
g_A = (\chi_0^1+i\bar{\chi}_0^1)/\sqrt{2},\
$
$
g_B = (\chi_0^2+i\bar{\chi}_0^2)/\sqrt{2}.
$
In this basis, the vacuum is a maximally entangled state:
\begin{equation}
    |\Omega\rangle=|0_i, 0_o\rangle
= (|0_A 0_B\rangle-i|1_A 1_B\rangle)/\sqrt{2}.
\end{equation}
Below, we will see this  gives a contribution of $\ln{2}$ to the entanglement entropy.

\subsubsection{The NS-R sector}

Finally, we consider the NS-R sector which, as described above, describes a novel configuration in which we insert an anyon flux through one end of the cylinder which then exits through the entanglement cut. From Fig. \ref{fig:bipartition-fluxes}(c), we see that the fermions on the right cylinder braid around the anyon flux and so obey R boundary conditions, whereas the fermions on the left cylinder do not and hence are in the NS sector. In order to describe the gapped edge state at the entanglement cut, we must impose a modified boundary condition:
\begin{align}
    [\psi(\sigma) + i \mathrm{sgn}(\pi - \sigma) \chi(2\pi - \sigma)]\ket{V} = 0 ,
\end{align}
for $0 \leq \sigma < 2\pi$. Here, $\psi$ ($\chi$) obeys NS (R) boundary conditions.  
Formally, the sign function is needed to ensure the above expression is well-defined under shifts of $\sigma \to \sigma + 2\pi$. Physically, it represents the fact that an anyon flux is piercing the entanglement cut. Indeed, the Ising twist field is precisely the operator at the level of the edge CFT which introduces such a ``kink" for the Majorana fields.


To the best of our knowledge, the vertex state
in this case was first constructed in
\cite{jevicki1988supersymmetry}.
In the NS-R sector, we only need to introduce the branch cut for the second
string. The branch cut factor $g^{IJ}$ is chosen as
\cite{jevicki1988supersymmetry}:
\begin{equation}
g^{IJ}(\sigma,\sigma')=\frac{1}{2}\left[ \sqrt{\frac{\omega_I-\omega_{2,0}}{\omega'_J-\omega_{2,0}}}+\sqrt{\frac{\omega'_J-\omega_{2,0}}{\omega_I-\omega_{2,0}}}\right]. 
\end{equation}
Explicitly,
\begin{equation}
\begin{aligned}
g^{11}&=\frac{1}{2}\left(\sqrt{\frac{1-z'}{1-z}}+\sqrt{\frac{1-z}{1-z'}} \right),\\
g^{12}&=\frac{i}{2}\left(\sqrt{\frac{1-z'}{1-z}}\frac{1}{\sqrt{z'}}-\sqrt{z'}\sqrt{\frac{1-z}{1-z'}} \right),
\\
g^{22}&=\frac{1}{2}\left(\sqrt{\frac{1-z'}{1-z}}\sqrt{\frac{z}{z'}}+ \sqrt{\frac{1-z}{1-z'}}\sqrt{\frac{z'}{z}}\right).
\end{aligned}
\end{equation}
$R^{IJ}$ satisfies $R^{IJ}(\sigma,\sigma')=-iR^{I+1,J}(2\pi-\sigma,\sigma')$ for $0\leq\sigma\leq\pi$. 
The mode expansion of $R^{IJ}$ needed to extract the $R^{IJ}_{rs}$ in the definition of the vertex state, takes a more complicated form than that of the preceding two cases:
\begin{equation}
  \begin{aligned}
    R_{rs}^{11}&=\frac{r-s}{2(r+s)}u_{2r-1}u_{2s-1},
    \\
    R_{rn}^{12}&=-R^{21}_{nr}=\frac{n+r}{2(n-r)}u_{2r-1}u_{2n},
    \\
    R_{nm}^{22}&=\frac{n-m}{2(n+m)}u_{2n}u_{2m},
  \end{aligned}
\end{equation}
where $u_n$ is the expansion coefficients of $u(x)$:
\begin{equation}
u(x)=\sqrt{\frac{1+x}{1-x}}=\sum_{n=0}^{\infty}u_n x^n.
\end{equation}
Making use of this mode expansion and separating the oscillator and zero-mode contributions, we can write out the vertex state of Eq.~\eqref{gauss ansatz} as
\begin{widetext}
\begin{equation}
\begin{aligned}
|V\rangle
&=\exp\Big(
\sum_{r,s\geq 1/2}\frac{1}{2}\psi_{-r} R^{11}_{rs}\psi_{-s}
+\sum_{m,n\geq 1}\frac{1}{2}\chi_{-n}R^{22}_{nm}\chi_{-m}
\\
&\qquad
+\sum_{r\geq 1/2, n\geq 1}\psi_{-r}R^{12}_{rn}\chi_{-n}+\sum_{r\geq 1/2}2\psi_{-r}R^{12}_{r0}\chi_0
+\sum_{n\geq 1}2\chi_{-n}R^{22}_{n0}\chi_0
\Big)|\Omega\rangle.
\end{aligned}
\end{equation}

Now, as in the R-R sector, to fix the form of the vacuum $\ket{\Omega}$, we must treat the zero-mode sector carefully. Indeed, due to the $\pi$ flux through one half of the cylinder, we have another zero mode, $\bar{\chi}_0$, on the outer edge of the left cylinder [Fig.\ \ref{fig:bipartition-fluxes}(c)]. We can combine them
to define the complex fermion operator $g_0$:
\begin{equation}
g_0=\frac{1}{\sqrt{2}}(\chi_0-i\bar{\chi}_0),\tb
g_0^\dg =\frac{1}{\sqrt{2}}(\chi_0+i\bar{\chi}_0).
\label{eqn:Majorana-zero-mode}
\end{equation}
Now, prior to making the entanglement cut, this flux configuration is again topologically equivalent to a sphere supporting a pair of Ising anyons, corresponding to the $\chi_0$ and $\bar{\chi}_0$ zero modes, yielding a double degeneracy associated with the occupation of $g_0$. (Note that, in contrast to the R-R case, cutting the system along the entanglement cut does not introduce additional zero modes). We must again make a choice of which state in which to compute the entanglement. 
We can fix the state by choosing a value for the occupation number of $g_0$ of the reference state $\ket{\Omega}$; for simplicity, we take $g_0$ to be unoccupied, so that $\ket{\Omega} = \ket{0}$.
Finally, to simplify the expression for the vertex state, we observe that
$X\equiv \sqrt{2}(\sum_{r\geq 1/2}\psi_{-r}R_{r0}^{12}+\sum_{n\geq 1}\chi_{-n}R^{22}_{n0})g_0^\dg$
and
$Y\equiv \sqrt{2}(\sum_{r\geq 1/2}\psi_{-r}R_{r0}^{12}+\sum_{n\geq 1}\chi_{-n}R^{22}_{n0})g_0$,
commute, $[X,Y]=0$, and hence $e^{X+Y}=e^X e^Y$.
The vertex state thus takes the form 
\begin{equation}
\begin{aligned}
  |V\rangle=&\exp \Big(\sum_{r,s\geq 1/2}\frac{1}{2}\psi_{-r}R_{rs}^{11}\psi_{-s}+\sum_{m,n\geq 1}\frac{1}{2}\chi_{-n}R^{22}_{nm}\chi_{-m}
  \\
  &\qquad
  +\sum_{r\geq 1/2,n\geq 1}\psi_{-r}R^{12}_{rn}\chi_{-n} 
  +\sqrt{2}(\sum_{r\geq 1/2}\psi_{-r}R_{r0}^{12}+\sum_{n\geq 1}\chi_{-n}R^{22}_{n0})g_0^\dg \Big)|0\rangle.
\end{aligned}
\end{equation}
\end{widetext}

\subsubsection{Entanglement entropy}
\label{Entanglement entropy}

Having constructed the relevant boundary states for
the NS-NS, R-R, NS-R sectors, 
we now proceed to compute the entanglement entropy
$S_A$ after tracing out one half of the cylinder. 
Let us start with the NS-NS sector.
We recall that the ground state of the entanglement interface is given by a regularized version of the boundary state, as stated in Eq.\ \eqref{GS edge}; this
amounts to replacing $\psi_{-r}^I\ra\psi_{-r}^I e^{-\epsilon r}$
in Eq.\ \eqref{fermion ishibashi}.
The entanglement entropy can directly be evaluated as 
\begin{equation}
\begin{aligned}
S 
&=
\left(  1- \epsilon\frac{d}{d\epsilon} \right)
\ln{[\prod_{r\geq 1/2}(1+q^r)]}
\end{aligned}
\end{equation}
where $q=e^{2\pi i\tau}= e^{-4 \epsilon}$ and $\tau=\frac{2i\epsilon}{\pi}$.
We can write the argument of the logarithm in terms of the Dedekind $\eta$ function and a Jacobi $\theta$ function:
\begin{align}
\prod_{r\geq 1/2}(1+q^r)=q^{1/48}\sqrt{\frac{\theta_3(\tau)}{\eta(\tau)}}.
\end{align}
Under the modular $\mathcal{S}$ transformation
and taking the limit $\epsilon\ra0$ limit (which corresponds to taking the bulk gap to be very large), we have:
\begin{equation}
\frac{\theta_3(\tau)}{\eta(\tau)}=\frac{\theta_3(-\frac{1}{\tau})}{\eta(-\frac{1}{\tau})}
\ra \frac{1}{(e^{-\frac{2\pi i}{\tau}})^{1/24}}=e^{\frac{\pi^2}{24\epsilon}}.
\end{equation}
We thus find,
\begin{equation}
  S_{NS-NS}^{\mathrm{Real.}}\ra
  \frac{\pi (1/2)}{24} \frac{L}{\epsilon}
  \quad
  \mbox{as}\quad
  \frac{L}{\epsilon} \ra \infty,
\end{equation}
as expected.
Here, we reinstated the IR length scale
$L$ (which has been set to $2\pi$ for simplicity)
to make the area law form of the entropy
more explicit and so that the dimensions are correct.
We also make the chiral central charge
$c=1/2$ dependence explicit.


The entanglement entropy in the R-R sector
can be computed similarly.
However, the presence of the zero modes make the calculations
slightly more subtle.
Let us first compute the contribution from the oscillator modes $n\geq 1$.
With the regulator $\epsilon$, it can be computed as 
\begin{equation}
  S_{\mathrm{oscil.}}=
  \left( 1- \epsilon \frac{d}{d\epsilon}\right)
  \ln{[\prod_{n\geq 1}(1+q^n)]}.
\end{equation}
The product can be identified with $\theta_2$ function:
\begin{equation}
\prod_{n\geq 1}(1+q^n)=\frac{1}{\sqrt{2}}\sqrt{\frac{\theta_2(\tau)}{\eta(\tau)}}q^{-1/24}.
\end{equation}
Under the modular $\mathcal{S}$ transformation and again taking the limit $\epsilon\ra0$, we have:
\begin{equation}
\frac{\theta_2(\tau)}{\eta(\tau)}=\frac{\theta_4(-\frac{1}{\tau})}{\eta(-\frac{1}{\tau})}
\ra \frac{1}{(e^{-\frac{2\pi i}{\tau}})^{1/24}}=e^{\frac{\pi^2}{24\epsilon}}.
\end{equation}
This gives
\begin{equation}
S_{\mathrm{oscil.}}=\frac{\pi (1/2)}{24}\frac{L}{\epsilon}-\frac{1}{2}\ln{2}.
\end{equation}
For the zero mode part, after the basis transformation, the vacuum takes the form of a maximally entangled state,
$|\Omega\rangle=|0_i, 0_o\rangle
= (|0_A 0_B\rangle-i|1_A 1_B\rangle)/\sqrt{2},
$
which gives a contribution of $\ln{2}$. Summing up these two terms, the total entanglement entropy is:
\begin{equation}
S_{R-R}^{\mathrm{Real.}}
=\frac{\pi (1/2)}{24}\frac{L}{\epsilon}
+\ln{\sqrt{2}}.
\end{equation}
Compared with $S_{NS-NS}^{\mathrm{Real.}}$, the extra contribution $\ln{\sqrt{2}}$ is exactly the topological entanglement entropy from the $\sigma$ anyon, as expected \cite{yao2010entanglement}. 

\begin{table}
\centering
\begin{tabular}{ c | c c c c }
\hline \hline
$\epsilon$ & 0.005 & 0.008 & 0.01 & 0.02\\
 \hline
 \hline
$S^{\mathrm{Real.}}_{R-R}$ & 82.5933 &    51.7508 &  41.4699  & 20.9082\\
$S^{\mathrm{Real.}}_{R-NS}$ & 82.3433 & 51.5008 & 41.2199 &  20.6582\\
$\Delta S$ & 0.2500 &  0.2500    & 0.2500   & 0.2500\\
\hline \hline
\end{tabular}
\caption{$S^{\mathrm{Real.}}_{R-R}$ and $S^{\mathrm{Real.}}_{R-NS}$ for various choices of $\epsilon$. For each fixed $\epsilon$, we increase cutoff $N$ until $S$ saturates. We observe that the difference $\Delta S= S^{\mathrm{Real.}}_{R-R}-S^{\mathrm{Real.}}_{R-NS}$ is a constant.}
\label{tbl:RNSMajorana}
\end{table}

We now proceed to the NS-R case. 
Since the entanglement entropy
in this case is not amenable to analytical calculations,
we will perform a numerical computation using the correlation matrix method introduced in Section \ref{sec:fermonic-Gaussian} with a cutoff of mode $N_c$. 
For a given value of $\epsilon$, we take $N_c$ to be sufficiently large such that $S_A$ does not appreciably change with further increases in $N_c$. 
We collect the results in Table~\ref{tbl:RNSMajorana}. 
We observe that the area law contributions
($O(L/\epsilon)$) to $S^{\mathrm{Real.}}_{R-R}$ and $S^{\mathrm{Real.}}_{NS-R}$
cancel out exactly, and the difference
\begin{align}
  \Delta S= S^{\mathrm{Real.}}_{R-R}-S^{\mathrm{Real.}}_{R-NS}=0.2500
\end{align}
appears to be remarkably well quantized. Now, we recall that, in the R-R sector, the presence of the anyon flux passing through the cylinder (i.e. the presence of Ising anyons on the ends of the cylinder) led to a contribution of
$\Delta S_0 = S^{\mathrm{Real.}}_{R-R}-S^{\mathrm{Real.}}_{NS-NS}=\ln{\sqrt{2}}=0.3466$ 
to the entanglement entropy over the NS-NS case, in which there was no flux. 
We see that $0<\Delta S<\Delta S_0$.
This seems reasonable, as one expects the two halves of the cylinder in the present NS-R case where one Ising anyon straddles entanglement cut to somehow be less entangled than the R-R case, where the Ising anyons are located deep in the bulks of the two subregions. Evidently, $\Delta S_0-\Delta S=0.0966$ corresponds to a contribution to the entanglement from the anyon flux which pierces the entanglement cut. We should, however, perhaps be careful in identifying this as a universal contribution, as this cut-and-glue approach likely corresponds to a particular choice of regularization of how the anyon flux pierces the cut. The value of this new topological contribution may depend on this regularization.
Additionally, we note that the examination of the entanglement spectrum in the NS-R sector shows that levels are all equally spaced with no degeneracy. 
The equal spacing structure encodes the CFT signature. 

Finally, we consider the entanglement entropy
for the case of complex fermion,
i.e., the edge theory of a Chern insulator
with unit Hall conductivity,
and make a comparison with the above results.
In the NS-NS sector, 
the entanglement entropy for the complex fermion
is simply twice as large as the real fermion case,
\begin{equation}
  S_{NS-NS}^{\mathrm{Cplx.}} = 2S_{NS-NS}^{\mathrm{Real.}}
  = \frac{\pi}{24} \cdot 
  \frac{L}{\epsilon}.
\end{equation}
In the R-R sector,
we need to include the effect of
the fermion zero modes
properly, while
the treatment for the oscillator part is essentially the same.
For the zero mode part, since $\chi_0$ is already a well-defined degree of freedom, paired with $\chi_0^\dg$, we can only consider the inner edges. The vacuum $|\Omega\rangle$ needs to satisfy $(\chi_0^1-i\chi_0^2)|\Omega\rangle=0, (\chi_0^{1,\dg}-i\chi_0^{2,\dg})|\Omega\rangle=0$, which can be chosen as
$
|\Omega\rangle=(\chi_0^{2,\dg}+i\chi_0^{1,\dg})|0\rangle.
$
This is a maximally-entangled pair state and gives $S_0=\ln{2}$ contribution to $S$. To sum up, 
\begin{equation}
  S_{R-R}^{\mathrm{Cplx.}}=2S_{\mathrm{oscil.}}+S_0
  =\frac{\pi}{24}\cdot \frac{L}{\epsilon}.
\end{equation}
There is no topological contribution for the complex fermion. 
Furthermore, 
the numerical calculation of the NS-R case shows
$
    S_{NS-NS}^{\mathrm{Cplx.}} = S_{NS-R}^{\mathrm{Cplx.}}=S_{R-R}^{\mathrm{Cplx.}}.
$
This is desired since we expect $S_{NS-R}^{\mathrm{Cplx.}}$ to lie between $S_{NS-NS}^{\mathrm{Cplx.}}$ and $S_{R-R}^{\mathrm{Cplx.}}$. 
Once again, 
the NS-R entanglement spectrum shows equal spacing behavior with no degeneracy.

\section{Tripartite vertex states and entanglement}
\label{Tripartition solution}

Having illustrated how the Neumann coefficient
method reproduces the expected boundary states and entanglement entropy for a bipartition on the cylinder with and without flux threading it, as well as having derived a new result for the entanglement in the case where a flux pierces the cut, we turn to the main focus of this work, namely the entanglement for a tripartition [Fig.\ \ref{partition}(b)]. 
We will again focus primarily on the case of a chiral $p$-wave superconductor and consider the effect of inserting $\pi$-fluxes through the cylinders. 
In particular, we investigate the entanglement when no fluxes are inserted 
and when two fluxes are inserted through two cylinders such that one flux exits through the remaining cylinder and the other flux through the entanglement cut [Fig.~\ref{fig:tripartition-map-RRRfluxes}(a)]. At the level of the edge theories, these correspond to the NS-NS-NS 
and R-R-R sectors, respectively. 
We construct the vertex states for each case next before discussing the tripartite entanglement measures introduced in Sec.\ \ref{sec:measure}. 

As a complement to the Neumann coefficient approach, we also introduce a direct calculation method for computing the vertex state in Sec.\ \ref{subsec:direct}. We show these two methods give identical results for the vertex state solution numerically.

\subsection{Vertex states}

\label{subsec:3-string-vertex}
\subsubsection{The NS-NS-NS sector}

We first consider the simplest case in which no fluxes are inserted through the cylinders. The required vertex state $\ket{V}$ is given by the Gaussian ansatz of Eq.\ \eqref{gauss ansatz}, the construction of which is outlined in Sec.~\ref{The Neumann coefficient method}. 
All that remains is to determine the explicit form of the Neumann coefficients
from the correlation function, Eq.\ \eqref{eqn:neumann}.
The conformal factor in Eq.~\eqref{eqn:neumann}
is given explicitly as
$(\frac{d\omega_I}{id\sigma})^{1/2}=\frac{1}{\omega_I^{1/4}}(\frac{(\omega_I^3+1)}{3})^{1/2}$.
We choose the branch such that
$\omega_I^{1/4}(2\pi-\sigma)= i\omega_{I+1}^{1/4}(\sigma)$. This can be achieved by the following choice:
\begin{align}
  &
    \omega_I^{1/4}(\sigma)
    = \tilde{\omega}_I \left(\frac{1+e^{i\sigma}}{1-e^{i\sigma}}\right)^{1/6}
    \nonumber \\
  &
    \mbox{with}
    \quad
    \tilde{\omega}_1 =e^{i \pi /12},
    \quad 
    \tilde{\omega}_2 =e^{-i 7 \pi /12},
    \quad
    \tilde{\omega}_3 =e^{i 3 \pi /4}.
\end{align}
The explicit form of the
Neumann coefficients $K^{IJ}_{rs}$ is technically involved and not particularly physically illuminating, and so we relegate it to 
Appendix \ref{app:K-matrix-coeffi}. 

\subsubsection{The R-R-R sector}
\label{Subsec:R-R-R}

\begin{figure}[tb]
\includegraphics[width=0.45\textwidth]{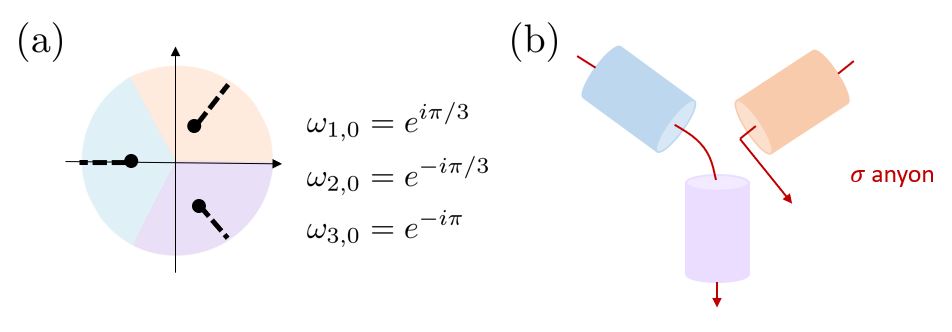}
\caption{
(a) The $\sigma$ anyon flux insertion in the R-R-R sector. One $\sigma$-anyon is forced to be exited at the junction. 
(b) The choice of the branch cuts for the R-R-R sector 
tripartition vertex state. The branch cuts connect $\omega_{I,0}$ to $\infty$. 
}
\label{fig:tripartition-map-RRRfluxes} 
\end{figure}

Next we consider the case in which all fermions are in the R sector.
Similar to the NS-R sector
discussed for the case of a bipartition,
in the R-R-R sector,
the conservation of topological charge
enforces the presence of an Ising anyon
at the junction where all three 
entanglement boundaries meet.
From the edge theory point of view,
we must again compute the Neumann functions for periodic fermions, which takes the form in Eq.~\eqref{eqn:branch-cut-neumann-fn}  with the factor $g^{IJ}$ accounting for the branch cuts. We choose
to work with the branch cut configuration
in Fig.~\ref{fig:tripartition-map-RRRfluxes}(b). 

%
To determine the branch cut factor $g^{IJ}(\sigma,\sigma')$, the following
general properties should be satisfied
\cite{ginsparg1988applied}: (i) $\omega$ and $\omega'$ are symmetric in $g^{IJ}$ (thus anti-symmetric in $R^{IJ}$); (ii) The branch points include $\omega_{1,0}$, $\omega_{2,0}$, and $\omega_{3,0}$; (iii) $g$ reduces to 1 when $\omega'\ra \omega$, so $R^{IJ}$ reduces to $K^{IJ}$ in this limit. Furthermore, for our specific problem, $R^{IJ}$ should also satisfy: (iv) The singular term in $R^{IJ}$ must be $\delta^{IJ}\sum_{n\geq 1}e^{-in(\sigma-\sigma')}$ to ensure the boundary condition is properly satisfied, as we show in Appendix \ref{app:boundary-NS-R}. This extra requirement is non-trivial, and may rule out some of the candidates that satisfy (i-iii). 

We propose to use the following branch cut factor:
\begin{equation}
\begin{aligned}
g^{IJ}(\sigma,\sigma')=\frac{1}{2}&\left[
\left(\frac{(\omega_I-\omega_{1,0})(\omega_I-\omega_{2,0})(\omega_I-\omega_{3,0})}{(\omega_J'-\omega_{1,0})(\omega_J'-\omega_{2,0})(\omega_J'-\omega_{3,0})}\right)^{1/2}\right.\\
+&\left.\left(\frac{(\omega_J'-\omega_{1,0})(\omega_J'-\omega_{2,0})(\omega_J'-\omega_{3,0})}{(\omega_I-\omega_{1,0})(\omega_I-\omega_{2,0})(\omega_I-\omega_{3,0})}\right)^{1/2}
\right],
\end{aligned}
\end{equation}
where 
$\omega_{1,0}=e^{i\pi/3}, \omega_{2,0}=e^{-i\pi/3},
\omega_{3,0}=e^{-i\pi}$,
and 
$\omega_I(\sigma),\omega_J'(\sigma')$ are defined in Eq.\ \eqref{eqn:conformal_map}.
It is easy to check that this candidate
fulfills the requirements (i-iii). The branch points also include $\infty$. The
branch cuts can be chosen from $\omega_{1,0}$ to $\infty$, $\omega_{2,0}$ to
$\infty$, and $\omega_{3,0}$ to $\infty$, as shown
in Fig.~\ref{fig:tripartition-map-RRRfluxes}(b). We will compute the singular terms explicitly later, which verifies requirement (iv). 
It turns out that $g^{IJ}$ is the same for any $I,J$, and
the mode expansion of $g^{IJ}$ in powers of $z=e^{i\sigma}$ is given by
\begin{align}
  g^{IJ}
  &= \frac{1}{2}
    \left[
    \big(\frac{1}{\sqrt{z'}}-\sqrt{z'}\big)\sum_{r\geq 1/2}z^r
    +
    \big(\frac{1}{\sqrt{z}}-\sqrt{z}\big)\sum_{r\geq 1/2}z^{\prime r} \right]
    \nonumber \\
  &=\frac{1}{2}\sum_{m=0}^{\infty}\left[e^{i\sigma(\frac{1}{2}+m)-i\frac{\sigma'}{2}}-e^{i\sigma(\frac{1}{2}+m)+i\frac{\sigma'}{2}}\right.
    \nonumber \\
  &\qquad \qquad
    +\left.e^{i\sigma'(\frac{1}{2}+m)-i\frac{\sigma}{2}}-e^{i\sigma'(\frac{1}{2}+m)+i\frac{\sigma}{2}}\right].
\end{align}
It is worth noting that this expression is valid for the vertex state of an $N$-junction with
arbitrary $N$ and the insertion of $N$ twist operators.
As an example, we give the construction of the vertex state
for $N=2$ using this branch cut factor
in Appendix
\ref{app:another-twist}, which reproduces the result for the R-R sector bipartition calculation of the preceding section.

We are now ready to examine requirement (iv).
Combining the singular term
of the Neumann coefficient in the NS-NS-NS sector
$K^{IJ}_{\mathrm{sing.}}=\delta^{IJ}\sum_{r\geq 1/2}e^{-ir(\sigma-\sigma')}$
with the branch cut factor $g^{IJ}$, we obtain: 
\begin{equation}
\begin{aligned}
K^{IJ}_{\mathrm{sing.}}g^{IJ}
&=\delta^{IJ}\sum_{r\geq 1/2}^\infty e^{-i(r+1/2)(\sigma-\sigma')}\\
&\tb+\frac{\delta^{IJ}}{2}
\Big[\sum_{m=0}^\infty e^{im\sigma}-\sum_{m\geq 1}e^{im\sigma'}\Big].
\end{aligned}
\end{equation}
The first term gives the correct singular term
in the R-R-R sector,
$R^{IJ}_{\mathrm{sing.}}=\delta^{IJ}\sum_{m\geq 1}^\infty
e^{-im(\sigma-\sigma')}$,
and the second term contributes to
the zero mode parts $R_{0,m}, R_{m,0}, R_{0,0}$.
This shows that our choice of $g^{IJ}$ is indeed a valid one. 
We thus verified the Neumann function has the following expansion:
\begin{align}
  &R^{IJ}(\sigma,\sigma')=
  \sum_{m\geq 0,n\geq 0}
  e^{im\sigma}R^{IJ}_{mn}e^{in\sigma'}
  +
  \delta^{IJ}
  \sum_{n\geq 1}e^{-in\sigma}e^{in\sigma'}.
\end{align}
The non-singular terms can be worked out easily in a similar way. We summarize the expansion coefficients below:
\begin{equation}
\begin{aligned}
&R^{IJ}_{r'+1/2,s'+1/2}
=\frac{1}{2}
\Big[\sum_{r=1/2}^{r'}(K_{r,s'+1}^{IJ}
-K_{r,s'}^{IJ})
\\ &
\qquad
\qquad 
\qquad
\qquad 
+\sum_{s=1/2}^{s'}
(K_{r'+1,s}^{IJ}-K_{r',s}^{IJ})
\Big],
\\
&R^{IJ}_{0,s'+1/2}=\frac{1}{2}\sum_{s=1/2}^{s'}K_{1/2,s}^{IJ}-\frac{1}{2}\delta^{IJ},
\\
&R^{IJ}_{s'+1/2,0}=\frac{1}{2}\sum_{s=1/2}^{s'}K_{s,1/2}^{IJ}+\frac{1}{2}\delta^{IJ},
\\
&R_{00}^{IJ}=
\frac{1}{2} \delta^{IJ}.
\end{aligned}
\end{equation}

Finally, using the Neumann coefficients,
the vertex state can be constructed as
\begin{equation}
\begin{aligned}
&|V\rangle=\\
&\exp{\left(\sum_{m,n\geq 1}\frac{1}{2}\chi_{-m}^I R_{mn}^{IJ}\chi_{-n}^J+\sum_{m\geq 1}2\chi_{-m}^I R_{m0}^{IJ}\chi_0^J \right)}|\Omega\rangle.
\end{aligned}
\label{eqn:realfermionansatz}
\end{equation}
We show this state satisfies the boundary condition explicitly in Appendix
\ref{app:boundary-NS-R}.

As discussed in Sec.\ \ref{subsubsec:RR},
to have a well-defined Hilbert space, we need to combine the $\chi^I_0$ zero modes
with $\bar{\chi}^I_0$ at the outer edges.
Indeed, physically speaking, prior to physically cutting the system along the entanglement cut, the R-R-R sector configuration is topologically equivalent to a sphere with one Ising anyon placed on the entanglement cut and three Ising anyons in the three regions $A$, $B$, and $C$. These correspond to the three outer edge Majorana fermion zero modes and one of the zero modes that appears at the inner edge when we physically cut along the entanglement cut. This results in a four-fold degeneracy and we must choose one of these states for which to compute the entanglement. To do so,
we define the complex fermion as in Eq.\ \eqref{eqn:Majorana-zero-mode}, 
$\chi^I_0=(g^I_0+g_0^{I,\dg})/\sqrt{2}$. 
Denoting $X=\sqrt{2}\sum_{m\geq 1}\chi_{-m}^I R_{m0}^{IJ}g_0^{\dg,J}, Y = \sqrt{2}\sum_{m\geq 1}\chi_{-m}^I
R_{m0}^{IJ} g_0^J$,
one can show $[X,Y]=0$ and hence $e^{X+Y}=e^X e^Y$.
In order to fix a state within the four-fold degenerate subspace, we must fix the occupations of the zero modes. For simplicity, we choose the reference state $\ket{\Omega}$ to be one of definite fermion parity take $\ket{\Omega} = \ket{000}$,  
which is annihilated by all $g_0^I$. 
Under this choice, the solution is simplified to:
\begin{equation}
\begin{aligned}
&|V\rangle=\\
&\exp \Bigg( \sum_{m,n\geq 1}\frac{1}{2}\chi_{-m}^I R^{IJ}_{mn} \chi_{-n}^J+\sum_{m\geq 1}\sqrt{2}\chi_{-m}R_{m0}^{IJ}g_0^{\dg,J} \Bigg)|0\rangle.
\end{aligned}
\end{equation} 

Finally, by combining two copies of real fermions,
we can construct the complex fermion vertex state as 
\begin{equation}
\begin{aligned}
|V\rangle=&\exp\left(\sum_{m,n\geq 1}g^I_{-m}R^{IJ}_{mn}g_n^{\dg,J}\right.\\
&\qquad\left.+\sum_{m\geq 1}2R^{IJ}_{m0}(g_{-m}^I g_0^{\dg,J}+g_m^{\dg,I}g_0^J) \right)|\Omega\rangle.
\end{aligned}
\label{eqn:R-R-R-complex}
\end{equation}
Again, we postpone the verification of boundary condition
in Appendix \ref{app:boundary-NS-R}. We choose $|\Omega\rangle$ to be the vacuum that is annihilated by $g_0^I$. 
Identifying $X=2\sum_{m\geq 1}R_{m0}^{IJ}g_{-m}^I g_0^{\dg,J}$, $Y=2\sum_{m\geq 1} R^{IJ}_{m0} g_m^{\dg,I} g_0^J$, 
and $[X,Y]=4\sum_{m,n\geq 1}R_{m0}^{IJ} R_{n0}^{I'J} g_n^{\dg,I'} g_{-m}^I$, 
the solution is simplified to
\begin{equation}
\begin{aligned}
&|V\rangle=\\
&\exp\left(\sum_{m,n\geq 1}g_{-m}^I\tilde{R}_{mn}^{IJ}g_n^{\dg,J}+\sum_{m\geq 1}2g_{-m}^I R_{m0}^{IJ}g_0^{\dg,J} \right)|0\rangle,
\end{aligned}
\end{equation}
with $\tilde{R}_{mn}^{IJ}=R_{mn}^{IJ}-2R_{m0}^{IK}R_{n0}^{JK}$.

\subsection{Entanglement entropy, negativity and reflected entropy} 
\label{Entanglement entropy, negativity and reflected entropy}

\begin{table*}[!t]
\centering
\begin{tabular}{ c | c c c | c c c }
\hline \hline
 & $a_{-1}$ & $a_0$ & $a_1$  & $b_{-1}$ & $b_0$ & $b_1$\\
 \hline
 \hline
Majorana (NS-NS-NS)& 
0.0654
&  0.0299 & 
$-0.0232$
& 
0.0491
& 0.0310 & 
$-0.4021$
\\
Majorana (R-R-R)   & 
0.0654
&  0.6227 & 
$-5.3746$
& 
0.0491
& 0.3341 & 
$-0.2984$
\\
Dirac (NS-NS-NS)   
& 
0.1309
&  0.0597 & 
$-0.0119$
& 
0.0982
& 0.0600 & 
0.3657
\\
Dirac (R-R-R)      
& 
0.1309
&  $-0.1139$  & 
22.9493
& 
0.0982
& 0.0025 &
15.1538
\\
\hline \hline
\end{tabular}
\caption{
The scaling of the entanglement entropy
and negativity with respect to $L/\epsilon$
from the numerical analysis.
\label{tbl:coeff}}
\end{table*}


With the tripartite vertex states in hand,
we now proceed to the calculations of
the correlation measures, namely, 
the entanglement entropy $S_A$ and spectrum
when tracing out $B$ and $C$,
and negativity $\mathcal{E}_{A:B}$ and the spectra
when tracing out $C$,
and the reflected entropy $R_{A:B}$ 
when tracing out $C$.
Once again, the regularization
$|V\rangle \to |G\rangle = \mathcal{N}e^{- \epsilon H_0}|V\rangle$
amounts to multiplying the Neumann coefficients by
an exponential factor,
e.g.,
$
R_{mn}^{IJ}\ra R_{mn}^{IJ} e^{-\epsilon(m+n)}
$.
As the resulting state $|G\rangle$
is Gaussian, we can use the correlator method
to compute various entanglement measures,
as described in Sec.\ \ref{sec:fermonic-Gaussian}. 
The technical details are left to Appendix\ \ref{app:correlation}. 
To evaluate the correlators (covariance matrices)
numerically, we need to introduce
a cutoff $N_c$ to truncate the Neumann coefficients.
The correlation measures
(for a given $L/\epsilon$)
are then computed
for different $N_c$ and
the results are extrapolated to $N_c\to \infty$.
We typically take $N_c \ge 200 - 800$.


We first present our results for the entanglement entropy and negativity.
For both cases,
we find that they scale with $L/\epsilon$ as
\begin{equation}
\begin{aligned}
  &
  S_A=a_{-1} \frac{L}{\epsilon}+a_0+a_1 \frac{\epsilon}{L}+\cdots,
  \\
  &
  \mathcal{E}_{A:B}
  = b_{-1} \frac{L}{\epsilon}+b_0+b_1 \frac{\epsilon}{L}+\cdots,
\end{aligned}
\label{eqn:S-scale}
\end{equation}
for both the NS-NS-NS and R-R-R sectors.
The numerically extracted coefficients are summarized 
in Table \ref{tbl:coeff}.
The coefficients $a_{-1}$ and $b_{-1}$
are the same for 
the NS-NS-NS and R-R-R sectors.
The numerical result 
for $a_{-1}$
is consistent with
$a_{-1} = \pi c /24$
(see Sec.\ \ref{Entanglement entropy}).
On the other hand, 
the numerically computed 
$b_{-1}$ is consistent with
$
b_{-1} = 3\pi c /96
$. 
These may be 
understood
as commonly appearing
coefficients
in the entanglement
entropy and negativity 
in topological liquids.
For example, 
for the mutual information
and negativity 
on the torus,
when
$A, B, C$ are non-contractible 
and $A$ and $B$ are adjacent,
the area law terms of these quantities 
are proportional to
$
(1/n+1) (\pi c/12)
$
($n\to 1$),
and
$(4/n_e - n_e) (c\pi/48)$
($n_e \to 1$)
\cite{wen2016edge}.
We also note that
the area law terms 
should cancel 
in $\mathcal{E}_3=2\mathcal{E}_{A:B}-\mathcal{E}_{A\cup C:B}$,
and we know 
$\mathcal{E}_{A\cup C:B}
=
S^{(1/2)}_{A\cup C:B}
\sim 
(3\pi c/48) (L/\epsilon)$.
The constant term $a_0$ in the NS-NS-NS sector 
is small compared with $\ln 2 \sim 0.693$,
and may be consistent 
with $a_0=0$, the result we expect 
from the calculation for a bipartition.
On the other hand, in the R-R-R sector, 
$a_0$ is an order of magnitude larger.
We may attribute it to the extra $\sigma$ anyon 
positioned at the junction. 
We recall that we obtained a similar result 
in the NS-R sector for a bipartition.

%
%
%
%

\begin{figure}[t]
\centering
\includegraphics[width=\columnwidth]{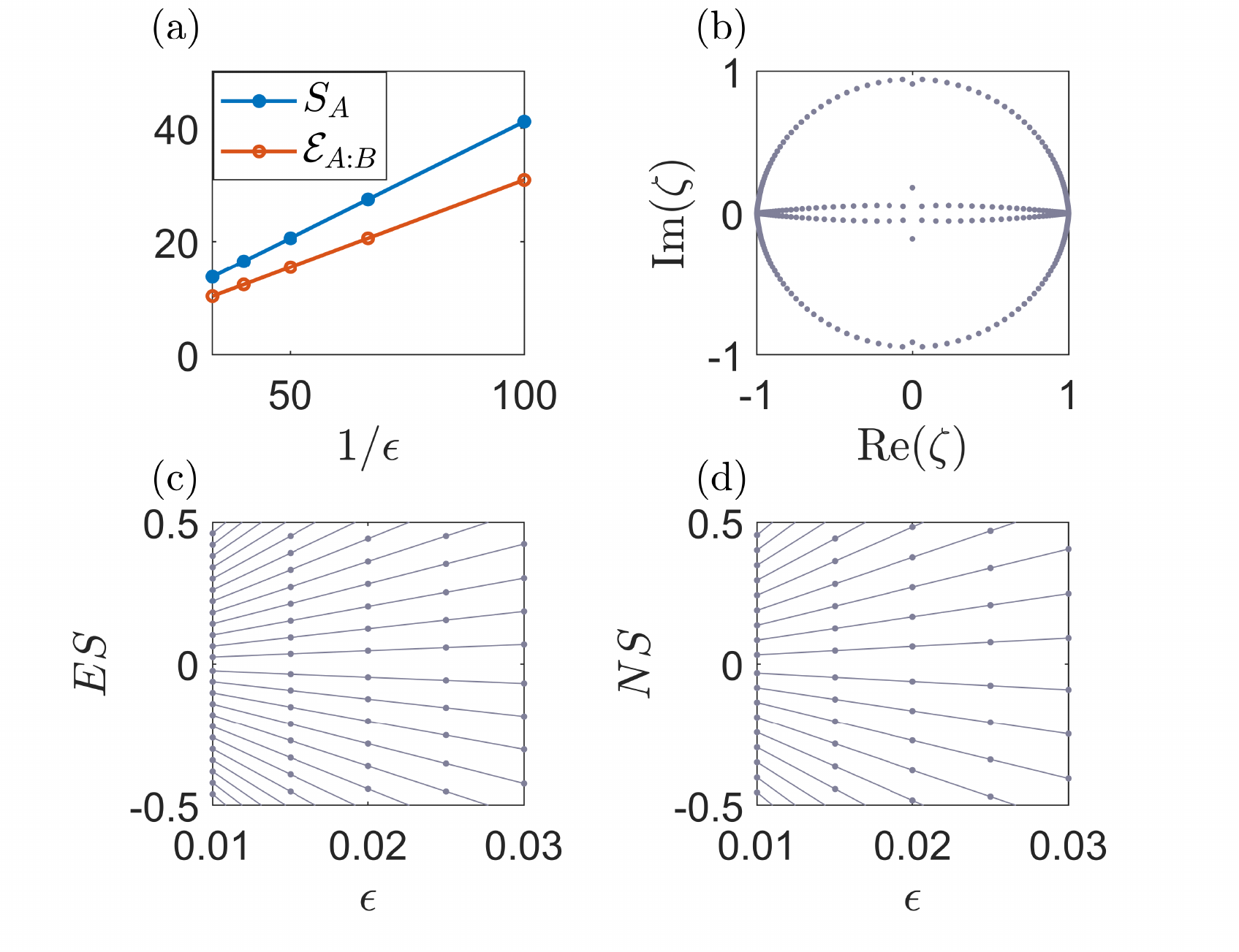}
\caption{(a)
  The evolution of $S_A$ and negativity $\mathcal{E}_{A:B}$
  with different regulator $\epsilon$ at $N_c\ra\infty$ limit, in the NS-NS-NS sector for the Majorana fermion. 
  (b) Distribution of the eigenvalues of $\Gamma_+$, at $N_c=500$ and $\epsilon=0.02$ for the complex fermion. 
  (c,d) Entanglement spectrum and negativity spectrum for $N_c=200$ at different
  $\epsilon$, which shows equal spacing behavior. 
  }
\label{Fig:fv}
\end{figure}

In Fig.\ \ref{Fig:fv}
we plot the entanglement and negativity spectra.
Here, we focus on the NS-NS-NS sector
(as the R-R-R sector shows the same features).
Both the entanglement and negativity spectra
exhibit an equal-spacing structure.
For the entanglement spectrum, this is expected
as it is given by the spectrum of
the CFT realized on a physical edge
\cite{li2008entanglement}.
Similarly, 
the equal-spacing structure of the negativity
spectrum may suggest that it is described by some CFT. 
For the Majorana fermions, the entanglement spectrum is non-degenerate while the negativity spectrum is two-fold degenerate. 
For the complex fermions, the degeneracy of the entanglement spectrum is two-fold,
while that for the negativity spectrum is four-fold.
We will see in the next section 
that  the degeneracy matches 
with the lattice calculation result deep in the topological region.

Plotted in Fig.\ \ref{Fig:fv}(b) is
the single-body spectrum $\lbrace\zeta\rbrace$ of $\rho^{T_A}_{A\cup B}$
(the spectrum of the  correlation matrix $\Gamma_+$). 
The eigenvalues appear to come in
various branches; 
those that are circularly distributed and
those that are clustered near the real axis.
The non-trivial distribution of 
the spectrum over the complex plane
can be regarded as a
smoking gun of topological non-triviality of the bulk.
As a comparison, we note that
for a simple product state
the spectrum $\{\zeta\}$ 
consists of just two eigenvalues,
$\zeta = 1$ and $\zeta = -1$.
We also note that such non-trivial 
distribution of the eigenvalues $\{\zeta\}$
was found previously 
in (1+1)d fermionic CFTs
\cite{shapourian2019twisted},
and 
(1+1)d SPT phases (the Kitaev chain)
\cite{inamura2020non}.
In these examples, 
the many-body spectrum of  
$\rho^{T_A}_{A\cup B}$ 
has a 8-fold rotation symmetry. 
On the other hand, we do not find 
such a symmetric pattern 
for the case of our (2+1)d topological liquids.
In the next section, 
we will see that a similar distribution of $\{\zeta\}$
is also found in the lattice Chern insulator calculation.

\begin{figure}[t]
    \centering
    \includegraphics[width=\columnwidth]{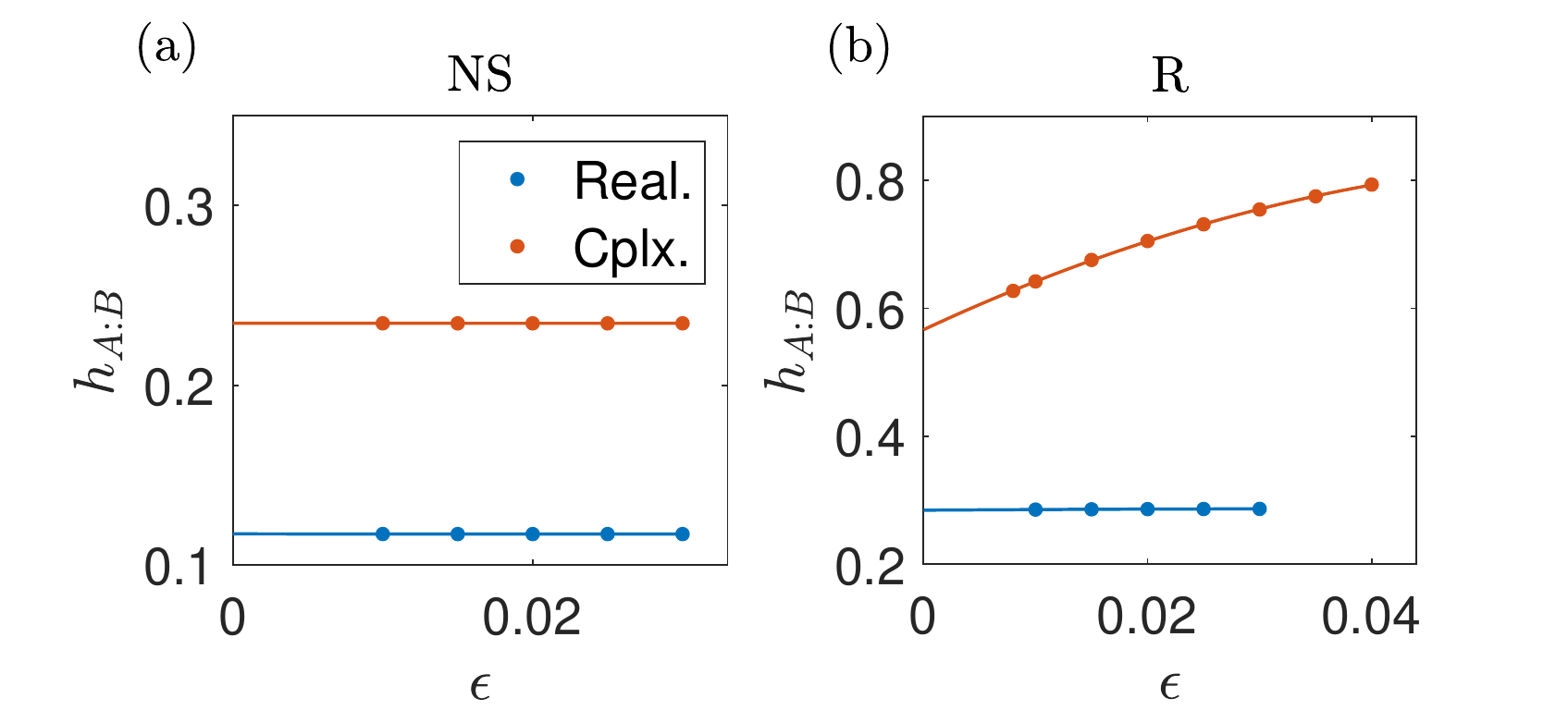}
    \caption{
    The difference between
    reflected entropy and mutual information $h_{A:B}=R_{A:B}-I_{A:B}$. 
    (a) The NS-NS-NS sector for Majorana fermion and complex fermion. The intercept (0.2344) is twice of that of the Majorana fermion (0.1172). 
    (b) 
    The R-R-R sector for Majorana fermion and complex fermion. Using a power two polynomial fit, the intercept (0.5698) is almost the twice of that of the Majorana fermion (0.2850). In (a) and the real fermion case of (b), $h_{A:B}$ does not change with $\epsilon$. }
    \label{fig:RE_MI}
\end{figure}

Finally, we turn to the reflected entropy and the conjecture
\eqref{eq: conj for h}. 
We study this difference for the four aforementioned cases and show the results in Fig.\ \ref{fig:RE_MI}. 
For the Majorna and Dirac fermion edge theories
in the NS-NS-NS sector, and 
the Majoran fermion edge theory 
in the R-R-R sector,
$h_{A:B}$ does not change with $\epsilon$, 
with the values being 0.1172, 0.2344, 0.2850 respectively.
The results for the NS-NS-NS
sector are consistent with
the prediction
$(c/3)\ln 2= 0.1155, 0.2310$
for $c=1/2$ and $c=1$,
respectively.
(Alternatively, if we extract the central charge
from our numerics,
we obtain $c=0.5073,1.0145,1.2335$, respectively.) 
For the R-R-R sector, 
the numerics suggests that $h_{A:B}$ is 
slightly bigger than $(c/3) \ln 2$,
which once again  may be attributed to the Ising anyon 
at the junction.
Finally, for the Dirac fermion in 
the R-R-R sector, $h_{A:B}$ changes with $\epsilon$ and the polynomial fit up to second order gives the intercept 0.5698.
Notice that 0.2344 is twice 
as large as 0.1172, and 0.5698 is (almost) twice as large as 0.2850.
We note that to get the universal result
in the edge theory  calculations, 
we do not have to consider 
a local unitary that remove 
short-range correlations 
at the junction(s).

\section{Lattice model approach}
\label{lattice}

\begin{figure}[t]
\centering
\includegraphics[width=\columnwidth]{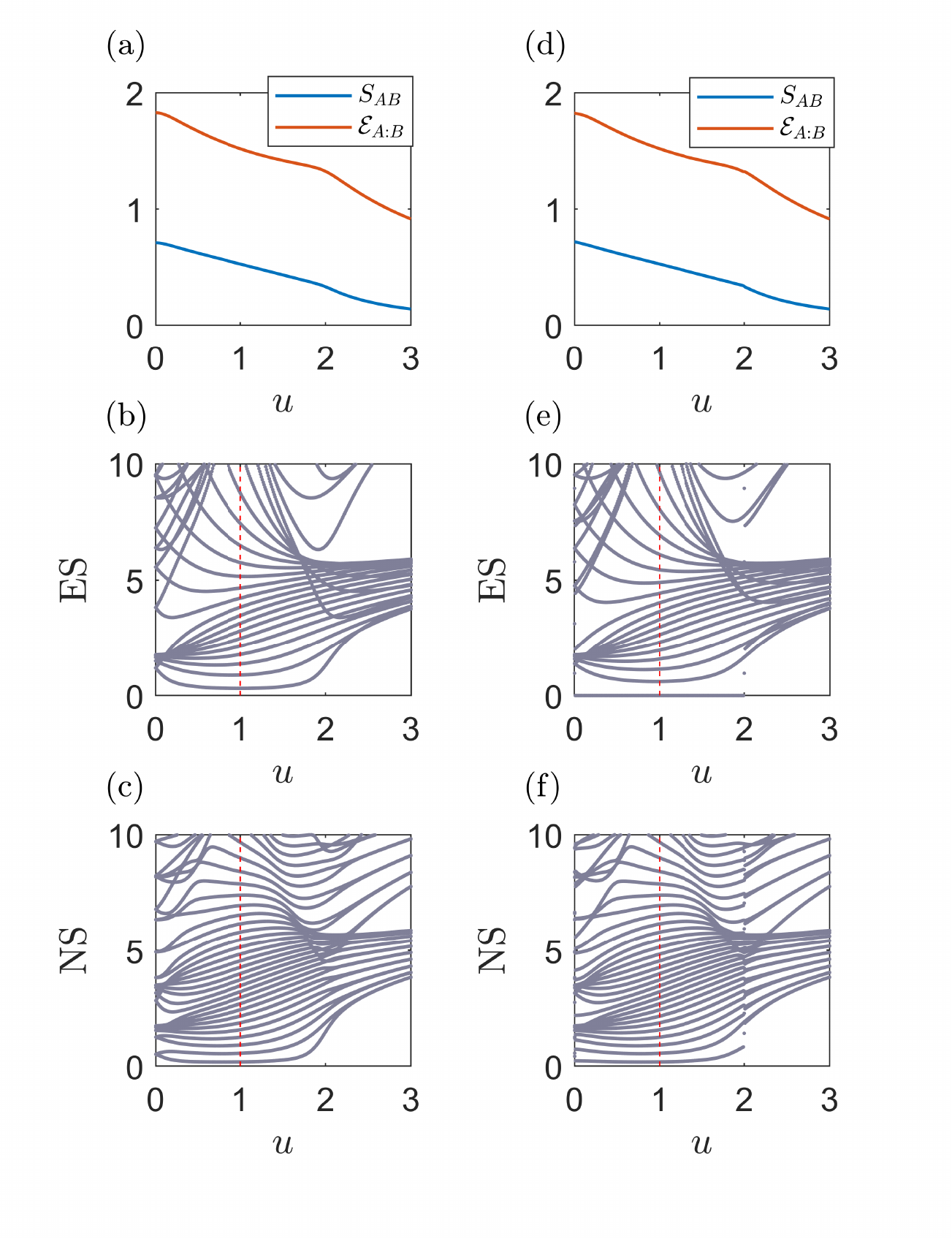}
\caption{
  The von Neumann entanglement entropy and logarithmic negativity
  for the Chern insulator model on a $20\times 20$ lattice ($l=20$) for (a-c) anti-periodic boundary condition and (d-f) periodic boundary condition.
  The lattice partition is shown in Fig.\ \ref{partition}(b)
  where  $l_A=10$;
  both $A$ and $B$ are $10\times 10$ blocks.
  (a,d)
  The von Neumann entanglement entropy $S_{AB}$ and logarithmic negativity $\mathcal{E}_{A:B}$.
  (b,e)
  Entanglement spectrum $\ln(\frac{2}{1+\gamma}-1)$. $\gamma$ is four-fold degenerate for both topological and trivial region, which is observed for both of the boundary conditions. For the periodic boundary condition (e), there exist four-fold degenerate zero modes. (c,f)
  Negativity spectrum $\ln(\frac{2}{1+\gamma_{\times}}-1)$.
  Note at $u=1$, the low lying spectrum shows equal spacing pattern, which is a CFT signature. $\gamma_{\times}$ is 4-fold degenerate in the topological region and becomes 8-fold degenerate in the trivial region, which is observed for both the boundary conditions. }
\label{CI}
\end{figure} 

Though the edge theory, or ``cut-and-glue" approach provides a theoretically appealing way of computing entanglement measures in the thermodynamic limit, it is limited by the fact that it is only applicable to systems deep in the topological phase. It is natural to ask how the entanglement properties of a system change closer to and across a topological phase transition.

To that end and as a check on the conclusions we have drawn from the edge theory
approach,
in this section, we study
a tight-binding model on the square lattice
that realizes a Chern insulator phase. 
The Hamiltonian is given by
\begin{align}
  H
  &= \frac{-i}{2}\sum_{\bm{r}} \sum_{\mu=x,y}
    \left[
     f_{\bm{r}}^\dag\tau_\mu  f^{\ }_{\bm{r}+\bm{a}_\mu}
    - f_{\bm{r}+\bm{a}_\mu}^\dag\tau_\mu f^{\ }_{\bm{r}} \right]
    \nonumber \\
  &\quad
    +\frac{1}{2}\sum_{\bm{r}}\sum_{\mu=x,y}
    \left[ f_{\bm{r}}^\dag \tau_z f^{\ }_{\bm{r}+\bm{a}_\mu}
    + f^\dag_{\bm{r}+\bm{a}_\mu}\tau_z f^{\ }_{\bm{r}}\right]
    \nonumber \\
  &\quad
    +u\sum_{\bm{r}} f_{\bm{r}}^\dag \tau_z f^{\ }_{\bm{r}},
\end{align}
where the two-dimensional integer vector
$\bm{r}$ labels sites on the square lattice, 
and
$\bm{a}_x=(1,0)$ and $\bm{a}_y=(0,1)$;
$ f^{\dag}_{\bm{r}}/ f^{\ }_{\bm{r}}$
are two-component fermion creation/annihilation operators
at site $\bm{r}$,
and $\tau_{\mu=x,y,z}$ are the Pauli matrices. 
In momentum space, the corresponding Bloch Hamiltonian is, 
\begin{align}
h(\mathbf{k})=\sin{k_x}\tau_x+\sin{k_y}\tau_y+(u+\cos{k_x}+\cos{k_y})\tau_z,
\end{align}
with $k_{x,y}\in[-\pi, \pi]$. 
The parameter $u$ tunes the model across insulating phases with different Chern numbers: the Chern number
${\it Ch}=0$ for $|u|>2$,
${\it Ch}=1$ for $0<u<2$ and
${\it Ch}=-1$ for
$-2<u<0$.
The many body ground state $|{\it GS}\rangle$
is obtained by filling the lower band.
On an $N\times N$ square lattice,
the correlation matrix elements
are given by
$\langle {\it GS}| f^\dagger_{\bm{r},s} f^{\ }_{\bm{r}',s'}
|{\it GS}\rangle=
{N}^{-1}\sum_{\mathbf{k}}v^*_{(\mathbf{k},s,-)}v_{(\mathbf{k},s',-)}
e^{-i\mathbf{k}\cdot(\bm{r}-\bm{r}')}$,
where $s=1,2$
and $v_{\mathbf{k},s,-}$ is the $s$-th component of the Bloch eigen vector of the lower band.
Since it is a particle number conserving model, the correlation matrix $\Gamma$ is simply given by $\Gamma=(\mathbbm{1}-2C)\otimes \sigma_y$.

\begin{figure}[t]
\centering
\includegraphics[width=\columnwidth]{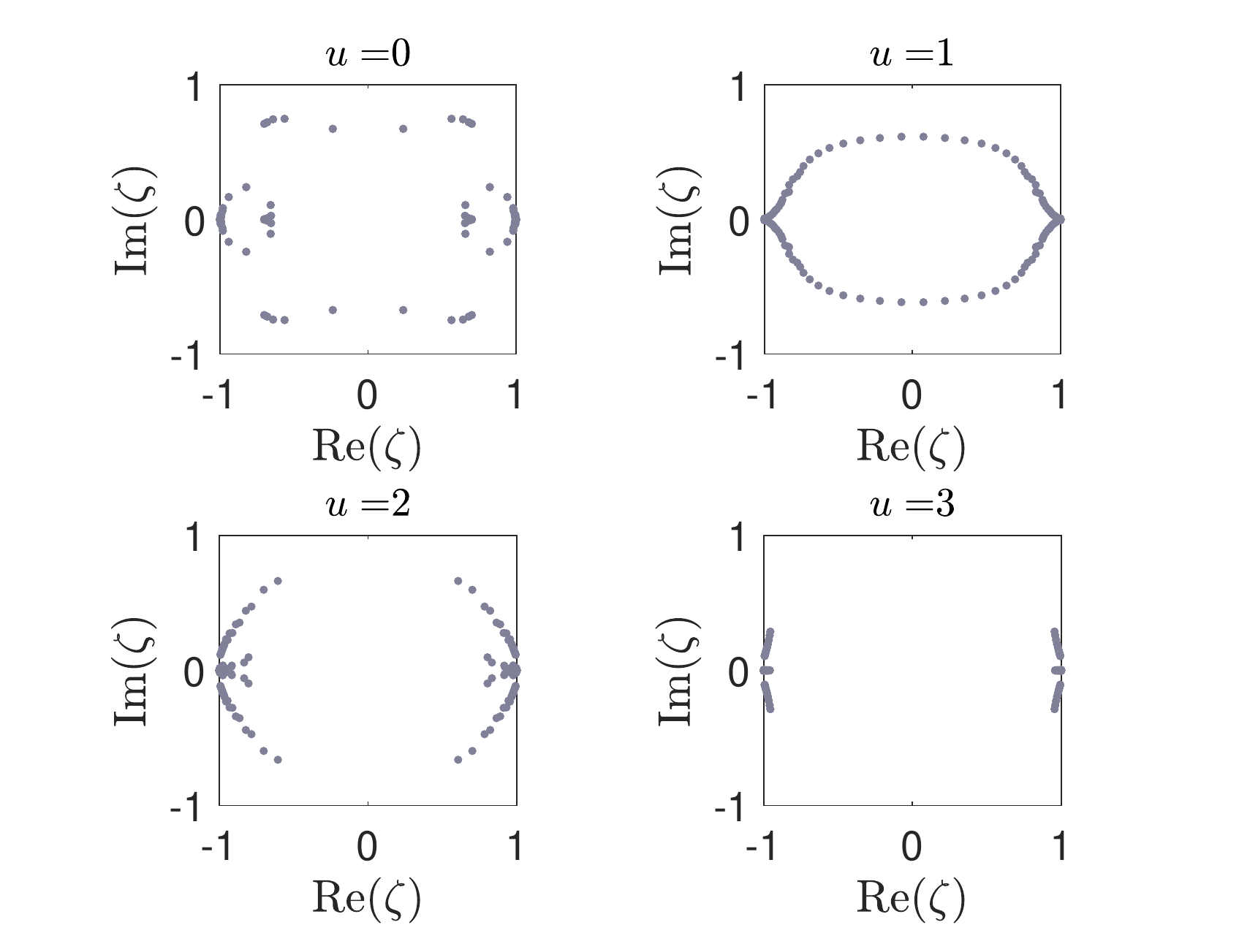}
\caption{Single-particle spectrum
  $\lbrace\zeta\rbrace$ of $\rho^{T_A}_{A\cup B}$'s correlation matrix $\Gamma_+$ for various values of $u$ on $20\times 20$ lattice with anti-PBC. Note that the $\zeta$ spectrum is complex, with real and imaginary parts. 
  }
\label{Fig:CIrhoTAs}
\end{figure}

\begin{figure*}[!htb]
    \centering
    \includegraphics[width=\textwidth]{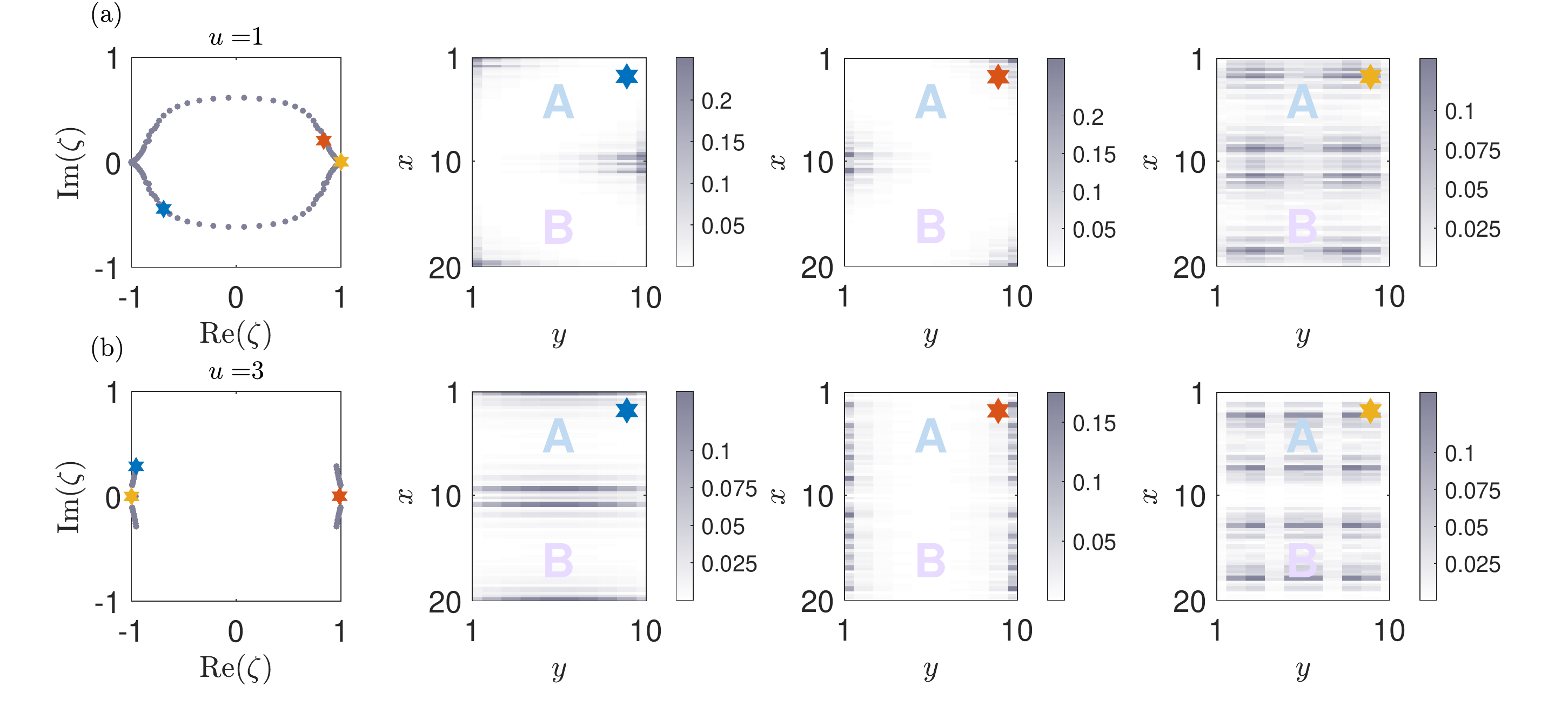}
    \caption{Eigenvectors of $\Gamma_+$ at (a) $u=1$ and (b) $u=3$ using anti-PBC for a $20\times 20$ lattice. For each $u$, we take three different eigenvalues, as indicated using the blue, orange and yellow stars, and plot the corresponding eigenvectors supported on $A\cup B$. The eigenvectors exhibit differing patterns of spatial localization for difference phases.
    }
    \label{fig:rhoTAev}
\end{figure*}

\begin{figure*}[!htb]
\centering
\includegraphics[width=0.8\textwidth]{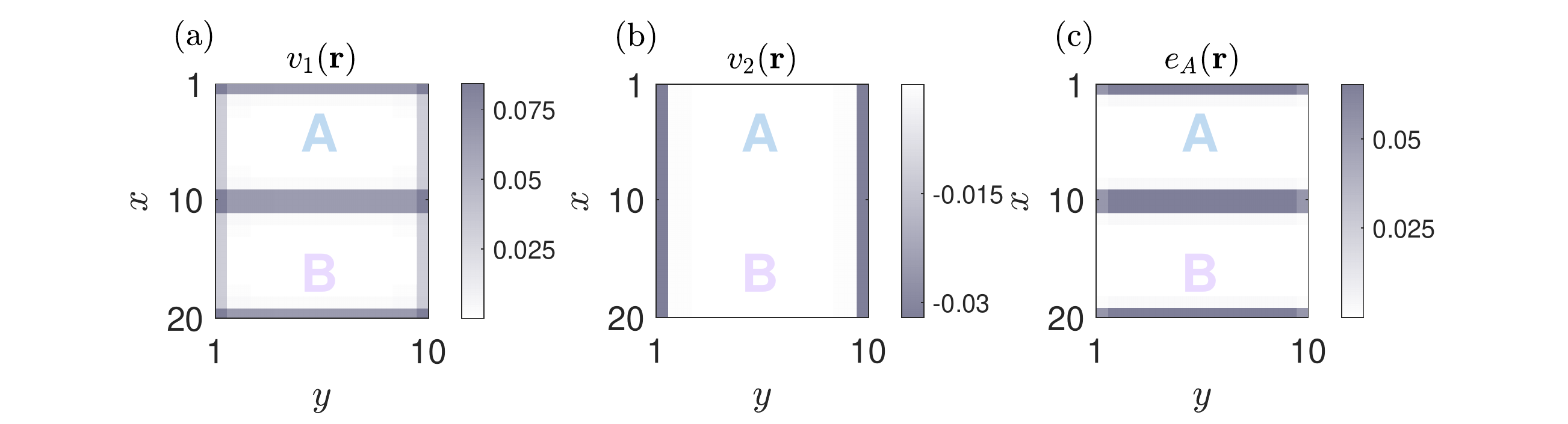}
\caption{Distribution of (a) $v_1(\bm{r})$, (b) $v_2(\bm{r})$ and (c) negativity contour $e_{A:B}(\bm{r})$ at $u=1$ for a $20\times 20$ lattice, supported on region $A\cup B$.
}
\label{contour}
\end{figure*}
 

We consider tripartitioning the square lattice
into three regions $A,B,C$,
and trace out the region $C$
(Fig.\ \ref{partition}(b)).
Since this is a non-interacting system, the reduced density matrix $\rho_{A\cup B}$ is
Gaussian. We can then use the correlator method reviewed in Sec. \ref{sec:fermonic-Gaussian} to construct the partially
transposed density matrix $\rho_{A\cup B}^{T_A}$. The entanglement spectrum of this model was first studied in Ref. \cite{Ryu_2006}. 
 
\paragraph{Entanglement entropy and negativity}

The numerically computed entanglement entropy $S_{AB}$ and negativity $\mathcal{E}_{A:B}$, and the corresponding spectra $\lbrace\gamma\rbrace$ and $\lbrace\gamma_\times\rbrace$ are shown in Fig.\ \ref{CI}. 
We first verify that both $S_{AB}$ and $\mathcal{E}_{A:B}$ obey area law scaling with the size of lattice $l$, 
as expected (not shown in the figure). 
In addition, we see that the phase transition at $u=2$ appears to manifest as a small ``bump" in $\mathcal{E}_{A:B}$. A similar though less pronounced change in the slope of $S_{AB}$ as a function of $u$ at $u=2$ is somewhat visible.

Clearer signatures of this phase transition, as well as the topological nature of the phases, are provided by the entanglement and negativity spectra.
Indeed, for periodic boundary conditions, both the entanglement spectrum and negativity spectrum exhibit discontinuous behavior at the phase transition point $u=0,\pm2$ , as we can see in Fig.\ \ref{CI}(e,f). 
For anti-periodic boundary conditions, the spectra are no longer discontinuous across the phase transition. However, the transition still appears to manifest in the spectra 
by lifting of low lying modes 
and change in the degeneracy (see the discussion below) when crossing from the topological phase to the trivial phase.
The discontinuous behavior also does not exist for more general twisted boundary conditions.

Moving on to the properties of the phases themselves, we see that deep inside the topological phase, around $u=1$ where the bulk gap is the largest,
the entanglement spectrum is evenly spaced, at least for the ``low-energy"
regime. This is consistent with the expectation that the low-energy part of the reduced density matrix
is well described by $\rho_{A\cup B}\sim \exp{(-\xi H_{{\it CFT}})}$
where $H_{{\it CFT}}$ is the (physical) CFT Hamiltonian for the edge state, namely the free complex fermion CFT with $c=1$. 
Here, $\xi$ is a non-universal parameter,  controlled by the bulk correlation length, for example.
Similarly, around $u=1$, the negativity spectrum is also evenly spaced.
This likewise suggests that $\rho_{\times,A\cup B}$
is given by $\rho_{\times,A\cup B}\sim\exp{(-\xi H'_{{\it CFT}})}$,
where $H'_{{\it CFT}}$ is a Hamiltonian
of CFT, which may differ from $H_{{\it CFT}}$. 

Moreover, the degeneracy of the entanglement and negativity spectra reveal signatures of the two phases and the boundary conditions. 
One the one hand, every eigenvalue $\gamma$ is four-fold degenerate in the ES for $S_{AB}$ and two-fold degenerate in the ES for $S_A$. On the other hand, the negativity spectrum $\gamma_{\times}$ is four-fold degenerate in the topological region and becomes eight-fold degenerate deep in trivial region, which is observed for both of the boundary conditions. 
We thus see that the degeneracy of the NS provides a signal for the topology of the ground state, in contrast to the ES.
The degeneracies deep in the topological region (two-fold for ES and four-fold for NS) match up with the edge theory results presented earlier in
Sec.\ \ref{Entanglement entropy, negativity and reflected entropy}.

 
To compare with the results from the conformal field theory calculation, let's compare the entanglement entropy and logarithmic negativity at $u=1$ for anti-periodic boundary condition (i.e., the N-N-N sector) and periodic boundary condition (i.e., the R-R-R sector). 
 
When taking $AB$ as the subsystem to compute the entanglement entropy, the entanglement spectra for PBC and APBC are different (due to the zero mode), but they give the same entanglement entropy. This is similar to our previous experience in bipartition boundary state, where 
the NS-NS and R-R sectors give the same entanglement entropy.  

When taking $A$ as the subsystem to compute the entanglement entropy, we find the entanglement spectra for $S_A$ are the same when deep in the topological region $u=1$, and deep into the trivial region $u=3$. When coming closer to the critical point, these two spectra become different. 
(We note, in contrast, in the edge theory calculation,
the NS-NS-NS and R-R-R sectors give 
different entanglement entropies, $S_A$.
The precise reason for the disagreement between
the lattice and edge theory calculations is 
unclear. We however note that 
the configurations are not exactly the same --
for example, there are two junctions 
in the edge theory calculations whereas
there are four junctions in the lattice calculation.)

For negativity spectrum, we also find that the PBC and APBC give the same spectrum $\gamma_{\times}$ at $u=1$. This is only true deep in the topological region. For example, if we take $u=1.9$ or $u=0.1$, we can see the vast difference between the two spectra. Furthermore, when going deep into the trivial region $u=3$, the two spectra become identical again.

\paragraph{Spectrum of $\Gamma_+$}

We now move on to the numerically obtained spectrum $\lbrace\zeta\rbrace$ of $\Gamma_+$,
plotted in Figs.\ (\ref{Fig:CIrhoTAs})-(\ref{fig:rhoTAev}), for various $u$ with anti-PBC. 
We see that they provide clear signatures of the topology of the phase. Indeed, 
in the Chern insulator phases, the eigenvalues $\lbrace\zeta\rbrace$ are non-trivially distributed over the complex plane. In the trivial insulator phases, on the other hand, the eigenvalues $\lbrace\zeta\rbrace$ are localized near $\zeta=-1;1$. In the atomic limit $u\ra\infty$, we expect that the spectrum collapses to two points $\zeta=-1;1$.
The distribution of $\lbrace\zeta\rbrace$ is also non-trivial at the critical points $u=0,\pm 2$. However, we defer the discussions for the critical points, and focus on the Chern insulator phase.

In particular, 
in the Chern insulator phase, we can identify two types (branches) of eigenvalues, those that are away from the real axis ($\mathrm{Im}(\zeta)\neq 0$); 
and those that are exactly on $\zeta=-1$ and $\zeta=1$, which are highly degenerate. 
We believe that the appearance of these states  is closely tied to the topological properties of the Chern insulator phase, in the same way that midgap states in the regular entanglement spectrum indicate nontrivial topology. 



  Moreover, the eigenstates corresponding to these two types of eigenvalues are distinguished by their real space profiles, as shown in
  Fig.\ \ref{fig:rhoTAev}.
  For the first type of eigenvalues,
  the corresponding eigenstates are localized near the points
  where the regions $A$, $B$ and $C$ all meet.
  On the other hand, for the eigenvalues at $\zeta=-1,1$,
  the eigenstates are distributed throughout the bulk. 
In contrast, in the trivial phase $u=3$, from Fig.\ \ref{fig:rhoTAev}, there do not exist eigenstates localized at the intersection of $A,B$ and $C$.

\begin{figure}
    \centering
    \includegraphics[width=\columnwidth]{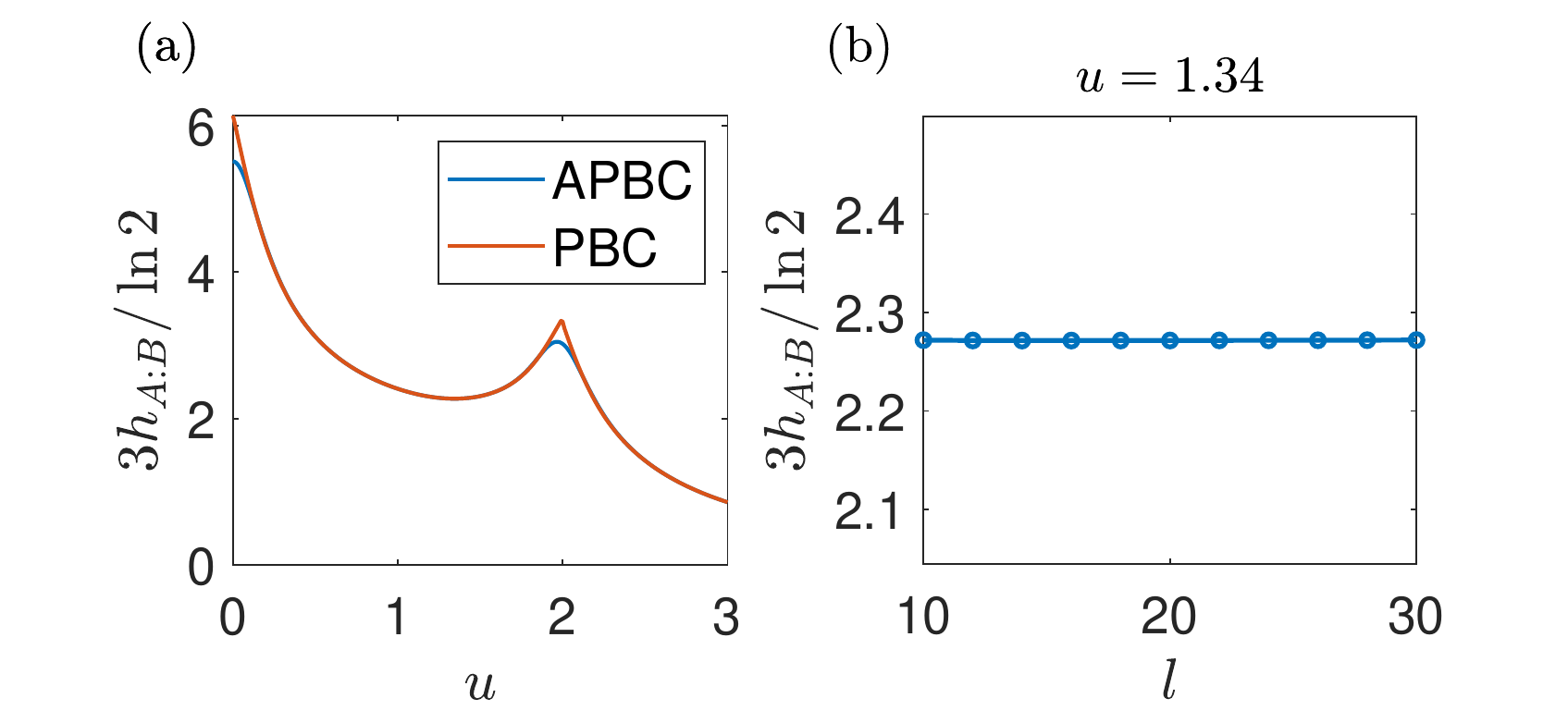}
    \caption{(a) The difference between reflected entropy and mutual information $h_{A:B}=R_{A:B}-I_{A:B}$, computed on $20\times 20$ lattice ($L=20$). (b) Scaling of $h_{A:B}$ with the size of lattice $l$ at $u=1.34$, which shows that $h_{A:B}$ is a constant 2.272. $u=1.34$ is where $h_{A:B}$ is minimal in the topological phase.}
    \label{fig:CI_REMI} 
\end{figure}

\paragraph{Negativity contour}

To better understand 
the spatial decomposition 
of the negativity,
we plot the negativity contour 
\eqref{neg cont}
of a $20\times 20$ lattice at $u=1$ (Fig.\ \ref{contour}). 
From (c), the negativity contour is only supported near the boundary 
between $A$ and $B$, but not the boundary between $AB$ and their complement,  which is as expected. From (a)(b), we find this is because adding $v_1,v_2$ together makes the non-zero values on the boundary between $AB$ and their complement cancel.

\paragraph{Reflected entropy}
We finally examine the reflected entropy and mutual information, and show their difference $h_{A:B}$ in Fig.\ \ref{fig:CI_REMI} (in units of $\ln(2)/3$). 
As the entanglement entropy and negativity,
it is peaked at the phase transitions
and takes smaller values in gapped phases.
In the ${\it Ch}=1$ phase,
$h_{A:B}$ takes its minimum around $u\sim 1.34$
-- we focus on this point and test the conjecture 
\eqref{eq: conj for h}.
There, $h_{A:B}$ is 
independent of $l$, and 
$h_{A:B}\sim 2.272 \times (c/3)\times \ln 2$
(with $c=1$).
We should first note that the 
setup in the lattice calculations
has four junctions where all 
the three regions meet,
whereas in our edge theory calculations
there are two junctions
[see Fig.\ \ref{fig:ct_close}(a)].
This may result in a factor of two 
difference between
the edge theory and lattice calculations.
Even taking into account the difference 
in the number of junctions, 
$h_{A:B}$ is not quantized 
to $(c/3)\ln 2$.
We expect this to be a consequence of non-universal contributions coming from the sharp corners at the trijunction. This would suggest that the edge theory approach provides a reliable way of extracting universal topological contributions to the reflected entropy (and other entanglement measures) without being obscured by non-universal and/or geometric effects. 
Similar to the entanglement entropy, 
both APBC and PBC give the same result when $u$ is not so close to the critical point.
Once again this may be attributed to the 
different configurations adopted
in the edge theory and lattice calculations.

\section{Conclusion}

We have investigated correlation measures,
i.e., 
entanglement entropy, entanglement negativity,
and reflected entropy,
in the ground states of
topological liquid in (2+1) dimensions,
in the multipartition setting
(Fig.\ \ref{partition}).
This was done 
by constructing vertex states 
explicitly in various configurations
with or without fluxes.

In the bipartition case, we study the entanglement entropy 
in the NS-NS, R-R, and NS-R sectors, 
and unveil a new topological contribution in the NS-R case. This contribution is due to the non-trivial configuration where a $\sigma$-anyon exits from the entanglement cut. 

In the tripartition case, 
we find the correlation measures 
capture various universal characteristics of topological liquids.
For example, we found that the spectrum of
the partially transposed density matrix
is non-trivially distributed over the complex plane.
This is somewhat similar to the spectrum
previously computed for (1+1)d fermionic conformal field theory
and symmetry-protected topological phases.
There, a non-trivial dependence of the spectrum
on the spin structures was observed  
\cite{shapourian2019twisted,inamura2020non}. 
We also found universal 
topological contribution 
to negativity and $h_{A:B}$.
In the NS-NS-NS case, 
we verified the conjecture
\eqref{eq: conj for h}
for the reflected entropy,
while there exists 
an additional contribution to $h_{A:B}$ in the R-R-R sector due to the $\sigma$-anyon.

%
%

There are a number of open questions to be discussed.
First of all,
our tripartition setup is different from the ones considered previously 
(except for the original Kitaev-Preskill setup
 \cite{kitaev2006topological}),
and more complicated in the sense that
the entangling boundaries are not smooth, but 
have a singular point where all spatial regions meet.
One may wonder if  the correlation measures
depend not just
on topological but also on
geometrical properties of entangling boundaries.
For example, entanglement entropy is known to have
a non-trivial corner contribution
when the entangling boundary has a sharp corner in critical theories \cite{Hirata_2007,kallin2013,Kallin_2014,Bueno_2015,Faulkner_2016,Bueno_2016,Whitsitt_2017,Seminara_2017,Bueno_2019,Stoudenmire2014}; similar behavior was recently found in the context of integer quantum Hall states \cite{Sirois2021}.
One could imagine that there is a similar contribution 
to quantities that we studied in our work.
It is unclear at this moment if our method is capable of
capturing non-trivial geometry at the point where all spatial
regions meet. 
Also, as we mentioned,
in the R-R-R sector,
we expect that a non-trivial flux (anyon) should
be located just at the junction because of the conservation
of topological charge. 
Understanding 
how precisely correlation measures depend on
such excitation 
is an important open question.

Putting our work in a slightly broader context,
one of the important questions is to understand
what kind of underlying (topological/geometrical) data
can appear in entanglement measures.  
While we took chiral $p$-wave superconductors and
Chern insulators as examples,
in order to get more general pictures, 
it is desirable to extend
our analysis to more generic topological liquids.
In the future, we plan to study Abelian fractional quantum Hall states 
by constructing vertex states for
multi-component compactified boson theories.
We can also discuss cases 
where the different spatial regions 
$A,B,C$ 
have different topological orders.
Such configurations involving gapped interfaces between distinct phases have garnered much attention recently due to the possibility of trapping parafermion zero modes at domain walls along these interfaces \cite{BarkeshliQi-2012,Lindner-2012,Clarke-2013,Cheng-2012,BarkeshliJianQi-2013-a,BarkeshliJianQi-2013-b,Mong-2014,khanteohughes-2014,SantosHughes-2017,santos2019}. The entanglement entropy for an interface between two distinct arbitrary Abelian phases \cite{cano2015interfaces,fliss2017interface} and for particular classes of non-Abelian phases \cite{fliss2020nonabelianinterface,sohal2020nonabelian} has already been computed. In the former case, the entanglement was subsequently shown to signal the presence of an emergent one-dimensional topological phase along the interface \cite{Santos2018}. It is natural to expect more exotic outcomes could occur in the trijunction configurations we have considered. 

Finally, while we took in this paper an approach from the edge theory,
it is interesting to study the entanglement negativity
using complementary bulk approaches.
For example, we can study entanglement negativity
in lattice models such as string net models.
Also, it is interesting to formulate surgery calculations for the entanglement measures we have considered \cite{witten1989jones,dong2008surgery,wen2016surgery,fliss2020nonabelianinterface,berthiere2021}. 
These alternative bulk calculations
can clarify precisely what kind of topological data
can be captured by the entanglement negativity
in the setup studied in this work.

\acknowledgements

We would like to acknowledge 
Roger Mong, 
Karthik Siva,
Tomo Soejima,
Mike Zaletel, and
Yijian Zou,
for insightful discussions,
and 
for sharing their manuscript
\cite{BerkeleyPaper} prior to arXiv submission.
S.R.~is supported by the National Science Foundation under 
Award No.\ DMR-2001181, and by a Simons Investigator Grant from
the Simons Foundation (Award No.~566116).
R.S. acknowledges the support of the Natural Sciences and Engineering Research Council of Canada (NSERC) [funding reference number 6799-516762-2018] and of the US National Science Foundation under Grant No.\ DMR-1725401 at the University of Illinois.

\appendix

\section{Direct calculation method}
\label{subsec:direct}

The Neumann function method provides an elegant way of deriving the form of the
conformal boundary state for free theories, which extends straightforwardly to
the tripartition case (and, indeed, more general $n$-partition).
As a check on our results using this method, we rederive
the
vertex states in this section using a more direct approach.
In this section, we will work with
the Majorana and complex fermion fields.
We recall that
the edge state Hamiltonian including gapping potential terms is given by
\begin{align}
  H_0
  &= \int_0^{2\pi}d\sigma
        \sum_I f^{I\dag} i\partial_\sigma f^{I},
        \nonumber \\
  H_{{\it int}}
   &= \int_0^{2\pi}d\sigma
     \boldsymbol{f}^{\dag}(\sigma) M \boldsymbol{f}(2\pi-\sigma)
     + {\it h.c.}
\end{align}
where in the last line we used a vectorial notation
and the mass matrix $M$ is given by
\begin{align}
M=m\left(
\begin{array}{ccc}
0 & 1 & 0\\
0 & 0 & 1\\
1 & 0 & 0
\end{array}
\right).
\end{align}
Corresponding to this situation, we seek for a state $|V\rangle$ which satisfies, for $0<\sigma<\pi$,
\begin{align}
    \left[ f^I(\sigma)-i f^{I+1}(2\pi-\sigma)\right]|V\rangle =0.
\end{align}
Solving the constraint, the state $|V\rangle$ is given in the form of a
fermionic coherent state.
A major simplification for the case of complex fermion
is that we can diagonalize the mass matrix $M$ by a unitary rotation
$U$
as
$M=U^\dag \Lambda U$, where
\begin{align}
  &
U=\frac{1}{\sqrt{3}}\left(
\begin{array}{ccc}
1 & 1 & 1\\
\omega^* & \omega & 1\\
\omega & \omega^* & 1
\end{array}
                    \right), \nonumber\\
                    &
    \Lambda=\mathrm{diag}(1,\omega,\omega^*),
    \quad
   \omega=e^{2\pi i/3}.
\label{rot}
\end{align}
In the rotated basis $\bm{\eta} :=U\bm{ f}$,
the edge Hamiltonian is diagonal and given by
\begin{align}
  &
    H_0=\int_0^{2\pi} d\sigma \sum_{a=1}^3 \eta^\dag_a i\partial_\sigma \eta^{\ }_a,
    \nonumber \\
  &
    H_{int}=\int_0^\pi d\sigma
    \sum_{a=1}^3 m e^{ i\theta_a}
    \eta_a^\dag(\sigma)\eta^{\ }_a(2\pi-\sigma),
\end{align}
where
$\theta_1=0$,
$\theta_2=2\pi/3$,
and
$\theta_3=-2\pi/3$.
We take the spatial boundary conditions (Ramond or Neveu-Schwarz)
for $I=1,2,3$ being all identical, so the rotation does not affect the spatial boundary condition. Thus, in the rotated basis, we have three copies of the single fermion problem,
\begin{gather}
\left[\eta_a(\sigma)+g_a(\sigma)\eta_a(2\pi-\sigma)\right]|V\rangle=0,
\nonumber \\
\mbox{where}\quad
g_a(\sigma)=-is(\sigma)e^{i s(\sigma)\theta_a}, \label{v state single copy 1}
\end{gather}
and $s(\sigma)$ is the sign
function: $s(\sigma)=1$ for $0<\sigma<\pi$ and $s(\sigma)=-1$ for
$\pi<\sigma<2\pi$.
Similarly, the boundary condition for $\eta^\dag$ is given by
\begin{gather}
  \left[\eta^{\dag}_a(\sigma)
  +\tilde{g}_a(\sigma)\eta^{\dag}_a(2\pi-\sigma)\right]|V\rangle=0
    \nonumber \\
  \mathrm{where}
  \quad
  \tilde{g}_a(\sigma)\equiv-g_a(-\sigma)=-is(\sigma)
    e^{-is(\sigma)\theta_a}. \label{v state single copy 2}
\end{gather}

For now, we focus on the vertex state for a given copy and
omit the subscript $a$.
We will restore the subscript later when it is necessary. 
By mode expansion,
$\eta(\sigma)=\sum_{r}e^{-i\sigma r}\eta_r$ and
$g(\sigma)=\sum_{n\in\mathbb{Z}}e^{in\sigma}g_n$,
the gluing condition can be written as
\begin{align}
[\eta_r+ N_{r,s}\eta_s]|V\rangle=0,
\end{align}
where $N_{r,s}:=g_{-r-s}$ and 
the Fourier components of $g(\sigma)$ are
given by
\begin{align}
  g_n
     &=
       \begin{cases}
         0&  n\neq0\text{, $n$ is even}\\
         \frac{-2\cos{\theta}}{n\pi}& \text{$n$ is odd}\\
         \sin{\theta} & n=0
       \end{cases}.
\end{align}
In the next subsections,
we discuss the construction of the vertex state for
each copy, for the Majorana and Dirac fermion fields
separately.
Here, we summarize the result.
We separate $N_{r,s}$ into four blocks, 
\begin{align}
  &
  N^{++}_{r,s}=N_{r,s},\quad N^{--}_{r,s}=N_{-r,-s},\tb
  \nonumber \\
  &
    N^{-+}_{r,s}=N_{-r,s}, \quad N^{+-}_{r,s}=N_{r,-s},
    \quad 
    r,s>0
\end{align}
The vertex state solution is
\begin{align}
  &
    |V\rangle
    \propto
    \exp
    \Big(\sum_{r,s\geq 1/2}K_{rs}\eta^\dag_r \eta^{\ }_{-s}\Big)
    |0\rangle,
    \nonumber \\
  &
    \mathrm{with}\quad K\equiv -(\mathbbm{1}+N^{++})^{-1}(N^{+-})
    \nonumber \\
  &\hspace{1.35cm}
    =-(N^{-+})^{-1}(\mathbbm{1}+N^{--}),
    \label{K and V}
\end{align}
where $|0\rangle$ is the Fermi sea annihilated by $\eta_r,r>0$ and
$\eta_r^\dag,r<0$. The equivalence of the two expressions of $K$ comes from the
fact that $\sum_{s}N_{rs}N_{st}=\delta_{r,t}$.
We give the detailed derivation in 
the next subsections.
Denoting $\eta_r=u_r$ and $\eta_{-r}=v_r^\dag$ for $r>0$,
the solution can be written in the familiar Gaussian state form,
\begin{align}
  |V\rangle
   &\propto \exp{
     \Big(\sum_{r,s\geq 1/2}K_{rs} u^\dag_r v^\dag_s\Big)}|0\rangle
    \nonumber \\
   &=\exp{
     \Big(\frac{1}{2}\sum_{r,s\geq 1/2}
     \left[
     K_{rs} u^\dag_r v^\dag_s-
     (K^T)_{rs}v_r^\dag u_s^\dag
     \right]
     \Big)}|0\rangle.
\end{align}


Combinining
the three copies 
$\eta_1,\eta_2,\eta_3$
and restoring the subscript $a=1,2,3$ for $K,u,v$,
the vertex state in the rotated basis is
\begin{align}
  &
  |V\rangle
    =
    \mathcal{N}
    \exp{\left[\frac{1}{2}(\bm{V}^\dag)^T \bm{K}\bm{V}^\dag\right]}|0\rangle
\end{align}
where
\begin{align}
&(\bm{V}^\dag)^T=
\left(
\bm{u}_1^\dag,
\bm{v}_1^\dag,
\bm{u}_2^\dag,
\bm{v}_2^\dag,
\bm{u}_3^\dag,
\bm{v}_3^\dag
\right),
\nonumber \\\
&\bm{K}=
\left(
\begin{array}{cccccc}
0 & K_1 & 0 & 0 & 0 &0\\
-K_1^T & 0 & 0 & 0 & 0 &0\\
0 & 0 & 0 & K_2 & 0 &0\\
0 & 0 & -K_2^T & 0 & 0 &0\\
0 & 0 & 0 & 0 & 0 & K_3\\
0 & 0 & 0 & 0 & -K_3^T &0\\
\end{array}
\right).
\end{align}
We may use an inverse rotation to write the solution in the original basis $f,f^\dg$ (See Appendix \ref{app:compare}). 


\subsection{Majorana fermion}

Let us now
discuss the type of state presented
in
\eqref{v state single copy 1}
and
\eqref{v state single copy 2}
in more detail.
As a warm up, we first consider 
a similar problem 
for the Majorana fermion edge mode,
following
Ref.\ \cite{imamura2008boundary}.
Let us consider the Majorana fermion field,
and the boundary condition
\begin{equation}
\label{maj bdy cond}
  \big[
  \psi(\sigma)+g(\sigma)\psi(-\sigma)
  \big]|V\rangle =0 
  \quad
  \mathrm{for}
  \quad
  -\pi<\sigma<\pi.
\end{equation}
We focus on the NS sector. 
As a specific example, we can take $g(\sigma)=\pm i\mathrm{sign}(\sigma)$.
We however proceed with 
a generic choice of $g(\sigma)$.
$g(\sigma)$ is subject to 
a consistency condition:
Assuming $g(\sigma)\neq 0$,
we note that 
the condition \eqref{maj bdy cond}
can be rewritten as
\begin{align}
  &
    \big[
g(\sigma)^{-1}\psi(\sigma)+\psi(-\sigma)
    \big]|V\rangle =0
  \nonumber \\
  &
\Longrightarrow
    \big[
    g(-\sigma)^{-1}\psi(-\sigma)+\psi(\sigma)
    \big]|V\rangle=0.
\end{align}
Comparison with Eq.\ \eqref{maj bdy cond}
implies
\begin{equation}
g(\sigma)g(-\sigma)=1.
\end{equation}
In terms of the Fourier components
of $g(\sigma)$,
$
g(\sigma)=\sum_{n\in\mathbb{Z}}e^{in\sigma}g_n,
$
the consistency condition reads
$
    \sum_n g_n g_{n+p}=\delta_{p,0}.
$
%

Let us now proceed to the construction of $|V\rangle$.
Defining a matrix $N_{n,m}=g_{-n-m}$,
the boundary condition and
the consistency relation 
can be written as 
\begin{align}
  \label{bc}
  &
  \big[\psi_r+\sum_s N_{r,s}\psi_s\big]|V\rangle =0,
  \nonumber \\
  &
  \sum_m N_{n,m}N_{m,l}
  =\delta_{nl}
  \quad
  ( N^2=\mathbbm{1}),
\end{align}
respectively.
For convenience, we use fermionic creation/annihilation operators to write
\begin{equation}
\left(
\begin{array}{c}
\psi_{1/2}\\
\psi_{3/2}\\
\vdots
\end{array}
\right)\equiv \bm{b},
\quad
\left(
\begin{array}{c}
\psi_{-1/2}\\
\psi_{-3/2}\\
\vdots
\end{array}
\right)\equiv \bm{b}^\dag.
\end{equation}
We also introduce a block structure
\begin{gather}
N=\left(
\begin{array}{cc}
N^{++} & N^{+-}\\
N^{-+} & N^{--}
\end{array}
         \right),
          \\
  N^{++}_{r,s}=N_{r,s}=g_{-r-s},
  \quad
  N^{+-}_{r,s}=N_{r,-s}=g_{-r+s},
    \nonumber \\
  N^{-+}_{r,s}=N_{-r,s}=g_{r-s},
  \quad
  N^{--}_{r,s}=N_{-r,-s}=g_{r+s}.
  \nonumber
\end{gather}
From the consistency condition $N^2=\mathbbm{1}$,
these blocks satisfy
\begin{align}
  (i)&\quad N^{++}N^{++}+N^{+-}N^{-+}=\mathbbm{1},
       \nonumber \\
  (ii)&\quad N^{++}N^{+-}+N^{+-}N^{--}=0,
        \nonumber \\
  (iii)&\quad N^{-+}N^{++}+N^{--}N^{-+}=0,
         \nonumber \\
(iv)&\quad N^{-+}N^{+-}+N^{--}N^{--}=\mathbbm{1}.
\end{align}
We also note $N^T=N$, which is the consequence of $N_{r,s}=g_{-r-s}$. This leads to
\begin{align}
  &
  (N^{++})^T=N^{++},
  \quad
  (N^{--})^T=N^{--},
  \nonumber \\
  &
  (N^{+-})^T=N^{-+},
  \quad
  (N^{-+})^T=N^{+-}.
\end{align}
The boundary condition \eqref{bc}
can now be written two different ways 
as
\begin{align}
  &
    \big[
    \psi_r+\sum_{s>0}N_{r,s}\psi_s+\sum_{s>0}N_{r,-s}\psi_{-s}
    \big]|V\rangle =0
    \nonumber \\
  &\quad
    \Longrightarrow
    \begin{cases}
    \big[\bm{b}+(\mathbbm{1}+N^{++})^{-1}N^{+-}\bm{b}^\dag
    \big]|V\rangle =0\\
    \big[
    \bm{b}+(N^{-+})^{-1}(\mathbbm{1}+N^{--})\bm{b}^\dag\big]
    |V\rangle=0.
  \end{cases}
\end{align}
The equivalence of 
the two conditions can be established by using the
consistency equations $(i)-(iv)$:
We first note that
\begin{equation}
  (\mathbbm{1}+N^{++})(\mathbbm{1}-N^{++})=\mathbbm{1}-(N^{++})^2=N^{+-}N^{-+}
\end{equation}
where we used $(i)$ in the first line.
This relation can be rewritten as
\begin{equation}
\begin{aligned}
&
(\mathbbm{1}-N^{++})^{-1}(\mathbbm{1}+N^{++})^{-1}=(N^{+-}N^{-+})^{-1}
\\
&\quad \Longrightarrow
\quad
(\mathbbm{1}+N^{++})^{-1}(N^{+-})=(\mathbbm{1}-N^{++})(N^{-+})^{-1}.
\end{aligned}
\end{equation}
Next, from $(iii)$, we note that $N^{-+}$ intertwines
the $+$ and $-$ sectors
as $N^{-+}N^{++}=-N^{--}N^{-+}$.
This leads to $N^{++}(N^{-+})^{-1}=-(N^{-+})^{-1}N^{--}$,
and 
$(N^{++})^k(N^{-+})^{-1}=(N^{-+})^{-1}(-N^{--})^k$. Then,
\begin{equation}
  (\mathbbm{1}-N^{++})(N^{-+})^{-1}
  =(N^{-+})^{-1}(\mathbbm{1}+N^{--}),
\end{equation}
which
completes
the proof of the equivalence. We now define
\begin{equation}
  K
  :=(\mathbbm{1}+N^{++})^{-1}(N^{+-})
  =(N^{-+})^{-1}(\mathbbm{1}+N^{--}),
\end{equation}
and write the boundary condition as
\begin{equation}
\left[\bm{b}+K\bm{b}^\dag\right]|V\rangle=0.
\end{equation}

We note that 
if $g(\sigma)$ is odd under
$\sigma\to -\sigma$, 
$g_n=-g_{-n}$,
then $K^T=-K$.
This can be seen 
by first noting
that $g_n=-g_{-n}$,
implies
$N^{++}=-N^{--}$
and
$N^{+-}=-N^{-+}$,
which make $N^{++}/N^{--}$ and $N^{+-}/N^{-+}$ commute. We then see
\begin{equation}
\begin{aligned}
K^T
  &=(N^{+-})^T[(\mathbbm{1}+N^{++})^{-1}]^T
  \\
&=(N^{-+})(\mathbbm{1}+N^{++})^{-1}\\
&=(\mathbbm{1}+N^{++})^{-1}(N^{-+})\\
&=(\mathbbm{1}+N^{++})^{-1}(-N^{+-})=-K.
\end{aligned}
\end{equation}

Finally, using the antisymmetry of $K$,
we can write down the solution of boundary condition $\left[\bm{b}+K\bm{b}^\dag\right]|V\rangle=0$:
\begin{equation}
  |V\rangle
  \propto
  \exp{\Big(-\frac{1}{2}\sum_{r,s\geq 1/2}K_{r,s}b_r^\dag b_s^\dag\Big)}
  |0\rangle.
\end{equation}
This can be checked by the Baker-Hausdorff formula.

\subsection{Dirac fermion}
\label{app:Diracfermion}
Let us now turn to the case of Dirac fermions
$f(\sigma), f^{\dag}(\sigma)$.
Consider a boundary condition
\begin{align}
  \label{vtx cond cmplx}
  &
   [f(\sigma)+g(\sigma) f(-\sigma)]|V\rangle =0,
  \nonumber \\
 & 
  [f^\dag(\sigma)+\tilde{g}(\sigma) f^\dag(-\sigma)]|V\rangle =0,
\end{align}
for $-\pi<\sigma<\pi$.
At this moment, $\tilde{g}(\sigma)$
appears to be an independent function, not related to
$g(\sigma)$. 
We however require the condition
\begin{equation}
  \label{cond on g and t tilde}
\tilde{g}(\sigma)=-g(-\sigma).
\end{equation}
We will see momentarily the implication of this condition on
the vertex state.
As a specific example, we consider
\begin{align}
  &
  g(\sigma)=-is(\sigma)e^{is(\sigma)\theta},
  \nonumber \\
  &
    \tilde{g}(\sigma)=-g(-\sigma)
    =-is(\sigma)e^{-is(\sigma)\theta},
\end{align}
where $s(\sigma)=\mathrm{sgn}(\sigma)$.

In the Fourier space,
the condition \eqref{vtx cond cmplx} reads
\begin{align}
  \label{vtx cond cmplx Fourier}
  &
   \big[f_r+\sum_s N_{r,s} f_s\big]|V\rangle =0,
    \nonumber \\
  &
    \big[
    \tilde{ f}_r+\sum_s \tilde{N}_{r,s}\tilde{ f}_s\big]
    |V\rangle =0,
  \nonumber \\
  &
  N_{r,s}=g_{-r-s},
    \quad
  g(\sigma)=\sum_{n\in\mathbb{Z}}e^{in\sigma}g_n.
    \nonumber \\
  &
  \tilde{N}_{r,s}=\tilde{g}_{-r-s},
  \quad
  \tilde{g}(\sigma)=\sum_{n\in\mathbb{Z}}e^{in\sigma}\tilde{g}_n.
\end{align}
where the Fourier decomposition of $f^{\dag}$
is given by 
$f^\dag(\sigma)=\sum_{s\in\mathbb{Z}+1/2}e^{i\sigma s} f_s^\dag=\sum_{s\in\mathbb{Z}+1/2}e^{-i\sigma s}\tilde{ f}_s$.
Namely, we introduced the new set of operators $\tilde{ f}_s$ by $\tilde{f}_s\equiv f_{-s}^\dag$.
Similarly the condition \eqref{cond on g and t tilde}
in the Fourier space is
\begin{equation}
\tilde{g}_n=-g_{-n}.
\end{equation}
We define the creation/annihilated operators as
\begin{align}
  &
\left(
\begin{array}{c}
 f_{1/2}\\
 f_{3/2}\\
\vdots
\end{array}
  \right)\equiv \bm{b},
  \quad
\left(
\begin{array}{c}
 f_{-1/2}\\
 f_{-3/2}\\
\vdots
\end{array}
  \right)\equiv \bm{c}^\dag,
  \nonumber \\
  &
\left(
\begin{array}{c}
 f_{1/2}^\dag\\
 f_{3/2}^\dag\\
\vdots
\end{array}
  \right)\equiv \bm{b}^\dag,
  \quad
\left(
\begin{array}{c}
 f_{-1/2}^\dag\\
 f_{-3/2}^\dag\\
\vdots
\end{array}
\right)\equiv \bm{c},
\end{align}
and we also define, similarly,
\begin{align}
\left(
\begin{array}{c}
\tilde{ f}_{1/2}\\
\tilde{ f}_{3/2}\\
\vdots
\end{array}
  \right)\equiv \bm{c},
  \quad
\left(
\begin{array}{c}
\tilde{ f}_{-1/2}\\
\tilde{ f}_{-3/2}\\
\vdots
\end{array}
\right)\equiv \bm{b}^\dag.
\end{align}
The conditions
in \eqref{vtx cond cmplx Fourier}
can be organized as
\begin{equation}
  \label{cond 1}
  \begin{aligned}
    &
    \big[f_r+\sum_{s\geq 1/2}N_{r,s} f_s+\sum_{s\geq 1/2}N_{r,-s} f_{-s} \big]|V\rangle=0\\
    &\quad \Longrightarrow
    \begin{cases}
      \big[\bm{b}+(\mathbbm{1}+N^{++})^{-1}N^{+-}\bm{c}^\dag \big]|V\rangle =0\\
      \big[\bm{b}+(N^{-+})^{-1}(\mathbbm{1}+N^{--})\bm{c}^\dag \big]|V\rangle =0
    \end{cases}
  \end{aligned}
\end{equation}
\begin{equation}
  \label{cond 2}
  \begin{aligned}
    &
    \big[\tilde{ f}_r+\sum_{s\geq 1/2}\tilde{N}_{r,s}\tilde{ f}_s+\sum_{s\geq 1/2}\tilde{N}_{r,-s}\tilde{ f}_{-s}\big]|V\rangle=0\\
    &\quad \Longrightarrow
    \begin{cases}
      \big[\bm{c}+(\mathbbm{1}+\tilde{N}^{++})^{-1}\tilde{N}^{+-}\bm{b}^\dag \big]|V\rangle =0\\
      \big[\bm{c}+(\tilde{N}^{-+})^{-1}(\mathbbm{1}+\tilde{N}^{--})\bm{b}^\dag \big]|V\rangle =0
    \end{cases}
  \end{aligned}
\end{equation}
As we have seen, the two conditions
in \eqref{cond 1} are equivalent by using $N^2=\mathbbm{I}$. 
Similarly,
the two conditions in \eqref{cond 2}
are equivalent by using $\tilde{N}^2=\mathbbm{I}$.

Now let us define
\begin{align}
  K&=(\mathbbm{1}+N^{++})^{-1}(N^{+-})
     =(N^{-+})^{-1}(\mathbbm{1}+N^{--}),
  \nonumber \\
  \tilde{K}&=(\mathbbm{1}+\tilde{N}^{++})^{-1}(\tilde{N}^{+-})
             =(\tilde{N}^{-+})^{-1}(\mathbbm{1}+\tilde{N}^{--}).
\end{align}
Then, the boundary conditions are written as
$\big[\bm{b}+K\bm{c}^\dag\big]|V\rangle =
  \big[\bm{c}+\tilde{K}\bm{b}^\dag\big]|V\rangle=0$,
  or
\begin{equation}
\left[
\left(
\begin{array}{c}
\bm{b}\\
\bm{c}
\end{array}
\right)+
\left(\begin{array}{cc}
0 & K\\
\tilde{K} & 0
\end{array}\right)
\left(
\begin{array}{c}
\bm{b}^\dag\\
\bm{c}^\dag
\end{array}
\right)
\right]|V\rangle=0.
\end{equation}
Here, we note that the condition
\eqref{cond on g and t tilde}
imposes 
\begin{align}
  K^T = - \tilde{K}.
\end{align}
This can be seen by first noting
\begin{equation}
  K^T=(N^{-+})(\mathbbm{1}+N^{++})^{-1}
  =(\mathbbm{1}-N^{--})^{-1}(N^{-+})
\end{equation}
where we use the intertwining relation $N^{-+}N^{++}=-N^{--}N^{-+}$.
Second,
\eqref{cond on g and t tilde} implies 
$\tilde{N}^{++}=-N^{--}$,
and
$\tilde{N}^{+-}=-N^{-+}$,
which leads to
$K^T=(\mathbbm{1}-N^{--})^{-1}(N^{-+})
=-(\mathbbm{1}+\tilde{N}^{++})^{-1}(\tilde{N}^{+-})
=-\tilde{K}$.

With this condition, the vertex state is given by
\begin{equation}
\begin{aligned}
|V\rangle &\propto \exp{\left[-\frac{1}{2}\sum_{r,s\geq 1/2}\left(K_{rs}b_r^\dag c_s^\dag+\tilde{K}_{rs}c_r^\dag b_s^\dag\right) \right]}|0\rangle\\
&=\exp{\left[-\sum_{r,s\geq 1/2}K_{rs}b_r^\dag c_s^\dag\right]}|0\rangle\\
&=\exp{\left[-\sum_{r,s\geq 1/2}K_{rs} f_r^\dag  f_{-s}\right]}|0\rangle
\end{aligned}
\end{equation}

 \begin{widetext}
\subsection{Comparison with the Neumann 
function method} 
\label{app:compare}

Let us now take the complex fermion as an example and compare the elements of 
the Neumann coefficient matrix $K$
in the NS-NS-NS sector, 
between the direct calculation and 
Neumann function method. 
For the direct calculation method, the matrix $K$ in Eq.\ \eqref{K and V} is in the rotated $\eta$ basis, so we need to rotate back to $f$ basis, namely,
\begin{equation}
K_f = U^\dg\left(
\begin{array}{ccc}
K_{\eta,1} & & \\
 & K_{\eta,2} & \\
 & & K_{\eta,3}
\end{array}
\right) U
\end{equation}
In the direction method,
we take the cutoff to be $N_c=400$ and compute 
the Neumann coefficients $K$ numerically. 
In the following tables, we take the first $8\times 8$ block from the $K^{12}$ matrix in both cases. The real and imaginary parts obtained from the direct calculation and Neumann function method are:
\begin{equation}
\mathrm{Re} \left[K\right]_{direct}=-\left(
\begin{array}{cccccccc}
0 & 0.2971 & 0 & 0.0945 & 0 & 0.0564 & 0 & 0.0406\\
-0.2964 & 0 & 0.3127 & 0 & 0.0990 & 0 & 0.0569 & 0\\
0 & -0.3124 & 0 & 0.3183 & 0 & 0.1047 & 0 & 0.0620\\
-0.0934 & 0 & -0.3178 & 0 & 0.3163 & 0 & 0.1033 & 0\\
0 & -0.0988 & 0 & -0.3159 & 0 & 0.3189 & 0 & 0.1060\\
-0.0549 & 0 & -0.1040 & 0 & -0.3184 & 0 & 0.3172 & 0\\
0 & -0.0568 & 0 & -0.1030 & 0 & -0.3168 & 0 & 0.3190\\
-0.0389 & 0 & -0.0612 & 0 & -0.1054 & 0 & -0.3185 & 0
\end{array}
\right),
\end{equation}
\begin{equation}
\mathrm{Re}\left[K\right]_{Neumann}=\left(
\begin{array}{cccccccc}
0 & 0.2963 & 0 & 0.0933 & 0 & 0.0548 & 0 & 0.0388\\
-0.2963 & 0 & 0.3128 & 0 & 0.0990 & 0 & 0.0570 & 0\\
0 & -0.3128 & 0 & 0.3177 & 0 & 0.1040 & 0 & 0.0611\\
-0.0933 & 0 & -0.3177 & 0 & 0.3163 & 0 & 0.1034 & 0\\
0 & -0.0990 & 0 & -0.3163 & 0 & 0.3184 & 0 & 0.1053\\
-0.0548 & 0 & -0.1040 & 0 & -0.3184 & 0 & 0.3173 & 0\\
0 & -0.0570 & 0 & -0.1034 & 0 & -0.3173 & 0 & 0.3184\\
-0.0388 & 0 & -0.0611 & 0 & -0.1052 & 0 & -0.3184 & 0
\end{array}
\right),
\end{equation}
\begin{equation}
\mathrm{Im}\left[K\right]_{direct}=\left(
\begin{array}{cccccccc}
-0.7699 & 0 & -0.0998 & 0 & -0.0638 & 0 & -0.0477 & 0\\
0 & -0.5703 & 0 & -0.0444 & 0 & -0.0330 & 0 & -0.0265\\
-0.0999 & 0 & -0.5537 & 0 & -0.0384 & 0 & -0.0302 & 0\\
0 & -0.0444 & 0 & -0.5322 & 0 & -0.0255 & 0 & -0.0212\\
-0.0639 & 0 & -0.0384 & 0 & -0.5291 & 0 & -0.0238 & 0\\
0 & -0.0330 & 0 & -0.0255 & 0 & -0.5210 & 0 & -0.0179\\
-0.0478 & 0 & -0.0303 & 0 & -0.0238 & 0 & -0.5199 & 0\\
0 & -0.0265 & 0 & -0.0212 & 0 & -0.0179 & 0 & -0.5156
\end{array}
\right),
\end{equation}

\begin{equation}
\mathrm{Im}\left[K\right]_{Neumann}=-\left(
\begin{array}{cccccccc}
-0.7698 & 0 & -0.0998 & 0 & -0.0638 & 0 & -0.0476 & 0\\
0 & -0.5702 & 0 & -0.0444 & 0 & -0.0329 & 0 & -0.0264\\
-0.0998 & 0 & -0.5536 & 0 & -0.0383 & 0 & -0.0302 & 0\\
0 & -0.0444 & 0 & -0.5321 & 0 & -0.0254 & 0 & -0.0211\\
-0.0638 & 0 & -0.0383 & 0 & -0.5291 & 0 & -0.0237 & 0\\
0 & -0.0329 & 0 & -0.0254 & 0 & -0.5209 & 0 & -0.0178\\
-0.0476 & 0 & -0.0302 & 0 & -0.0237 & 0 & -0.5199 & 0\\
0 & -0.0264 & 0 & -0.0211 & 0 & -0.0178 & 0 & -0.5155
\end{array}
\right).
\end{equation}
We see these two set of matrices are almost identical (up to a minus sign, which is presumably due to convention). The numerical check for other blocks $K^{11}$, etc shows the same results.

We also compare the $K$ matrices of closed string real fermion using direct calculation and Neumann coefficient method, and arrives at the same conclusion. Note that in the direct calculation, the rotation becomes
\begin{equation}
K_f = U^T \left(
\begin{array}{ccc}
K_{1,\eta}/2 & 0 & 0 \\
0&0 & -K_{2,\eta}^T/2 \\
0& K_{2,\eta}/2 & 0 
\end{array}
\right) U.
\end{equation}


\section{Details of the Neumann coefficient method}
\label{app:Neumann}

In this Section, we give 
some technical details for 
the Neumann coefficient method.


\subsection{Different choice 
of the branch cuts in the R-R sector}
\label{app:another-twist}

For the vertex state for bipartition
in the R-R sector,
we can work alternatively 
with the following 
choice of
the $g^{IJ}$ function:
\begin{equation}
    g^{IJ}_{\sigma-\sigma}=\frac{1}{2}
    \left[\sqrt{\frac{(\omega-\omega_{1,0})(\omega-\omega_{2,0})}{(\omega'-\omega_{1,0})(\omega'-\omega_{2,0})}}+(\omega\leftrightarrow\omega')
    \right].
\end{equation}
Both choices 
lead to the same vertex state
as we demonstrate below.
The choice we made in the main
text is somewhat simpler,
while 
this choice here is closer 
to the branch cuts we choose
in our calculations 
in the R-R-R sectors
for tripartition.
Using $\omega_I=\omega_{I,0}(\frac{1+z}{1-z})$, and $\omega_{1,0}=i,\omega_{2,0}=-i$, the Neumann function is given by
\begin{equation}
    \begin{aligned}
    R^{11} 
    = R^{22} 
    &= \frac{\sqrt{zz'}}{z-z'}
    \frac{1}{2}
    \left[\sqrt{\frac{z}{z'}}\frac{1-z'}{1-z}+\sqrt{\frac{z'}{z}}\frac{1-z}{1-z'}\right] 
     = \sum_{m\geq 1}
    \left(\frac{z'}{z}\right)^m
    +\frac{1}{2}
    \left[\sum_{n\geq 0}z^n
    -\sum_{n\geq 1}(z')^n
    \right],
    \\
    R^{12}  = -R^{21} 
    &= \frac{i\sqrt{zz'}}{1-zz'}
    \frac{1}{2}
    \left[\sqrt{\frac{z}{z'}}\frac{1-z'}{1-z}+\sqrt{\frac{z'}{z}}\frac{1-z}{1-z'}
    \right]
     =  (-i)\sum_{m>0}(zz')^m + \frac{i}{2}
    \left[\sum_{n>0}z^n+\sum_{n>0}(z')^n
    \right]. 
    \end{aligned}
\end{equation}
We note that we obtain
the desired singular term $\sum_{m\geq 1}(z'/z)^m = \sum_{m\geq 1}e^{-im(\sigma-\sigma')}$ in $R^{11}$ and $R^{22}$. 
From the expansion coefficients and use the same ansatz solution in Eq.\ \eqref{eqn:realfermionansatz}, we obtain the 
vertex state:
\begin{equation}
\begin{aligned}
    |V\rangle =\exp
    \left[
   -i\sum_{n\geq 1}\chi_{-n}^1\chi_{-n}^2 
   +\sum_{n\geq 1}
   \left(
   \chi_{-n}^1\chi_0^1+\chi_{-n}^2\chi_0^2
    +i\chi_{-n}^1\chi_0^2-i\chi_{-n}^2\chi_0^1 
    \right)
    \right]|\Omega\rangle. 
\end{aligned}
\end{equation}
This is the same solution as Eq.\ \eqref{eqn:solutionRR} 
with the additional requirement $(\chi_0^1+i\chi_0^2)|\Omega\rangle=0$. Similarly, 
for the Dirac fermion field in the R-R sector,
one can show the solutions 
from the two choices of the
branch cuts also match. 

\subsection{Verification of the boundary condition in the R sector}
\label{app:boundary-NS-R}

In this subsection, we verify that the
R-R-R sector vertex state ansatz 
satisfies the boundary condition for real and complex fermion. 
The verification for the NS-R two-string solution simply parallels the proof below\cite{jevicki1988supersymmetry}, which we shall omit. 

For the Majorana fermion case, the ansatz solution is:
\begin{equation}
\begin{aligned}
|V\rangle
&=
\exp
\left[
\frac{1}{2}\sum_{m,n\geq 1}
\chi_{-m}^I R^{IJ}_{mn} \chi_{-n}^J+2
\sum_{m,n\geq 1}\chi_{-m}^I R^{IJ}_{m0}\chi_0^J
\right]
|\Omega\rangle.
\end{aligned}
\label{eqn:MajoranaAnsatz}
\end{equation}
Let us denote $A=\sum_{m,n\geq 1}\frac{1}{2}\chi_{-m}^I R^{IJ}_{mn} \chi_{-n}^J+\sum_{m\geq 1}2\chi_{-m}^I R^{IJ}_{m0}\chi_0^J$. 
To show explicitly that this state satisfies the boundary condition, we define
\begin{equation}
D^I=\sum_{m\geq 1}2\chi_{-m}^{J}R_{m0}^{JI}.
\end{equation}
Using 
\begin{equation}
\begin{aligned}
&\chi_p^I|V\rangle
= \sum_{n\geq 1}R^{IJ}_{pn}\chi_{-n}^J|V\rangle+\exp{(A)}\left[2R^{IJ}_{p0}(\chi_0^J-\sum_{m\geq 1}\chi_{-m}^K R^{KJ}_{m0})\right]|\Omega\rangle,
\\
&\chi_0^I|V\rangle=\exp{(A)}\left[\chi_0^I-\sum_{m\geq 1}2\chi_{-m}^J R_{m0}^{JI}\right]|\Omega\rangle,
\end{aligned}
\end{equation}
one can check the following relation:
\begin{equation}
\begin{aligned}
\chi^I(\sigma)|V\rangle
&=\sum_{p\geq 1}\chi_p^I e^{-ip\sigma}|V\rangle+\chi_0^I|V\rangle+\sum_{p\geq 1}\chi_{-p}^I e^{ip\sigma}|V\rangle
\\
&=\sum_{p,n\geq 1}e^{-ip\sigma}
R^{IJ}_{pn}\chi_{-n}^J|V\rangle
+\sum_{p\geq 1}e^{-ip\sigma}\exp{(A)}\left[2R^{IJ}_{p0}(\chi_0^J-\sum_{m \geq 1}\chi_{-m}^K R^{KJ}_{m0})\right]|\Omega\rangle
+\chi_0^I|V\rangle+\sum_{p\geq 1}\chi_{-p}^I e^{ip\sigma}|V\rangle.
\end{aligned}
\end{equation}
On the other hand, defining
\begin{equation}
\tilde{\chi}^I_{cr.}=\sum_{n\geq 1}\chi_{-n}^I e^{in\sigma}+2\chi_0^I +D^I,
\end{equation}
and using
\begin{equation}
    (D^I+2\chi_0^I)|V\rangle=\exp{(A)}\left[2\chi_0^I-\sum_{m\geq 1}2 \chi_{-m}^J R_{m0}^{JI}\right]|\Omega\rangle,
\end{equation}
one can check,
\begin{equation}
\begin{aligned}
\int \frac{d\sigma'}{2\pi} R^{IJ}(\sigma,\sigma')\tilde{\chi}^J_{cr.}(\sigma')|V\rangle
& = \sum_{m,n\geq 1}
e^{-im\sigma}R^{IJ}_{mn}\chi_{-n}^J|V\rangle+\sum_{m\geq 1}e^{-im\sigma}R^{IJ}_{m0}(D^J+2\chi_0^J)|V\rangle\\
&
\quad 
+\sum_{n\geq 1}R^{IJ}_{0n}\chi_{-n}^J|V\rangle+\delta_{IJ}\frac{1}{2}(D^J+2\chi_0^J)|V\rangle+\sum_{m\geq 1}\chi_{-m}^I e^{im\sigma}|V\rangle
\\
&=\chi^I(\sigma)|V\rangle,
\end{aligned}
\end{equation}
where we exploited the fact $R^{IJ}_{m0}=-R^{JI}_{0m}$ and $R^{IJ}_{00}=\delta_{IJ}\frac{1}{2}$. Finally, the property $R^{I+1,J}(\sigma,\sigma')=iR^{I,J}(2\pi-\sigma,\sigma')$ ensures that $\chi$ satisfies the desired boundary condition
\begin{equation}
\chi^{I+1}(\sigma)|V\rangle=i\chi^{I}(2\pi-\sigma)|V\rangle.
\end{equation}

For the complex fermion, we start from the ansatz solution in Eq.~\eqref{eqn:R-R-R-complex}:
\begin{equation}
\begin{aligned}
|V\rangle
= \exp{\left(\sum_{m,n\geq 1}g^I_{-m}R^{IJ}_{mn}g_n^{\dg,J}+\sum_{m\geq 1}2R^{IJ}_{m0}(g_{-m}^I g_0^{\dg,J}+g_m^{\dg,I}g_0^J) \right)}|\Omega\rangle.
\end{aligned}
\end{equation}
We can verify the following relations:
\begin{equation}
\begin{aligned}
&g^I(\sigma)|V\rangle=\int \frac{d\sigma'}{2\pi}R^{IJ}(\sigma,\sigma')\tilde{g}^J_{cr.}(\sigma'
)|V\rangle,\\
&g^{\dg,I}(\sigma)|V\rangle=\int \frac{d\sigma'}{2\pi}R^{IJ}(\sigma,\sigma')\tilde{g}^{\dg,J}_{cr.}(\sigma')|V\rangle,
\end{aligned}
\end{equation}
where 
\begin{equation}
\begin{aligned}
&\tilde{g}^I_{cr.}(\sigma)=\sum_{n\geq 1}g_{-n}^I e^{in\sigma}+(2g_0^I+D^I),
\quad 
D^I=\sum_{m\geq 1}2R_{m0}^{JI}g_{-m}^J
\\
&\tilde{g}^{\dg,J}_{cr.}(\sigma')=\sum_{n\geq 1}g_{n}^{\dg,I} e^{in\sigma}+(2g_0^{\dg,I}+D^{\dg,I}),
\quad 
D^{\dg,I}=\sum_{m\geq 1}2R_{m0}^{JI}g_m^{\dg,J}.
\end{aligned}
\end{equation}
These relations allow us to verify the boundary condition:
\begin{equation}
    \begin{aligned}
    g^{I+1}(\sigma)|V\rangle&=\int \frac{d\sigma'}{2\pi}R^{I+1,J}(\sigma,\sigma')\tilde{g}^J_{cr.}(\sigma'
)|V\rangle\\
    & = i\int \frac{d\sigma'}{2\pi}R^{IJ}(2\pi-\sigma,\sigma')\tilde{g}^J_{cr.}(\sigma'
)|V\rangle\\
    & = ig^{I}(2\pi-\sigma)|V\rangle.
    \end{aligned}
\end{equation}
Similarly, for $g^\dg$, we 
can verify
$
g^{\dg I+1}(\sigma)|V\rangle
=i g^{\dg I}(2\pi-\sigma)|V\rangle. 
$

\subsection{Explicit form of 
the Neumann coefficients in the NS-NS-NS sector}
\label{app:K-matrix-coeffi}

The explicit form of 
the Neumann coefficient matrix $K$ in the NS-NS-NS sector is derived following the methods of Ref. \cite{gross1987operator} and is summarized below:
\begin{equation}
\begin{aligned}
&K=I_3\otimes K^{aa}+J_+ \otimes K^{a,a+1}+J_-\otimes K^{a,a-1},
\\
&I_{rs}=
\begin{cases}
\left(\frac{-m}{n+m+1}+\frac{-m}{n-m}\right)u_n u_m & n=\text{even},m=\text{odd}\\
\left(\frac{n}{n+m+1}-\frac{n}{n-m}\right)u_n u_m & n=\text{odd},m=\text{even}
\end{cases},
\\
&K^{aa}_{rs}=\frac{1}{3}I_{rs}+\left[\frac{M^+_{r-1/2,s-1/2}}{r+s}+\frac{M^-_{r-1/2,s-1/2}}{r-s}\right],
\\
&M^+_{nm}=-\left[(n+1)g_{n+1}(m+1)g_{m+1}-n g_n m g_m\right] \cdot\left[(-1)^n-(-1)^m\right],
\\
&M^-_{nm}=-\left[(n g_{n}(m+1)g_{m+1}-(n+1) g_{n+1} m g_m\right]\cdot\left[(-1)^n-(-1)^m\right],
\\
&K^{a,a+1}_{rs}=\frac{1}{2}I_{rs}-\frac{1}{2}K^{aa}_{rs}
 -\frac{(-i)}{2}\sqrt{3}
\left[\frac{\bar{M}^+_{r-1/2,s-1/2}}{r+s}+\frac{\bar{M}^-_{r-1/2,s-1/2}}{r-s}\right],
\\
&\bar{M}^+_{nm}=\left[(n+1)g_{n+1}(m+1)g_{m+1}-n g_n m g_m\right] \cdot\left[(-1)^n+(-1)^m\right],
\\
&\bar{M}^-_{nm}=\left[(n g_{n}(m+1)g_{m+1}-(n+1) g_{n+1} m g_m\right] \cdot\left[(-1)^n+(-1)^m\right],
\\
&K^{a,a-1}_{rs}=\frac{1}{2}I_{rs}-\frac{1}{2}K^{aa}_{rs}
+\frac{(-i)}{2}\sqrt{3}
\left[\frac{\bar{M}^+_{r-1/2,s-1/2}}{r+s}+\frac{\bar{M}^-_{r-1/2,s-1/2}}{r-s}\right],
\\
&I_3=\left(
\begin{array}{ccc}
1 & 0 & 0\\
0 & 1 & 0\\
0 & 0 & 1
\end{array}
\right),
\quad
J_+=\left(
\begin{array}{ccc}
0 & 1 & 0\\
0 & 0 & 1\\
1 & 0 & 0
\end{array}
\right),
\quad 
J_-=(J_+)^T,
\end{aligned}
\label{eqn:K_fstring}
\end{equation}
where $r=n+\frac{1}{2},s=m+\frac{1}{2}$. $u_n$ is the coefficient in the expansion of ($\frac{1+x}{1-x})^{1/2}=\sum_{n=0}^\infty u_n x^n$, which can be expressed compactly as $u_{2n}=u_{2n+1}={-\frac{1}{2} \choose n}(-1)^n$. 
We note $u_n$ satisfies the
recursion relation:
\begin{equation}
u_0=u_1=1,\tb 2n u_{2n}=(2n-1)u_{2n-2},\tb u_{2n}=u_{2n+1}.
\end{equation}
$g_n$ is the coefficient in $g(x)=(\frac{1+x}{1-x})^{1/6}=\sum_{n=0}^\infty g_n x^n$. Finally, $\Delta_n=\bar{M}^-_{nm}/(r-s)$ contained in the diagonal($r=s$) term should be evaluated using $\Delta_n=\frac{2}{3}\sum_{k=0}^n(-1)^{n-k}g_{n-k}^2$. 
We note, in addition, that the above coefficients differ from those appearing in Ref. \cite{gross1987operator} by factors of $i$. This is a consequence of the fact that we deal with free fermions with (anti-)periodic boundary conditions rather than open boundary conditions and hence different conformal maps $\omega_I$ [Eq. \eqref{eqn:conformal_map}] than those in Ref. \cite{gross1987operator}.
One can also show explicitly that the singular terms are indeed $\delta_{IJ}\sum_{r\geq 1/2}e^{-ir(\sigma-\sigma')}$, as required.

%

\end{widetext}

%



\section{Correlation matrix for the vertex state}
\label{app:correlation}

Once the vertex states are obtained,
we can compute various entanglement measures by the correlator method.
Here, we collect some details for
the numerical calculations of the
correlation matrices.
For numerical purposes, we need to truncate the matrix at size $N_c$, and in the direct calculation method we use
the second expression
in Eq.~\eqref{K and V}
to compute $K$ in order to avoid singularities (singularities become less problematic for larger $N_c$).
Then,
$\bm{K}$ is a $6N_c\times 6N_c$ real anti-symmetric matrix, so we can use an orthogonal matrix $Q$ to bring it to standard block diagonal form
\begin{align}
  \bm{K}=Q^T \Sigma Q,
  \quad
\Sigma=\oplus_{k=1}^{3N_c} \Sigma_k,
\quad
\Sigma_k=
\left(
\begin{array}{cc}
0 &\sigma_k\\
-\sigma_k & 0
\end{array}
\right).
\end{align}
In the block diagonal basis $\bm{b}^\dag=Q\bm{V}^\dag$,
the state $|G\rangle$ is
\begin{align}
  |G\rangle
  &=
  \mathcal{N}
  \exp{\left[\frac{1}{2}(\bm{b}^\dag)^T \Sigma\bm{b}^\dag \right]}|0\rangle
  \nonumber \\
  &=
  \mathcal{N}
  \exp{\left[\sum_{k=1}^{3N_c}\sigma_k b_{2k-1}^\dagger b_{2k}^\dagger\right]}|0\rangle.
\end{align}

In order to calculate
the entanglement entropy and negativity, we need to compute the correlation matrices $C$ and $F$. The non-zero elements are
\begin{align}
  &\langle G|b_{2k-1}^\dagger b_{2k}^\dagger|G\rangle
  =
  -\langle G|b_{2k}^\dagger b_{2k-1}^\dagger|G\rangle
    \nonumber \\
  &\quad =-\langle G|b_{2k-1} b_{2k}|G\rangle
    =
    \langle G|b_{2k} b_{2k-1}|G\rangle
    =\frac{\sigma_k}{1+\sigma_k^2},
    \nonumber \\
  &\langle G|b_{2k-1}^\dagger b_{2k-1}|G\rangle
    =
    \langle G|b_{2k}^\dagger b_{2k}|G\rangle
    =\frac{\sigma_k^2}{1+\sigma_k^2},
    \nonumber \\
  &\langle G|b_{2k-1} b_{2k-1}^\dagger|G\rangle
    =
    \langle G|b_{2k} b_{2k}^\dagger|G\rangle
    =\frac{1}{1+\sigma_k^2},
\label{fstringmode}
\end{align}
and the correlation matrices $C,F$ are expressed as
\begin{align}
  C_{rs}&=\langle G|V_r^\dag V_s|G\rangle
          =\langle G|b_{p}^\dagger b_{q}|G\rangle Q_{pr}Q_{qs}
  \nonumber \\
  &=\sum_{k=1}^{3N_c}\frac{\sigma_k^2}{1+\sigma_k^2}(Q_{2k-1,r}Q_{2k-1,s}+Q_{2k,r}Q_{2k,s}),
  \nonumber \\
  F_{rs}&=\langle G|V_r^\dag V^\dag_s|G\rangle
          =\langle G|b_{p}^\dagger b_{q}^\dagger|G\rangle Q_{pr}Q_{qs}
  \nonumber \\
  &
  =\sum_{k=1}^{3N_c} \frac{\sigma_k}{1+\sigma_k^2}(Q_{2k-1,r}Q_{2k,s}-Q_{2k,r}Q_{2k-1,s}).
\end{align}
These correlators need to
be rotated back to the original basis $ f_A, f_B, f_C$
by unitary transformation $U$. Noting that $ f^\dag$ transforms with $U^*$ rather than $U$, the full transformation matrix $U'$ is
\begin{equation}
\begin{aligned}
U'&=U^*\otimes\left(
\begin{array}{cc}
\mathbbm{1} & 0\\
0 & 0
\end{array}
\right)+
U \otimes\left(
\begin{array}{cc}
0 & 0\\
0 & \mathbbm{1}
\end{array}
\right)\\
\end{aligned}
\end{equation}
where $\mathbbm{1}$ is
the $N_c\times N_c$ identity matrix. The correlation matrices transform via
\begin{equation}
  C\to (U')^\dag C U',
  \quad
  F\to (U')^\dag F (U')^*.
\end{equation}
With $C,F$, we can obtain the correlation matrix $\Gamma$
using Eq.\ \eqref{eqn:GammaCF} and compute various entanglement measures.

\bibliographystyle{ieeetr}
\bibliography{vertex}


\end{document}